\documentclass[twocolumn]{aastex631}

\usepackage{epsfig}
\usepackage{graphicx}
\usepackage{amsthm}
\usepackage{amssymb}
\usepackage{rotating}
\usepackage{multirow}
\usepackage{longtable}
\usepackage{savesym}
\savesymbol{tablenum}
\restoresymbol{SIX}{tablenum}
\usepackage{booktabs}
\usepackage{xcolor}

\defcitealias{2019ApJ...873...51L}{Paper~I}
\defcitealias{2020ApJ...888...98L}{Paper~II}
\defcitealias{2021ApJ...923..198D}{Paper~III}
\defcitealias{2022ApJ...933...60D}{Paper~IV}
\defcitealias{2023ApJ...949...82D}{Paper~V}
\defcitealias{2024ApJ...963...55D}{Paper~VI}

\makeatletter
\renewcommand{\paragraph}{\@startsection{paragraph}{4}{0ex}%
   {-3.25ex plus -1ex minus -0.2ex}%
   {1.5ex plus 0.2ex}%
   {\normalfont\small\itshape\centering}}
\makeatother

\stepcounter{secnumdepth}
\stepcounter{tocdepth}

\accepted{by ApJ February 27, 2025}

\shorttitle{Surveying G\ion{H}{2} Regions: VII. Galactic Center Regions Sgr~B1, Sgr~B2, and Sgr~C}
\shortauthors{De Buizer et al.}

\begin{document}

\title{Surveying the Giant \ion{H}{2} Regions of the Milky Way with SOFIA: VII. Galactic Center Regions Sgr~B1, Sgr~B2, and Sgr~C}

\email{jdebuizer@seti.org}

\author[0000-0001-7378-4430]{James M. De Buizer}
\affil{Carl Sagan Center for Research, SETI Institute, 339 Bernardo Avenue, Suite 200, Mountain View, CA 94043, USA}

\author[0000-0003-4243-6809]{Wanggi Lim}
\affil{IPAC, Mail Code 100-22, Caltech, 1200 East California Boulevard, Pasadena, CA 91125, USA}

\author[0000-0003-0740-2259]{James T. Radomski}
\affil{\textit{SOFIA}-USRA, NASA Ames Research Center, Mail Stop 232-12, Moffett Field, CA 94035, USA}

\author[0000-0003-3682-854X]{Nicole Karnath}
\affil{Space Science Institute, 4765 Walnut Street, Suite B, Boulder, CO 80301, USA}

\begin{abstract}
This study examines the mid-infrared properties of Giant \ion{H}{2} (G\ion{H}{2}) regions in the Milky Way's Central Molecular Zone (CMZ) -- Sgr~B1, Sgr~B2, and Sgr~C -- using SOFIA-FORCAST imaging at 25 and 37\,$\mu$m. It compares these mid-infrared data with previous multi-wavelength observations to explore their present star formation activity and global properties. The study identifies 77 massive young stellar object (MYSO) candidates in and around the three regions. Sgr~B2 appears to host the youngest MYSOs and have much higher extinction than the other regions, containing several radio sources not detected in the mid-infrared even at 37\,$\mu$m. Meanwhile, cm radio continuum regions of Sgr~B1 shows remarkable correspondence to its mid-infrared emission. Sgr~C has fewer confirmed MYSOs, and seems to have a higher fraction of low-mass young stellar objects and contamination from more evolved interloper/foreground stars. Derived MYSO densities are consistent with G\ion{H}{2} regions elsewhere in the Galactic plane, though the CMZ G\ion{H}{2} regions appear to have less prolific present star formation overall. Unlike Sgr~B2, the cm continuum emission in Sgr~B1 and Sgr~C G\ion{H}{2} regions appears to be absent cold dust and molecular gas, suggesting environmental differences, possibly driven by turbulence and rapid dynamical changes near the Galactic Center. Furthermore, unlike typical G\ion{H}{2} regions, Sgr~B1 and Sgr~C are significantly ionized by evolved interloper stars, which likely did not form within these regions. In these ways, Sgr~B1 and Sgr~C deviate from classical G\ion{H}{2} region behavior, thus potentially representing a new category of G\ion{H}{2} region or challenging their classification as G\ion{H}{2} regions. 
\end{abstract}


\section{Introduction} 

Giant \ion{H}{2} (G\ion{H}{2}) regions are home to extremely massive young and forming OB stellar clusters. They contain a significant fraction of the most massive stars in a galaxy and therefore can dominate a galaxy's thermal emission \citep{1980A&A....90..246I}. The study of G\ion{H}{2} regions is not only important in understanding and interpreting observations of external galaxies, but they are crucial to the understanding of the formation and evolution of massive stars and their environments.

Massive stars provide critical insights into the processes of stellar evolution. Their strong winds and radiation influence the interstellar medium \citep[e.g.,][]{1980Sci...208....9C}, triggering star formation \citep[e.g.,][]{1994A&A...290..421W} and shaping the structure of galaxies \citep[e.g.,][]{2013ApJ...774...47L}. Understanding how massive stars form and evolve can shed light on the initial mass function and star formation theories. This knowledge helps in understanding the overall population of stars in different galactic environments.

To understand G\ion{H}{2} regions and their evolution, as well as their population of presently-forming massive young stellar objects (MYSOs), we have embarked on a survey of G\ion{H}{2} regions within the Milky Way using observations obtained from the Stratospheric Observatory For Infrared Astronomy (SOFIA) and its mid-infrared instrument FORCAST. The images from FORCAST are the highest spatial resolution (non-saturated) mid-infrared observations of the entirety of these G\ion{H}{2} regions yet obtained at wavelengths beyond $\sim$10\,$\mu$m (i.e., $\lesssim$3$\arcsec$). Most G\ion{H}{2} regions are optically obscured, and radio continuum observations do not have the ability to trace the very earliest stages of massive star formation or non-ionizing sources. Furthermore, because of typically high levels of extinction, often the extended emission of the G\ion{H}{2} regions, as well as the stars forming within them are undetectable at even near-infrared wavelengths. For these reasons the mid-infrared data from FORCAST are crucial to our analyses. 

Our first three papers on W51A (Lim \& De Buizer 2019; hereafter \citetalias{2019ApJ...873...51L}), M17 (Lim \& De Buizer 2020; hereafter \citetalias{2020ApJ...888...98L}) and W49A (De Buizer et al. 2021; hereafter \citetalias{2021ApJ...923..198D}) covered three of the top eight most-powerful G\ion{H}{2} regions in the Milky Way and established the analyses that would be applied throughout the rest of the papers in this survey, allowing us to compare and contrast regions effectively. Our previous papers have focused on understanding the massive stellar content of the presently-forming generation of stars in G\ion{H}{2} regions and trying to understand the internal evolution of the G\ion{H}{2} regions as well as their origins.

The original source list for our survey came from \citet{2004MNRAS.355..899C}, who performed an analysis of all-sky observations of bright and large cm-wavelength radio continuum sources which they cross-correlated with Midcourse Space Experiment (MSX) mid-infrared imaging data as well as Infrared Astronomical Satellite (IRAS) data. In order to be classified as a bonafide G\ion{H}{2} region \citep[according to][]{2004MNRAS.355..899C}, a source had to have a Lyman continuum photon rate in excess of 10$^{50}$ photons/s (as derived from its radio continuum emission) and stand out as a bright region in the mid-infrared. In order to try to understand the population of G\ion{H}{2} regions as a whole, we wanted to compare and contrast the properties of the most powerful G\ion{H}{2} regions (as covered in \citetalias{2019ApJ...873...51L}--\citetalias{2021ApJ...923..198D}) to those that were more modest. In De Buizer et al. (2022; hereafter \citetalias{2022ApJ...933...60D}), we investigated the properties of two G\ion{H}{2} regions (Sgr~D and W42) that we believed were near the Lyman continuum photon rate cut-off. However, we found out that updated distance measurements since the publication of the \citet{2004MNRAS.355..899C} survey placed both objects much closer to the Earth, disqualifying them from being classified as legitimate G\ion{H}{2} regions. Motivated by this, in this same paper we re-investigated all the G\ion{H}{2} regions from \citet{2004MNRAS.355..899C}, pouring through the intervening two decades of literature and extracting new data on the distances and electron temperatures of each source. This led to a new updated list of 31 legitimate Galactic G\ion{H}{2} regions, and 11 candidate G\ion{H}{2} regions.   

In De Buizer et al. (2023; hereafter \citetalias{2023ApJ...949...82D}), we were finally able to properly chose two modest G\ion{H}{2} regions just above the 10$^{50}$ photons/s Lyman continuum photon rate cut-off criteria (DR7 and K3-50) to compare to the most powerful G\ion{H}{2} regions of our previous studies. In this work, we discovered that perhaps the biggest difference between G\ion{H}{2} regions was not necessarily the Lyman continuum photon rate itself, but the contribution to the overall Lyman continuum photon rate by the presence (or lack) of a revealed stellar OB cluster from an earlier epoch of star formation. Those G\ion{H}{2} regions dominated by older OB clusters tended to carve cavities in their host molecular cloud, and the Lyman continuum photons predominantly arise from these ionized cavity walls instead of from internally ionized \ion{H}{2} or compact \ion{H}{2} regions surrounding nascent massive stellar clusters or individual MYSOs. In light of this, we introduced in De Buizer et al. (2024;  hereafter \citetalias{2024ApJ...963...55D}) two G\ion{H}{2} morphological types, one characterized by dispersed radio sub-regions (i.e., ‘distributed-type’) and the other marked by contiguous cavity structures (i.e., ‘cavity-type’), with both types being most easily discernible in their mid-infrared emission.

That brings us to the present paper, where we have chosen to compare and contrast our previous observations of G\ion{H}{2} regions to Sgr~B and Sgr~C, two regions residing in the Milky Way's Galactic Center. Sgr~B and Sgr~C lie at 100 and 90 pc, respectively, from our Galaxy's supermassive black hole \citepalias{2022ApJ...933...60D} and exist in an environment quite distinct from the G\ion{H}{2} regions farther out in the spirals arms of the Galactic disk. Massive stars forming in G\ion{H}{2} regions create swift-moving shock and ionization fronts and produce turbulence from powerful winds and outflows, leading to their classification as ``extreme star formation environments". However, the overall environmental conditions in the Galaxy's Central Molecular Zone (CMZ) are even more extreme. Gas temperatures are high in the CMZ \citep[$\gtrsim$50\,K;][]{2013A&A...550A.135A,2016A&A...586A..50G} as are thermal pressures \citep[$\sim$10$^{-10}$ erg cm$^{-3}$;][]{1996ARA&A..34..645M}. With values measured ranging between $\sim$10\,$\mu$G to $\gtrsim$1\,mG, the CMZ is pervaded with a stronger magnetic field than the Galactic disk which is typically $\sim$few $\mu$G \citep{2019A&A...630A..74M, 2009A&A...505.1183F}. With the addition of tidal forces, shearing forces, crossing orbits from material streams, and much higher overall levels of turbulence in the CMZ, this means that stars forming in these Galactic Center G\ion{H}{2} regions experience the most extreme conditions anywhere in the Galaxy. These turbulent motions translate to large line widths (i.e., $\sigma$) varying from 0.6 to 20 km/s over size scales of 0.2 to 2\,pc \citep{2018ApJ...868....7M}, which may affect how we can apply our evolutionary analyses which rely (in part) on these values. Additionally, metallicity ($Z_{\sun}$) is higher in the CMZ by almost a factor of two \citep{2023ASPC..534...83H, 2011ApJ...738...27B} which affects a G\ion{H}{2} region's electron temperature. Therefore, armed with the analysis tools and results from our prior work in this series of papers, our goal in this paper is to investigate the properties of the G\ion{H}{2} regions in Sgr~B and Sgr~C and compare and contrast them to the G\ion{H}{2} regions we have studied that exist out in the relatively calm backwaters of our Galaxy.

\section{Observations and Data Reduction} \label{sec:obs}

The SOFIA data used for this study were originally taken as part of the SOFIA/FORCAST Galactic Center Legacy Survey,  and not taken as part of our G\ion{H}{2} region observing program. The details regarding the data acquisition, reduction, and calibration are all described in \citet{2020ApJ...894...55H}. However, the observational techniques and data reduction processes employed on the data were, for the most part, identical to those that we have utilized in all of our G\ion{H}{2} survey papers to date (see, e.g., \citetalias{2019ApJ...873...51L}). That being said, we will highlight below some of observation and reduction details specific to these particular observations. 

Data were taken with SOFIA's FORCAST instrument in its dual-channel mode, which utilizes an internal dichroic to simultaneously observe at two mid-infrared wavelengths at once. Unlike the majority of the observations of our G\ion{H}{2} region survey, in which data were taken simultaneously at 20 and 37\,$\mu$m, the Galactic Center Legacy Survey used a setup that took simultaneous data at 25 and 37\,$\mu$m. The slight wavelength difference between 20 and 25\,$\mu$m should have no appreciable affect on our analyses or our ability to compare and contrast the results with the other G\ion{H}{2} regions we have already studied in our previous papers. The final FORCAST Galactic Center Legacy maps were made by mosaicking individual fields (or ``pointings'') together. While most fields in our G\ion{H}{2} survey had typical exposure times between 180 and 300s, the Galactic Center Legacy data were typically deeper exposures, varying from 200 to 600s. The nominal spatial resolutions of the 25 and 37\,$\mu$m data are 2.6$\arcsec$ and 3.4$\arcsec$, respectively\footnote{From the FORCAST Handbook for Archive Users, available at IRSA: https://irsa.ipac.caltech.edu/data/SOFIA/docs/instru-ments/forcast/}. From \cite{2020ApJ...894...55H} it is stated that the astrometry of the final SOFIA mosaic of the entire Galactic Center is at worst $\sim$2$\arcsec$, which is slightly higher but similar than the quoted astrometric accuracy of our previous G\ion{H}{2} region studies (i.e., 1.5$\arcsec$). The slightly higher error is due to the large size of the FORCAST Galactic Center mosaic map, as astrometric errors tend to compound for larger mosaics.

\begin{figure*}[tb!]
\epsscale{0.80}
\plotone{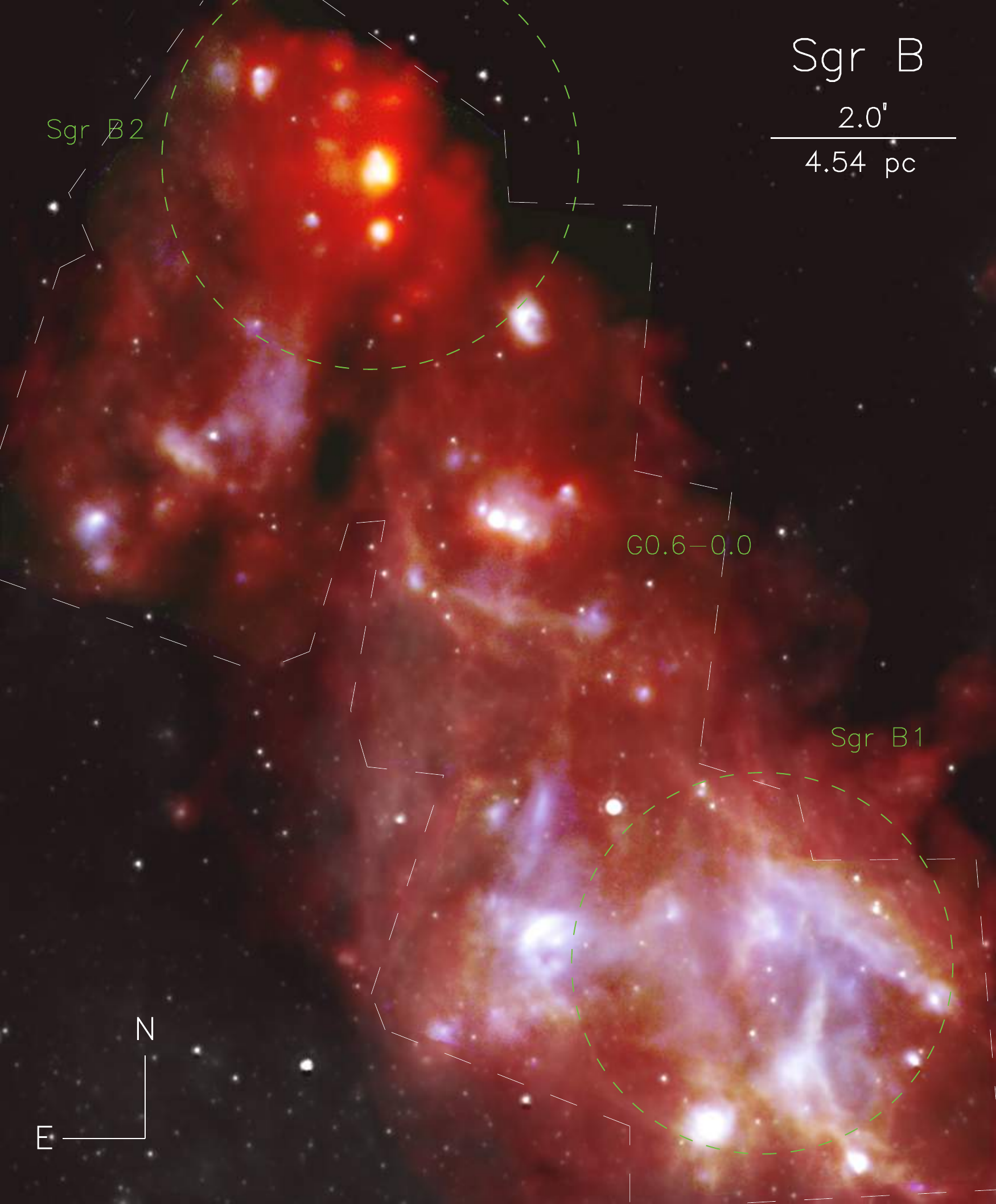}
\caption{ A 4-color image of the central $\sim$10$\farcm$8$\times$13$\farcm$0 (24.5$\times$29.5\,pc) region of Sgr~B. Blue is the SOFIA-FORCAST 25\,$\mu$m image, green is the SOFIA-FORCAST 37\,$\mu$m image, and red is the Herschel-PACS 70\,$\mu$m image. Overlaid in white is the Spitzer-IRAC 8.0\,$\mu$m image, which traces the revealed stars within Sgr~B, field stars, and hot dust. The green dashed circles
denote the locations and extent of the Sgr~B1 and Sgr~B2 \ion{H}{2} radio continuum regions as reported in \citet{1980A&AS...40..379D}. The emission extending between Sgr~B1 and B2 is called G0.6-0.0. The white dashed lines denote the areas covered buy the SOFIA mid-infrared map.
\label{fig:fig1}}
\end{figure*}

Our maps of Sgr~B were cropped from the final FORCAST Galactic Center Legacy maps from \citet{2020ApJ...894...55H}. We had to apply a slight positive flux offset to the cropped images (15\,mJy/pixel at 25\,$\mu$m and 13\,mJy/pixel at 37\,$\mu$m) in order to avoid having multiple areas with negative backgrounds. For Sgr~C, additional data of were taken as part of that SOFIA/FORCAST Galactic Center Legacy Survey, but not incorporated into the final published map. Therefore, for Sgr~C, we reduced the data ourselves to create the final image mosaics used in this study, again in a manner very similar to that described in \citetalias{2019ApJ...873...51L}. Data for Sgr~C were combined using the last released public version of SOFIA Redux software\footnote{https://sofia-usra.github.io/sofia\_redux/sofia\_redux/index.html}. However, all FORCAST data presently in the NASA/IPAC Infrared Science Archive (IRSA) were reduced with a newer version of the software than this. In particular, the last released public version does not contain the latest calibration values (CALFCTR) corrected for precipitable water vapor, so for Sgr~C the values had to be updated based on CALFCTR values from the header keywords in the data in the IRSA archive from when the data was initially processed. All images were obtained by the FORCAST instrument by employing the standard chop-nod observing technique used in ground-based thermal infrared observing, with chop throws of up to 5.75$\arcmin$ and nod throws of up to 7$\arcmin$ in order to be sufficiently large enough to sample clear off-source sky regions uncontaminated by the extended emission of Sgr~C. The mid-infrared emitting area of Sgr~C was mapped using four pointings. Each of these fields had an average on-source exposure time of between 240-400\,s at both 25\,$\mu$m and 37\,$\mu$m. The SOFIA Data Pipeline software produced the final mosaicked images (Level 4 data products) from the individual pointings, and these final mosaicked images are presented and used here in this work. For Sgr~C, these final mosaics also had their astrometry absolutely calibrated using Spitzer data by matching up the centroids of point sources in common between the Spitzer and SOFIA data using Aladin Sky Atlas\footnote{https://aladin.cds.unistra.fr}. Absolute astrometry of the final Sgr~C SOFIA data is assumed to be better than $1\farcs5$, which is slightly better than the Sgr~B astrometric accuracy quoted above. Flux calibration for the Sgr~C data was provided by the SOFIA Data Cycle System (DCS) pipeline and the final total photometric errors in the images were derived using the same process described in \citetalias{2019ApJ...873...51L}. It is assumed that the photometric errors are the same for Sgr~B and Sgr~C, and these errors are discussed more in Section \ref{sec:cps}.

\begin{figure*}[tb!]
\epsscale{0.8}
\plotone{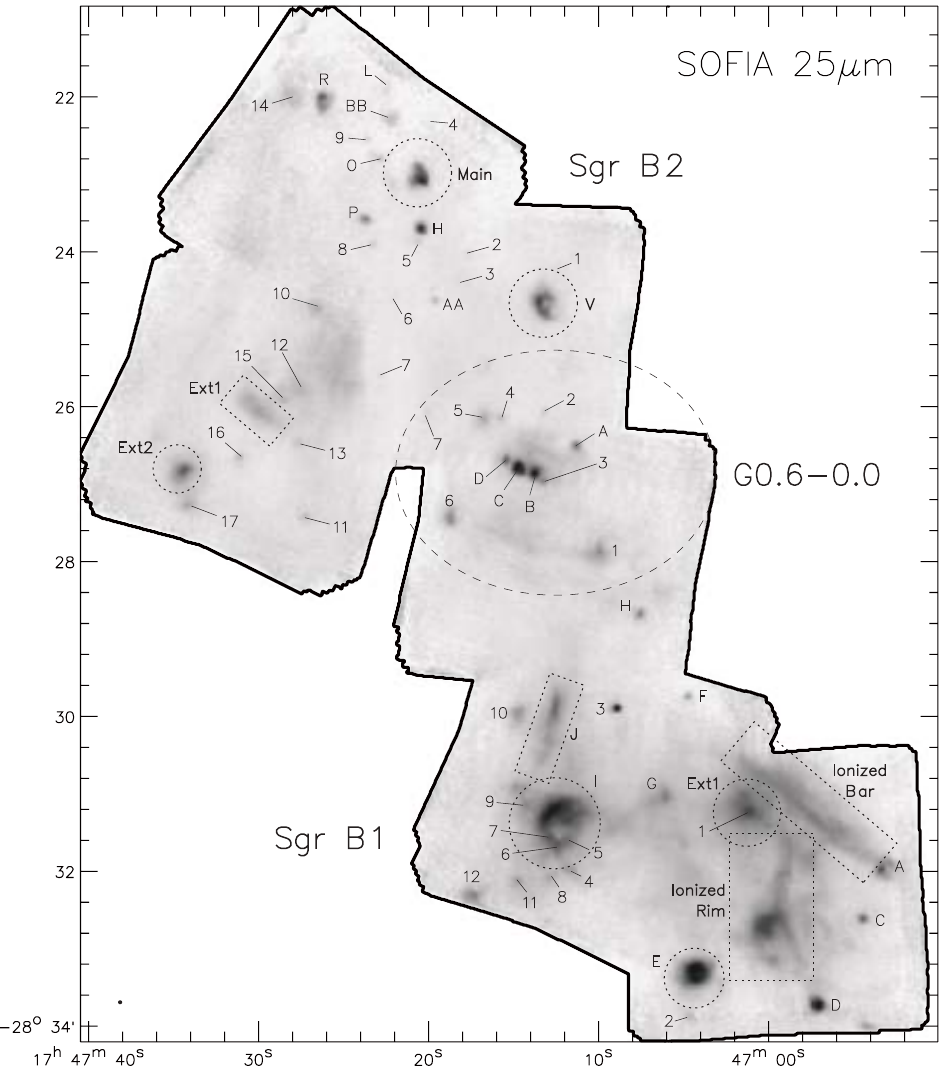}
\caption{Sgr~B image mosaic taken at 25\,$\mu$m by SOFIA shown in inverse color (i.e. brighter features are darker in color). The locations of radio continuum sources are labeled with letters and are from \citet{1992ApJ...401..168M} and \citet{1993ApJ...412..684M}. Sources labeled with numbers are the mid-infrared compact sources identified in this work. Extended infrared and/or radio sources are encompassed by dotted lines and are labeled in bold. A dashed ellipse surrounds the region known as G0.6-0.0, and Sgr~B2 lies north and east of this region, and Sgr~B1 lies to the south. The black dot in the lower left indicates the resolution of the image at this wavelength.\label{fig:fig2}}
\end{figure*}

In addition to the SOFIA data, we also utilize science-ready imaging data from the \textit{Spitzer Space Telescope} and \textit{Herschel Space Telescope} archives, which we will discuss more in Section~\ref{sec:cps}. For Sgr~B, we also utilized the 6\,cm maps of \citet{1992ApJ...401..168M} and \citet{1995ApJS...97..497M}, and downloaded additional 6\,cm and 20\,cm data from the \textit{Very Large Array (VLA)} archive, with 19.0 and 2.9$\arcsec$ spatial resolution, respectively. For Sgr~C, we used the 6 and 20\,cm radio data, also from the VLA archive, which has 4.5 and 11.0$\arcsec$ spatial resolution, respectively.

\begin{figure*}[tb!]
\epsscale{0.8}
\plotone{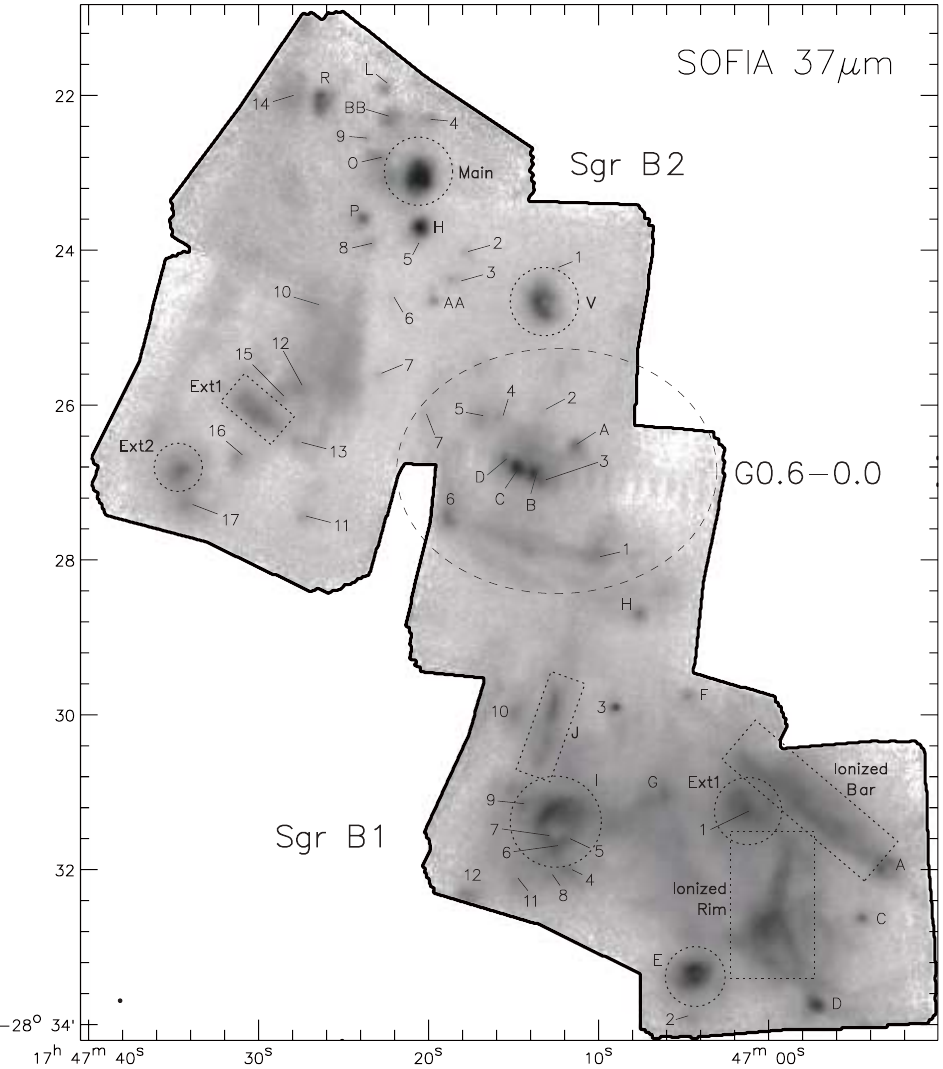}
\caption{Sgr~B image mosaic taken at 37\,$\mu$m by SOFIA. See caption of Figure~\ref{fig:fig2} for explanation of symbols and figure annotation.\label{fig:fig3}}
\end{figure*}

\begin{figure*}[tb!]
\epsscale{0.75}
\plotone{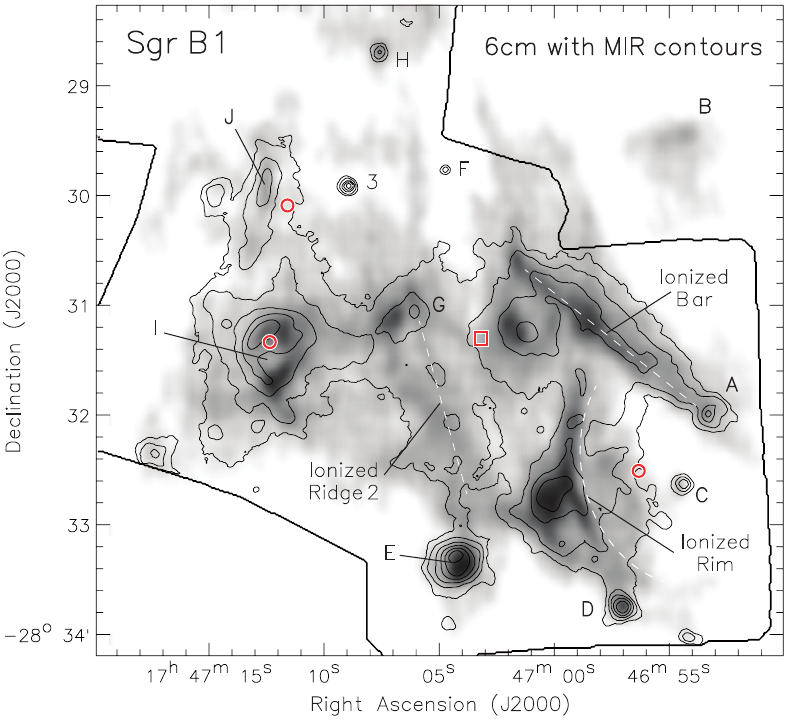}
\caption{The 6\, cm radio continuum image from \citet{1993ApJ...402L..69M} with the SOFIA 25\,$\mu$m contours overlaid. Labels indicate the Sgr~B1 radio sources identified by and discussed in \citet{1992ApJ...401..168M}, except for the source marked 3, which is a prominent infrared source with no detectable cm radio emission. The red circles denote the locations of the three Wolf-Rayet stars, and the red square denotes the location of the O supergiant, which were identified by \citet{2010ApJ...710..706M}. The dark lines around the 25\,$\mu$m contours show the field covered by the SOFIA data.  \label{fig:SgrB1radio}}
\end{figure*}

\begin{figure*}[t]
\epsscale{1.17}
\plotone{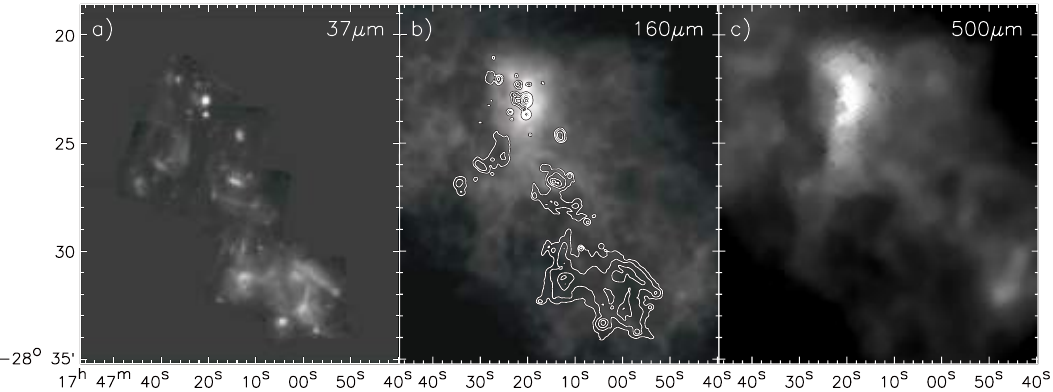}
\caption{Sgr~B region at a) 37\,$\mu$m, b) 160\,$\mu$m (with 37\,$\mu$m contours), and c) 500\,$\mu$m. The morphology of Sgr~B appears very similar to the 37\,$\mu$m image for wavelengths from 8 to 70\,$\mu$m. The morphology is also very similar from 160-500\,$\mu$m, but very different than the morphology from 8 to 70\,$\mu$m. Most of the emitting regions seen at 37\,$\mu$m appear to be devoid of cold dust at 160-500\,$\mu$m, except for the central region of Sgr~B2. In particular, the entire Sgr~B1 G\ion{H}{2} region appears to have no 37\,$\mu$m emission source or feature that is also emitting at 160\,$\mu$m or longer.\label{fig:SgrB_evo}}
\end{figure*}

\section{Comparing SOFIA Images to Previous Imaging Observations} \label{sec:results1}

As we have done in previous papers in this series, below we will review the (mainly qualitative) comparisons between the SOFIA mid-infrared images and those from prior infrared, radio, and/or submillimeter observations. However, in these previous studies we also dedicated a section that went into detail about each of the individual sources and sub-regions with each G\ion{H}{2} region, using the new and old observations to try to gain a better understanding of the nature of each individual source. Most of those previous papers concentrated on a single G\ion{H}{2} region, and since we are covering three G\ion{H}{2} regions in this work, we will not cover each at that same level of detail. Below we discuss the large scale comparisons of each region and forego much of the background discussion of individual sources.

\subsection{Sgr~B}
At radio wavelengths, Sgr~B is second brightness region in the Galactic Center with only Sgr~A being brighter \citep{2011AJ....141...82J}. Sgr~B is composed of two main sub-regions: Sgr~B1 and Sgr~B2 (see Figures \ref{fig:fig1}, \ref{fig:fig2}, and \ref{fig:fig3}). The radio continuum emission from Sgr~B1 mainly comes from diffuse \ion{H}{2} emission from many large and elongated bar-like and shell-like structures, though it does contain a number of modestly bright compact components. Sgr~B2, by comparison, is dominated by several dozens of compact \ion{H}{2} regions \citep{1995ApJ...449..663G, 2022A&A...666A..31M} with only a modest amount of extended radio continuum emission linking them together. \citet{1992ApJ...401..168M} define a third region, G0.6-0.0, which lies between the peaks of Sgr~B1 and Sgr~B2 (Figures \ref{fig:fig1}, \ref{fig:fig2}, and \ref{fig:fig3}), and displays components with velocities ($v_{lsr}\sim$ 55\,km/s) between the average velocities of Sgr~B1 and Sgr~B2 \citep[$\sim$45 and $\sim$65\,km/s, respectively;][]{1980A&AS...40..379D}. In reality, higher resolution observations, like those of \citet{1992ApJ...401..168M}, show that regions like Sgr~B1 are made up of multiple concentrations of molecular material with a wide range of velocities ($-40$ to $+80$\,km/s) which complicates the interpretation. Given the fact that these velocity measurements have similar values and/or have overlapping velocity ranges, the claim by \citet{1992ApJ...401..168M} that G0.6-0.0, Sgr~B1, and Sgr~B2 are all likely physically related is still valid.  

It appears that there is much less extinction towards Sgr~B1 than Sgr~B2, leading to speculations that Sgr~B2 may be located just behind Sgr~B1 \citep{1980A&AS...42..163B} and/or that Sgr~B1 is more evolved and has had time to disperse much of its molecular material \citep{1995ApJS...97..497M}. That said, there are more recent claims that Sgr~B1 is just behind Sgr~B2 \citep{2021ApJ...910...59S, 2021ApJ...921...33H}. Maser activity is thought to be linked to star formation activity, and in agreement with the idea that Sgr~B2 is younger than Sgr~B1, Sgr~B2 harbors much more intense maser activity than Sgr~B1 \citep{1993ApJ...402L..69M}.

Sgr~B2 has more than twice the 6\,cm radio continuum flux of Sgr~B1 \citep[53.9 vs. 25.5\,Jy;][]{1980A&AS...40..379D} and has a slightly larger emitting extent (4.5$\arcmin$ vs. 4.1$\arcmin$). Interestingly, however, Sgr~B1 is included in our list of bonafide G\ion{H}{2} regions \citepalias{2022ApJ...933...60D} while Sgr~B2 is not. This is because our list is a refinement of the compilation of G\ion{H}{2} regions made by \citet{2004MNRAS.355..899C}, which cross-correlated the all-sky radio continuum surveys of \citet{1997ApJ...488..224K} with MSX mid-infrared imaging data at 12 and 22\,$\mu$m (as well as IRAS data at 25, 60, and 100\,$\mu$m). Though Sgr~B2 is included as a bright radio continuum source in both the \citet{1997ApJ...488..224K} and \citet{1980A&AS...40..379D} radio surveys, it did not make the cut in the \citet{2004MNRAS.355..899C} compilation of G\ion{H}{2} regions. The reason why Sgr~B2 specifically was not in their final source list was not reported by \citet{2004MNRAS.355..899C}, though it was likely due to Sgr~B2's lack of strong and extended infrared emission. In fact, at 22\,$\mu$m in the MSX images, Sgr~B2 is not discernible as an extended infrared region and the total infrared flux of the few detectable sources (510\,Jy) within its radio emitting region is 7$\times$ fainter than the total infrared flux of Sgr~B1 (3550\,Jy) within its smaller radio emitting region. However, based upon the radio flux of Sgr~B2 alone, we calculate that the Lyman continuum photon rate would be $log(N_{LyC})=$51.04$\pm$0.22 s$^{-1}$, which would make it tied with G338.398+0.164 \citepalias[see][]{2022ApJ...933...60D} as the sixth most powerful G\ion{H}{2} region in the Galaxy (Sgr~B1 at $log(N_{LyC})=$50.87 s$^{-1}$ is twelfth). As we will touch on below (and discuss in more detail in Section \ref{sec:genuine}), we believe that Sgr~B2 is likely to be a genuine G\ion{H}{2} region, and so we will study it in the same depth as Sgr~B1 and Sgr~C.

\subsubsection{Sgr~B1} 

\citet{2009ApJ...705.1548R} determined the distance to the Sgr~B cloud using trigonometric parallax observations of masers, finding a value of $7.8^{+0.8}_{-0.7}$ kpc. Though this distance was actually measured toward Sgr~B2 specifically, Sgr~B1 and Sgr~B2 are believed to be physically related \citep{1992ApJ...401..168M, 2021ApJ...910...59S} since they have similar $v_{lsr}$ velocity ranges, and thus it is assumed here that the maser distance applies to Sgr~B1 as well.

\citet{1992ApJ...401..168M} made some of the first high-resolution ($\sim3-6\arcsec$) radio continuum observations of Sgr~B1 at 3.6 and 6~cm. Their observations reveal that the radio morphology of Sgr~B1 is quite complex and they identify several elongated and/or ridge-like features, including those they labeled as Ridge 1\footnote{Ridge 1 is not covered by our SOFIA maps, and is a weaker radio continuum feature that lies $\sim$1.2$\arcmin$ west of radio source C.} and 2, the Ionized Bar, and Ionized Rim (see Figure~\ref{fig:SgrB1radio}). Apart from these large ridge/rims structures, the radio regions E and I appear to be ionized partial shells with diameters of 0.55 and 1.4\,pc, respectively. \citet{1992ApJ...401..168M} also point out several compact sources or bright peaks, like A, C, D, and H. Interspersed among these larger structures and compact sources is considerable diffuse, extended radio continuum emission. \citet{1992ApJ...401..168M} claim that roughly 75\% of the radio flux density in Sgr~B1 comes from the diffuse and extended emission.

As can be seen in Figure~\ref{fig:SgrB1radio}, the extended radio continuum emission and the extended mid-infrared continuum emission as seen by SOFIA are fairly well-matched, with some minor differences. A couple of the peaks in the radio do not match the peaks seen in the mid-infrared SOFIA data (i.e., sources G and I) and no radio emission is seen at the location of the prominent mid-infrared source Sgr~B1~3 (Figure~\ref{fig:SgrB1radio}). The morphological similarities between radio and infrared emission also extends to the IRAC 5.8 and 8.0\,$\mu$m imaging data (whereas the emission from Sgr~B1 in the IRAC 3.6 and 4.5\,$\mu$m images is highly extinguished and this area of sky instead appears to be dominated by field stars). 

While this correspondence between the radio continuum emission and the mid-infrared is in good agreement, \citet{2021ApJ...910...59S} point out that Sgr~B1 is ``barely noticeable in molecular gas tracers or very cold dust". They also observed that the warm dust emission (as traced by IRAC, SOFIA, and even Herschel 70\,$\mu$m data) does not seem to be associated with molecular line emission, and therefore Sgr~B1 may lack a parent molecular cloud. In particular, they point out that many of the the bright rims and bars within Sgr~B1 are not correlated with high-density gas tracers. They therefore deduce that the infrared and radio structures are not tracing dense and optically thick volumes of material, but instead they are likely to appear bright because of a large amount of optically thin material spread out along the line of sight. Consistent with this idea, when looking at the Herschel $160-500$\,$\mu$m data (Figure \ref{fig:SgrB_evo}), we see that there are no sources/structures corresponding to those seen in the radio and at wavelengths $\le$70\,$\mu$m. The 800\,$\mu$m maps for Sgr~B1 and B2 by \citet{1994ApJ...424..189L} are also similar to the Herschel $160-500$\,$\mu$m data and show that, while Sgr~B2 has copious submillimeter continuum, Sgr~B1 is largely absent in the map. Furthermore, the $^{13}$CO maps by \citet{2024MNRAS.532.4187S} paint the picture of Sgr~B1 existing within the center of a bubble of molecular material.

That being said, Sgr~B1 is not completely devoid of molecular gas, as there are still modest levels present to discern something of its molecular nature. For instance, \citet{1995ApJS...97..497M} say that though it appears much of the molecular gas in Sgr~B1 appears to have dispersed, their observations of formaldehyde in absorption show that some of the ionized material on the eastern side of Sgr~B1 is more distant than that on the western side. 

\begin{figure*}[tb!]
\epsscale{0.75}
\plotone{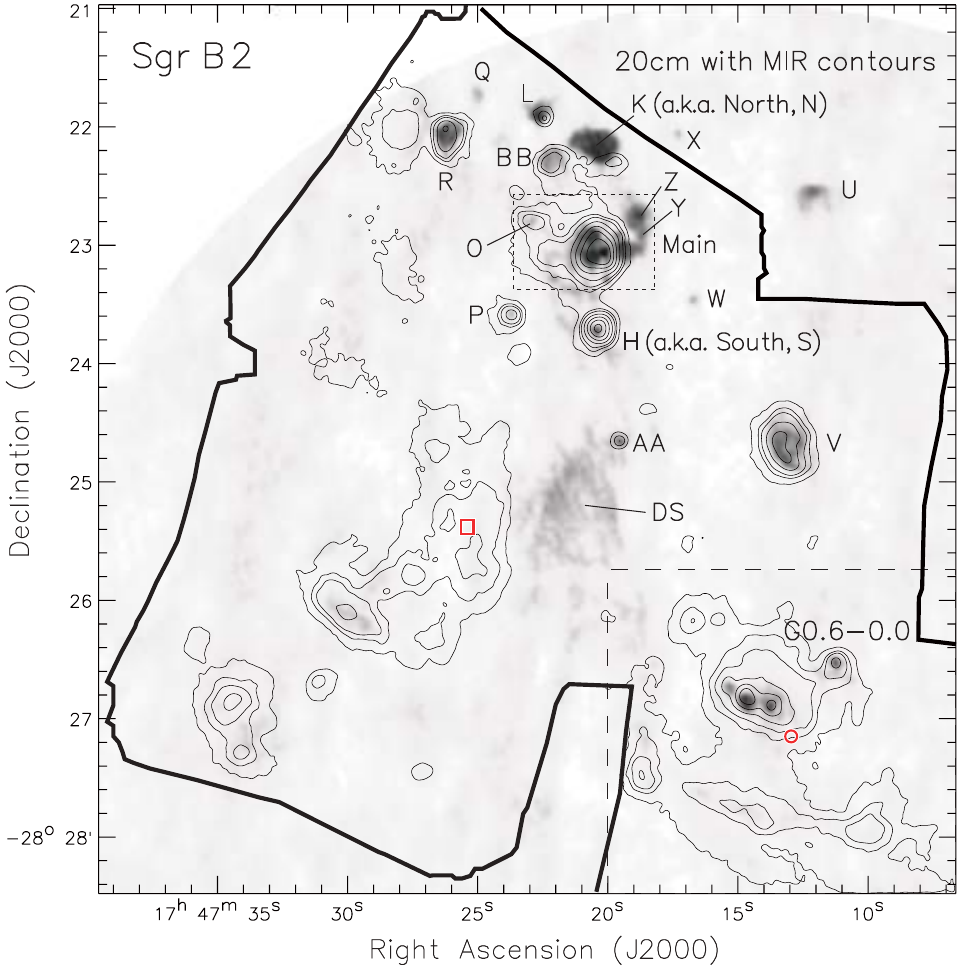}
\caption{The 20\, cm radio continuum image from \citet{1993ApJ...412..684M} with the SOFIA 37\,$\mu$m contours overlaid. Labels indicate the Sgr~B2 radio sources identified by and discussed in \citet{1993ApJ...412..684M}. The dark lines around the 37\,$\mu$m contours show the field covered by the SOFIA data. The dotted region is Sgr~B2~Main, which itself is broken up into many smaller radio sources, but all of which are either unresolved or undetected in the SOFIA data. The region designated G0.6-0.0 is encompassed in the dashed area of the lower right of the image. The red circles denote the locations of the Wolf-Rayet star in G0.6-0.0, and the red square denotes the location of the O supergiant in Sgr~B2, both taken from  \citet{2021A&A...649A..43C}. \label{fig:SgrB2radio}}
\end{figure*}

\subsubsection{Sgr~B2} \label{sec:sgrb}

There is some minor uncertainty about the distance to Sgr~B2. As stated above, we adopted the value of $7.8^{+0.8}_{-0.7}$ kpc which comes from the maser parallax and proper motion studies of \citet{2009ApJ...705.1548R}. In that work they also claim that Sgr~B2 is $\sim$130~pc closer to Earth than Sgr~A*. This result is consistent with other more recent studies that also place Sgr~B2 in front of Sgr~A* \citep{2015MNRAS.447.1059K, 2017MNRAS.469.2251R}. However, \citet{2022ApJ...927...97O} claim that Sgr~B2 lies $\sim$90~pc further way than Sgr~A* based upon spectroscopic measurements. As these differences are all within the margin of error of the trigonometric parallax observations, and since these relatively small distance differences do not considerably change the assessments we are making in this study, we will use the 7.8~kpc value here.  

\begin{figure*}[tb!]
\epsscale{0.8}
\plotone{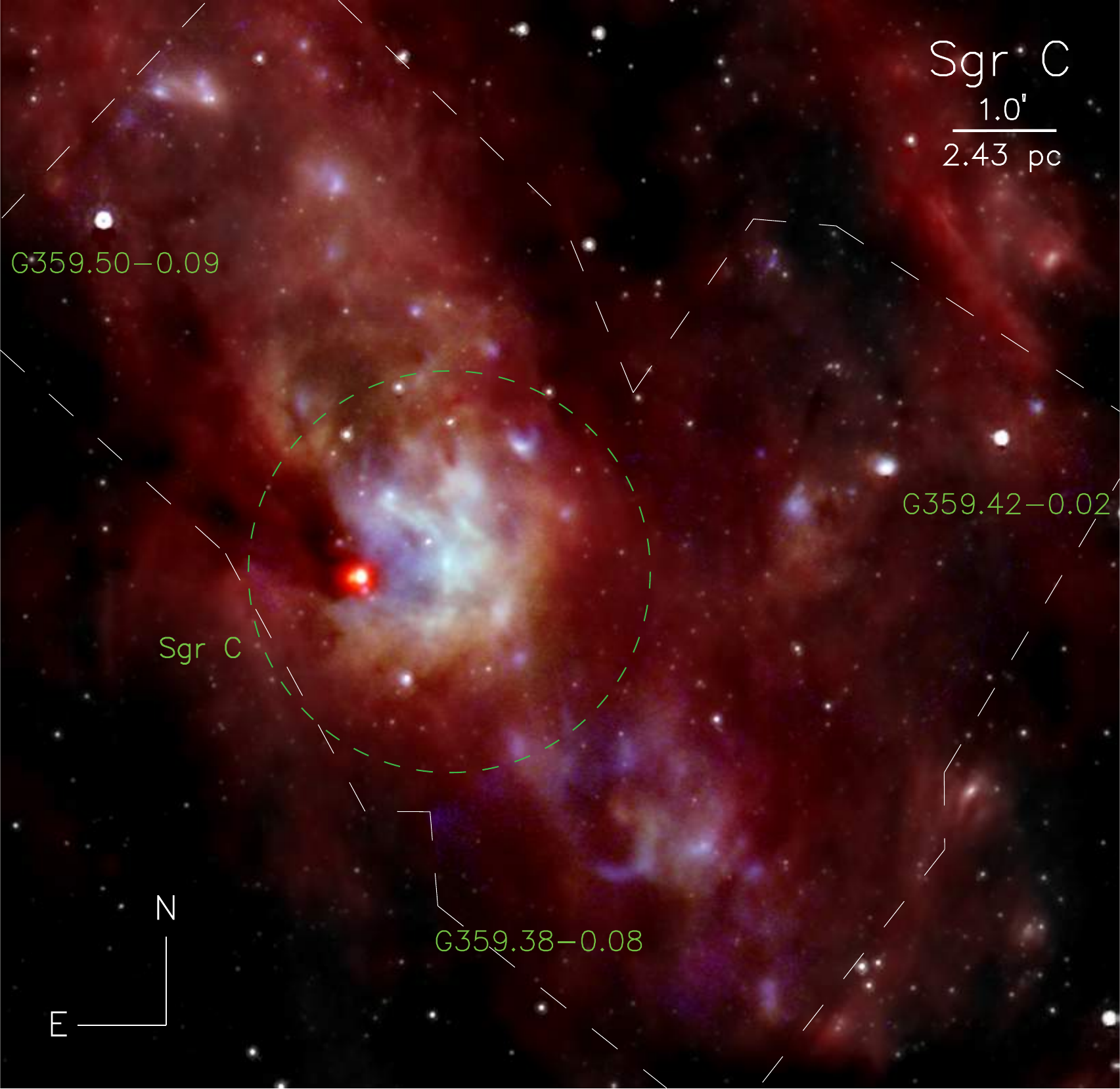}
\caption{ A 4-color image of the central $\sim$10$\farcm$4$\times$10$\farcm$1 (25.3$\times$24.5\,pc) region of Sgr~C. Blue is the SOFIA-FORCAST 25\,$\mu$m image, green is the SOFIA-FORCAST 37\,$\mu$m image, and red is the Herschel-PACS 70\,$\mu$m image. Overlaid in white is the Spitzer-IRAC 8.0\,$\mu$m image, which traces the revealed stars within Sgr~C, field stars, and hot dust. The green dashed circle denotes the location and extent of the \ion{H}{2} radio continuum region as reported in \citet{1980A&AS...40..379D}. The emission extending to the southwest of Sgr~C we refer to as G359.38-0.08. The white dashed lines denote the areas covered buy the SOFIA mid-infrared map.\label{fig:figC1}}
\end{figure*}

The first high-angular resolution radio continuum maps ($<$1$\arcsec$) produced of the Sgr~B2 region were from \citet{1984ApJ...277..181B}, who observed the region at 2 and 6\,cm. They identified the radio sources A$-$L, and were able to resolve Sgr~B2~Main into multiple compact radio components for the first time. This work and the later work of \citet{1990ApJ...351..538G} established the radio emission in Sgr~B2 as coming from many compact \ion{H}{2} regions. Most recently, \citet{2022A&A...666A..31M}, using sub-arcsecond 5\,cm imaging, find 54 ultracompact \ion{H}{2} regions in the central $4\arcmin\times4\arcmin$ of Sgr~B2. In the central 2\,pc of Sgr~B2 there are two well-known and well-studied hot cores, Sgr~B2~Main and Sgr~B2~K (a.k.a. Sgr~B2~North in more recent studies), which \citet{2022A&A...666A..31M} claim contain at least 70 high-mass stars with spectral types from B0$-$O5. These two cores are embedded in a larger molecular clump $\sim$40\,pc in diameter that contains 99\% of the mass of Sgr~B2. Indeed, the first $\sim$1$\arcmin$ observations at far-infrared to millimeter wavelengths (50 and 100\,$\mu$m from \citealt{1977ApJ...211..786H} and \citealt{1978ApJ...220..822G}; 1\,mm from \citealt{1976ApJ...209...94W}), as well as molecular lines ($^{13}$CO from \citealt{1975ApJ...201..352S}) show a similarly-shaped north-south elongated clump present here centered near the position of Sgr~B2~Main. Recently, at 3\,mm using ALMA with $\sim$0.5$\arcsec$ resolution, \citet{2018ApJ...853..171G} identified 271 compact cores across Sgr~B2. Though \citet{2022A&A...666A..31M} claim that there is no direct evidence of the existence of high-mass stars embedded in these 3\,mm cores, we do find two of them to coincident with SOFIA mid-infrared sources (Sgr~B2~4 and Sgr~B2~5). However, as pointed out by \citet{2022A&A...666A..31M}, the reason for such a low detection rate in the infrared may be the high levels of extinction present.

\begin{figure*}[tb!]
\epsscale{0.80}
\plotone{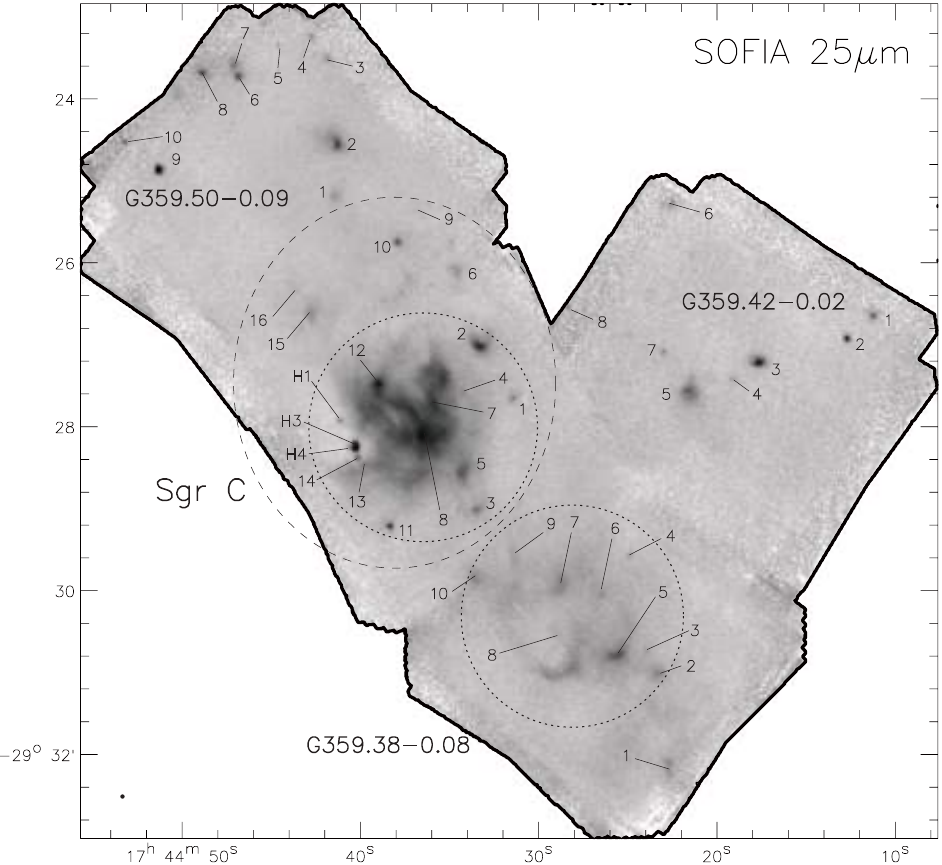}
\caption{Sgr~C image mosaic taken at 25\,$\mu$m by SOFIA shown in inverse color (i.e. brighter features are darker in color). We break up the region into several subregions: Sgr~C (whose bounds are shown by the dashed ellipse), G359.42-0.02 to the west, G359.38-0.08 to the southwest, and G359.50-0.09 to the northeast. The dotted circles denote the apertures used to measure the extended Sgr~C \ion{H}{2} and G359.38-0.08 \ion{H}{2} emission regions. The black dot in the lower left indicates the resolution of the image at this wavelength.\label{fig:C2}}
\end{figure*}

\begin{figure*}[tb!]
\epsscale{0.8}
\plotone{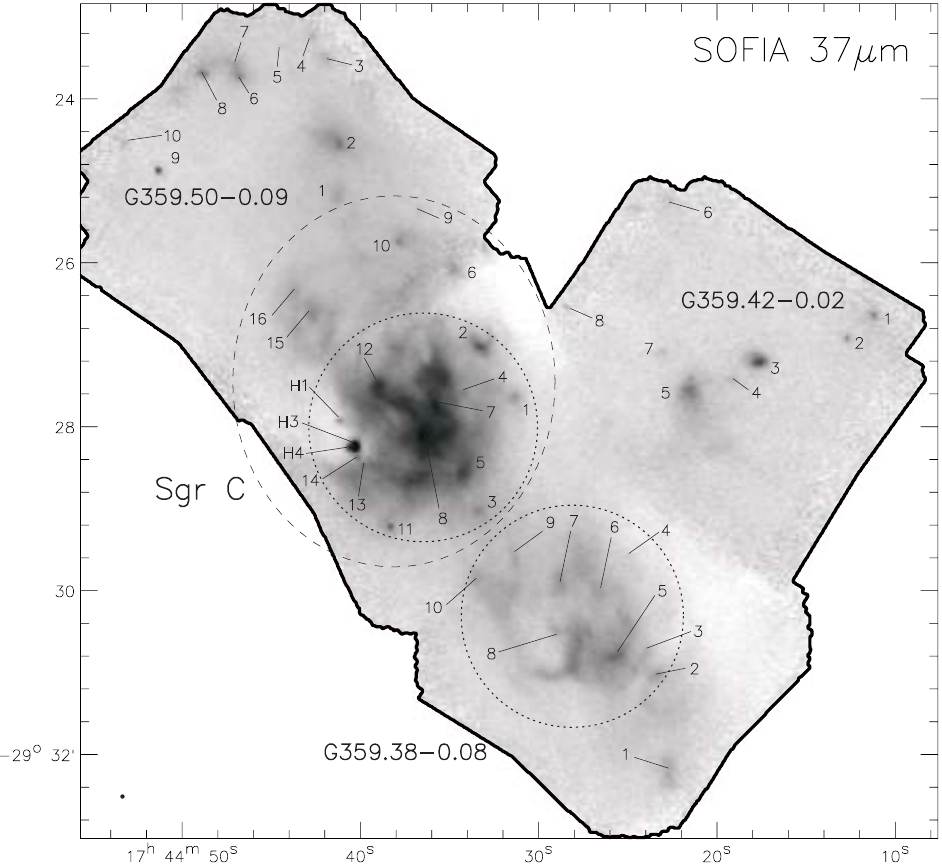}
\caption{Sgr~C image mosaic taken at 37\,$\mu$m by SOFIA. See caption of Figure~\ref{fig:C2} for explanation of symbols and figure annotation.\label{fig:C3}}
\end{figure*}

In Figure \ref{fig:SgrB2radio}, we present the 20\,cm radio continuum image from \citet{1993ApJ...412..684M} with the SOFIA 37\,$\mu$m contours overlaid. Like the case for Sgr~B1, we see that there is generally good morphological agreement between the two wavelengths, however there are several exceptions. Extinction appears to be extremely high to the west of Sgr~B2~Main and we do not see mid-infrared emission from prominent radio sources like K, U, W, Y, and Z. We also see that the diffuse radio continuum emission region south of AA (referred to as Sgr~B2~Deep South, or DS, in recent studies; e.g., \citealt{2018ApJ...853..171G}) is not present in the mid-infrared, and the diffuse mid-infrared region east of it has no radio continuum emission. A possible factor contributing to the non-detection of mid-infrared emission from Sgr~B2~DS is that \citet{2019A&A...630A..73M} show the region to have wide-spread and diffuse non-thermal radio emission, which \citet{2019A&A...630A..72P} claim may be due to thermal electrons getting accelerated up to relativistic energies within the \ion{H}{2} region. Modest radio continuum emission is present at the locations of Sgr~B2~Ext1 and mid-infrared source Sgr~B2~17. Sgr~B2~V also looks similar at both wavelengths, however as pointed out by \citet{1995ApJS...97..497M}, it exhibits a significantly higher velocity (+99\,km/s) compared to the 55$-$80\,km/s velocities observed in the rest of Sgr~B2, suggesting that it is likely not directly associated with the region.

\begin{figure*}[tb!]
\epsscale{0.75}
\plotone{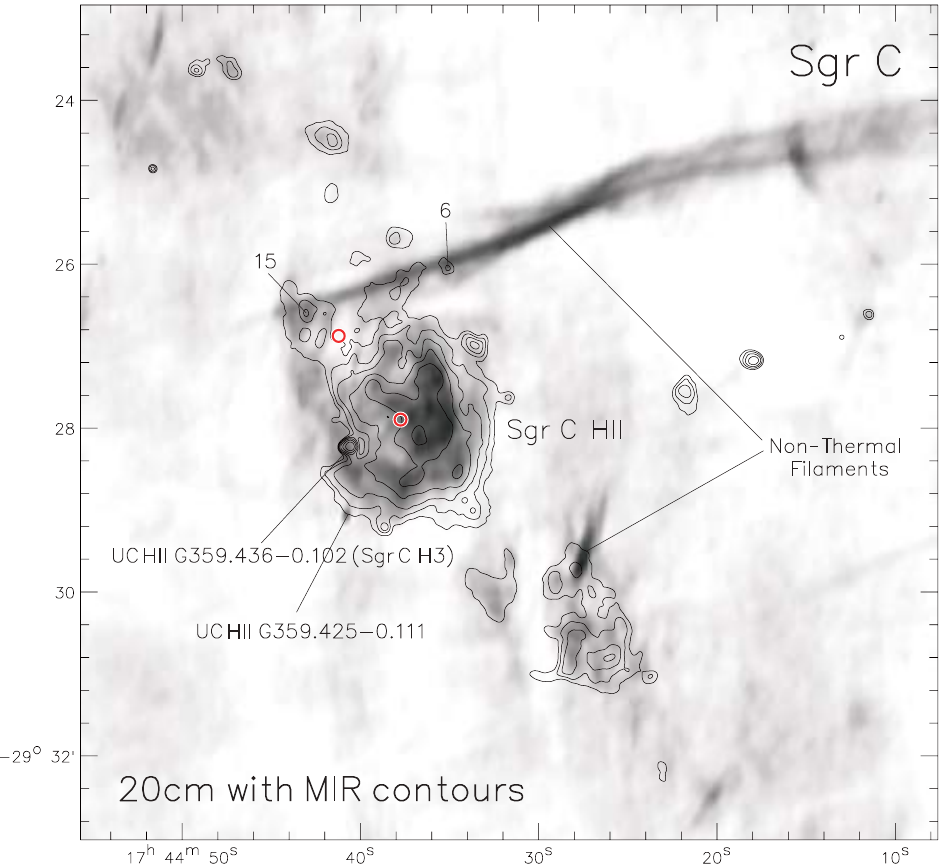}
\caption{The 20\, cm radio continuum image over overlaid with SOFIA 25\,$\mu$m contours. The two ultra-compact \ion{H}{2} regions from \citet{2000ApJ...530..371F}, are labeled, as well as the locations of some non-thermal filaments. The red circles denote the locations of Wolf-Rayet stars from \citet{2021A&A...649A..43C}. \label{fig:SgrCradio}}
\end{figure*}

At $160-500\,\mu$m, the region centered approximately on Sgr~B2~Main and having a radius of $1\farcm5$ appears very similar to the radio (and mid-infrared) morphologically. At wavelengths of 350\,$\mu$m and longer, the radio region south of AA, Sgr~B2~DS, begins to appear as well. This colder material running north-south at this location is designated as an infrared dark cloud (IRDC) by \citet{2024MNRAS.527.1275Y}. It runs through the region occupied by the diffuse radio region south of AA and runs north, passing to the west of Sgr~B2~Main. This extended cold dust structure is also see in the 1.3\,mm observations of \citet{1991ApJ...380..429L}, which displays two prominent peaks near Sgr~B2~K and Sgr~B2~Main with emission extending $\sim$1$\farcm$3 north of K, and $\sim$3$\farcm$3 south of Main. At $160-500\,\mu$m, there is no apparent emission from mid-infrared regions Sgr~B2~Ext1 and Sgr~B2~Ext2, signifying that these sources lack cold dust and possibly dense molecular material (analogous to most of Sgr~B1).

In the Spitzer IRAC data, at 8\,$\mu$m, we see evidence of the IRDC (due to the lack of emission), and the majority of what is visible at this wavelength is from scattered diffuse emission and sources east of the IRDC location. That being said, some emission is seen coming from Sgr~B2~Main, as well as H, BB, O, P, and R. Most of the sources seen at SOFIA wavelengths (including Sgr~B2~Main) have emission components at the shorter 3.6 and 4.5\,$\mu$m wavelengths, signifying either a steep change in extinction across the Sgr~B2 region from west to east or that these sources lie just in front of the IRDC. \citet{2009ApJ...702..178Y} further points out that Sgr~B2 has the highest density of sources in the Galactic center with excess emission in the IRAC 4.5\,$\mu$m band (a.k.a., extended green object, or EGO, emission), and this emission tends to be directly associated with high mass star formation \citep{2011ApJ...743...56C, 2010AJ....140..196D}.  

\begin{figure*}[t]
\epsscale{1.17}
\plotone{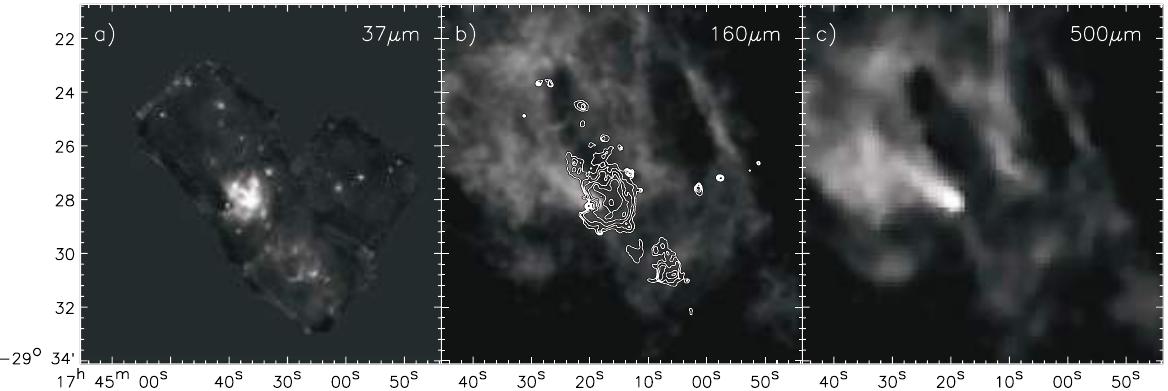}
\caption{Sgr~C region at a) 37\,$\mu$m, b) 160\,$\mu$m (with 37\,$\mu$m contours), and c) 500\,$\mu$m. The morphology of Sgr~C appears very similar to the 37\,$\mu$m image for wavelengths from 8 to 70\,$\mu$m. The morphology is also very similar from 160-500\,$\mu$m, but very different than the morphology from 8 to 70\,$\mu$m. Most of the emitting regions seen at 37\,$\mu$m appear to be devoid of cold dust at 160-500\,$\mu$m, except for the IRDC finger protruding into the \ion{H}{2} region from the east. \label{fig:SgrC_evo}}
\end{figure*}

\subsection{Sgr~C}

There are three main components to Sgr~C: a $\sim$10\,pc \ion{H}{2} region, an infrared dark cloud, and a prominent non-thermal radio filament. As pointed out by \citet{2013ApJ...775L..50K},  Sgr~C is the only known star-forming region in the western CMZ. Nonetheless, there are relatively few studies dedicated to Sgr~C at any wavelength. Save a few studies concentrating on radio emission from Sgr~C, most data that cover Sgr~C are included in studies of the entire CMZ of the Galactic center. 

Sgr~C was first mapped in the infrared at 100\,$\mu$m by \citet{1971ApJ...170L..89H} at 6$\arcmin$ resolution, and then identified as an \ion{H}{2} region at 3 an 6\,cm radio continuum with $\lesssim$1.3$\arcmin$ resolution by \citet{1979A&AS...35....1D}. Much higher spatial resolution radio continuum observations of Sgr~C were made by \citet{1995ApJS...98..259L}, at 18.5\,cm with $\sim$5$\arcsec$ resolution, resolving the structure of the \ion{H}{2} region and its nearby ionized environment. 

In the infrared, \citet{1978ApJ...220..822G} imaged Sgr~C at 30, 50, and 100\,$\mu$m, and showed that the emission peaks at wavelengths shorter than 100\,$\mu$m. This was later confirmed by ISO spectra \citep{2002A&A...381..571P} which showed a peak in brightness for Sgr~C at $\sim$50\,$\mu$m, falling drastically to relatively small flux values by 150\,$\mu$m.  Extremely low-resolution (4$\arcmin$) images of the entire CMZ were obtained using the IRAS Satellite at 12, 25, 60, and 100\,$\mu$m \citep{1984ApJ...278L..57G, 1989IAUS..136..121C} where Sgr~C can clearly be seen at all wavelengths, and \citet{1985AJ.....90.1812L} obtained at 25 and 27\,$\mu$m observations of Sgr~C using rocket-borne instruments with $\sim3.5\arcmin$ resolution. The first sub-arcminute infrared images of Sgr~C were the MSX data presented in \citet{2004MNRAS.355..899C}. Spitzer images with $\sim3\arcsec$ resolutions have been presented of Sgr~C, again in images showing the whole of the CMZ, e.g., by \citet{2009ApJ...702..178Y} and \citet{2015ApJ...799...53K}, in which it is hard to see details. Very recently, \citet{2024arXiv241009253C} have published James Webb Space Telescope (JWST) NIRCam observations of Sgr~C, creating sub-arcsecond images from 1-5\,$\mu$m that cover most, but not all of the G\ion{H}{2} region. In a contemporaneous companion paper, \citet{2024arXiv241210983B} present JWST-NIRCam Br~$\alpha$ images, that show exquisite detail within the large ($r\sim1\arcmin$) Sgr~C~\ion{H}{2} region and show that it is pervaded with filamentary structures. In our SOFIA 25 and 37\,$\mu$m images at $\sim$3$\arcsec$ resolution (Figures \ref{fig:figC1}, \ref{fig:C2}, and \ref{fig:C3}) we see the dominant emission is from the Sgr~C~\ion{H}{2} region, which appears clumpy in our images. The overall morphology and size of the Sgr~C \ion{H}{2} region appears very similar to the cm radio continuum emission (Figure \ref{fig:SgrCradio}), conspicuously bending around but avoiding the location of the IRDC. Several fainter infrared sources lie spread through the rest of our SOFIA field, including in a region $\sim$3.5$\arcmin$ south of the Sgr~C \ion{H}{2} region which we call G359.38-0.08, another region $\sim$4.0$\arcmin$ to the west which we call G359.42-0.02, and yet another region $\sim$4.2$\arcmin$ to the northeast which we call G359.50-0.09 (Figures \ref{fig:figC1}, \ref{fig:C2}, and \ref{fig:C3}). 

\citet{1991ApJ...380..429L} observed Sgr~C at 1.3 and 0.8\,mm with 0.5$\arcmin$ resolution, and were likely the first to isolate and resolve the IRDC component. This IRDC is also the dominant feature in the 350\,$\mu$m observations of \citet{2004dimg.conf..277S}. The location of the \ion{H}{2} region, however, appears to be devoid of cold dust traced by these wavelengths, in agreement with the conclusions of \citet{1978ApJ...220..822G} who, and based upon the low measured far-infrared optical depths, claimed that the Sgr~C \ion{H}{2} region is not embedded in a dense molecular cloud. This is further backed by a host of molecular line data, like CS \citep{1998IAUS..184..173K, 2013MNRAS.433..221J} and $^{13}$CO \citep{1995ApJS...98..259L}. Like Sgr~B1, when we look at Herschel imaging data tracing the location of cold dust, we once again see that the the mid-infrared and radio emitting regions lie in far-infrared emission voids at wavelengths $\gtrsim$160\,$\mu$m (Figures \ref{fig:SgrC_evo}).

The vast majority of the present star formation appears to be going on in or near the IRDC (Figures \ref{fig:C2}, \ref{fig:C3}, and \ref{fig:SgrC_evo}). The IRDC is shaped like a finger which protrudes into the side of the \ion{H}{2} region from the east. \citet{2013ApJ...775L..50K} speculate that the enhanced star formation at this specific location in the IRDC may be due to it being close to the interface of a cloud-cloud collision or feedback from the \ion{H}{2} region. The western tip of the IRDC is rich in tracers of present star formation. The sub-arcsecond radio continuum observations of \citet{2000ApJ...530..371F} were the first to find a UC\ion{H}{2} region here (G359.436-0.102, see Figure \ref{fig:SgrCradio}) coincident with H$_2$O maser emission, and nearby OH masers (to within $\sim3\arcsec$). Additional intense maser activity has been found by others, including \citet{2019ApJS..244...35L}, who find CH$_3$OH and H$_2$CO masers here. They argue that H$_2$CO masers are thought to trace a very short period in high mass star formation, and the occurrence of these masers in Sgr~C means there is an ongoing burst of star formation. They further state that Sgr~C is one of the most maser-rich sites in the entire Galaxy. Additionally, the tip of this IRDC is now known to house as many as 19 millimeter cores \citep{2013ApJ...775L..50K, 2019ApJS..244...35L} and several UC\ion{H}{2} regions \citep{2019ApJ...872..171L}. \citet{2020ApJ...894L..14L,2021ApJ...909..177L} also found multiple mm cores and molecular outflows from sources throughout the IRDC using data obtained with ALMA. Furthermore, \citet{2009ApJ...702..178Y} showed with IRAC data that there was 4.5\,$\mu$m EGO emission at the location of the stars forming in the tip of the IRDC. Again, this EGO emission is thought to be a tracer of predominantly massive star formation, and \citet{2024arXiv241009253C} interpret this entire region as a blue-shift outflow lobe. In our SOFIA data, this IRDC appears as a bright region in the reverse intensity of Figures \ref{fig:C2} and \ref{fig:C3}, and we clearly detect several of the UC\ion{H}{2} regions present in the tip of the IRDC.

Despite all of the evidence for star formation occurring in Sgr~C, the present generation of star formation is predominantly confined to the IRDC. Consequently, \citet{2013ApJ...775L..50K} claim that, when looking at the whole Sgr~C region, it appears to be relatively inactive compared to Galactic disk molecular clouds that exhibit similar physical conditions. However, \citet{2009ApJ...702..178Y} found several point sources throughout the Sgr~C region using Spitzer-MIPS 24\,$\mu$m imaging data, and claimed that they were YSO candidates. Based upon their infrared spectra, several of these sources were later ruled out as YSOs by \citet{2011ApJ...736..133A}, and \citet{2015ApJ...799...53K} showed that many could be main sequence stars in a high extinction environment. Further analysis by \citet{2022MNRAS.517..294Y} found four YSOs in Sgr~C, however two are only seen in Spitzer-IRAC data (i.e., not in MIPS 24\,$\mu$m), and all four lie off our covered SOFIA field. We find several sources that are MYSO candidates in the SOFIA data in and around the Sgr~C \ion{H}{2} region, which we will discuss further in Section \ref{sec:data}, as well as others in G359.38-0.08, G359.42-0.02, and G359.50-0.09, which are presented in Appendix C. 

Unsurprisingly, the majority of the prominent non-thermal radio filament does not radiate emission at any infrared wavelength we studied. The exception may be Sgr~C~6 and Sgr~C~15 (and perhaps also Sgr~C~16) just north of the Sgr~C \ion{H}{2} region and some faint extended emission in between them (see Figure \ref{fig:SgrCradio}), which appear coincident with the location of the non-thermal filament. Since detectable infrared emission is not expected from non-thermal radio filaments, these two infrared sources and/or the extended faint emission seen at 25 and 37\,$\mu$m, may be indicating an interaction between the filament and the \ion{H}{2} region.

Unfortunately, unlike Sgr~B2, we do not have any parallactic or other high-precision distance measurements directly of Sgr~C. As pointed out by \citet{2013ApJ...775L..50K}, the $v_{lsr}$ of Sgr~C is very similar to those of sources in the Near~3~kpc~Arm at a distance of $\sim$5.5\,kpc, which complicates kinematic interpretations of its distance. That being said, measured $v_{lsr}$ values, like those ($-60.0\pm1.0$\,km/s from the measured H109$\alpha$+H110$\alpha$ transitions) of \citet{1987A&A...171..261C} yield tangent point kinematic distances of $8.34^{+0.15}_{-0.17}$\,kpc \citepalias{2022ApJ...933...60D}. Historically, Sgr~C has been assumed to be at the same distance as Sgr~A$^*$, and interestingly, the distance to Sgr~C just quoted agrees with more rigorous calculations of the distance to the Galactic Center of $R_o=8.34\pm0.16$\,kpc by \citet{2014ApJ...783..130R}. We adopt the 8.34\,kpc value here for Sgr~C.

\begin{deluxetable*}{lccccccccc}
\tabletypesize{\scriptsize}
\tablecolumns{10}
\tablewidth{0pt}
\tablecaption{SOFIA Observational Parameters of Sources in Sgr~B1 \label{tb:SgrB_compact}}
\tablehead{\colhead{  }&
           \colhead{  }&
           \colhead{  }&
           \multicolumn{3}{c}{${\rm 25\mu{m}}$}&
           \multicolumn{3}{c}{${\rm 37\mu{m}}$} &
           \colhead{  }\\[-4pt]
           \cmidrule(lr){4-6} \cmidrule(lr){7-9}\\[-12pt]
           \colhead{ Source }&
           \colhead{ R.A.}&
           \colhead{ Decl. }&           
           \colhead{ $R_{\rm int}$ } &
           \colhead{ $F_{\rm int}$ } &
           \colhead{ $F_{\rm int-bg}$ } &
           \colhead{ $R_{\rm int}$ } &
           \colhead{ $F_{\rm int}$ } &
           \colhead{ $F_{\rm int-bg}$ } &
           \colhead{ Aliases }\\[-6pt]
	   \colhead{  } &
          \colhead{(J2000) }&
          \colhead{(J2000) }& 	   
	   \colhead{ ($\arcsec$) } &
	   \colhead{ (Jy) } &
	   \colhead{ (Jy) } &
	   \colhead{ ($\arcsec$) } &
	   \colhead{ (Jy) } &
	   \colhead{ (Jy) } &
           \colhead{  }\\[-12pt]
}
\startdata
\multicolumn{9}{c}{Compact Sources} \\
\hline
Sgr B1 A	&	17 46 53.28	&	-28 32 00.5	&	6.1	    &	30.9	&	25.2	&	6.1	    &	59.9	&	46.1 & SSTGC726327$^{\dagger}$	\\	
Sgr B1 C	&	17 46 54.36	&	-28 32 38.8	&	8.4	    &	26.1	&	15.3	&	9.2	    &	46.1	&	27.1 &	\\	
Sgr B1 D	&	17 46 57.04	&	-28 33 45.6	&	10.0	&	106	    &	87.8	&	10.7	&	190	    &	131	& \\	
Sgr B1 1	&	17 47 01.09	&	-28 31 15.7	&	10.0	&	126	    &	75.9	&	10.0	&	236	    &	126	& \\	
Sgr B1 2	&	17 47 04.56	&	-28 33 55.3	&	8.4	    &	13.8	&	4.77	&	9.2	    &	40.0	&	13.6 &	\\	
Sgr B1 F	&	17 47 04.63	&	-28 29 46.2	&	6.9	    &	7.22	&	5.99	&	9.2	    &	28.8	&	15.8 &	\\	
Sgr B1 G	&	17 47 06.04	&	-28 31 04.8	&	8.4	    &	41.3	&	18.6	&	8.4	    &	103	    &	41.3 &	\\	
Sgr B1 H	&	17 47 07.53	&	-28 28 42.6	&	9.2	    &	21.3	&	11.4	&	10.0	&	47.2	&	28.9 & SSTGC760679$^{\dagger}$	\\	
Sgr B1 3	&	17 47 08.88	&	-28 29 55.5	&	6.1	    &	33.5	&	28.4	&	7.7	    &	55.0	&	41.7 & OH 0.548-0.059$^{\ast}$\\	
Sgr B1 4	&	17 47 11.65	&	-28 32 00.6	&	7.7	    &	25.7	&	5.87	&	7.7	    &	62.2	&	16.8 &	\\	
Sgr B1 5	&	17 47 11.82	&	-28 31 37.6	&	6.1	    &	36.9	&	7.05	&	6.1	    &	68.0	&	13.0 &	\\	
Sgr B1 6	&	17 47 12.34	&	-28 31 42.9	&	6.1	    &	33.8	&	17.4	&	6.1	    &	73.3	&	34.3 &	\\	
Sgr B1 7	&	17 47 12.81	&	-28 31 36.0	&	6.1	    &	47.1	&	10.2	&	6.1	    &	75.5	&	14.6 &	\\	
Sgr B1 8	&	17 47 12.87	&	-28 32 06.7	&	7.7	    &	21.6	&	3.57	&	7.7	    &	48.2	&	3.81 & SSTGC772981$^{\dagger}$	\\	
Sgr B1 9	&	17 47 14.26	&	-28 31 10.6	&	6.1	    &	24.4	&	4.73	&	6.9	    &	57.1	&	12.1 &	\\	
Sgr B1 10	&	17 47 14.65	&	-28 30 00.7	&	11.5	&	42.6	&	20.7	&	12.3	&	84.1	&	32.5 &	\\	
Sgr B1 11	&	17 47 14.74	&	-28 32 09.7	&	10.0	&	33.6	&	8.50	&	10.7	&	68.3	&	15.2 &	\\	
Sgr B1 12	&	17 47 17.25	&	-28 32 21.1	&	11.5	&	48.6	&	33.7	&	11.5	&	58.5	&	48.3 & SSTGC782872$^{\dagger}$	\\	
\hline
\multicolumn{9}{c}{Extended Sources} \\
\hline
Ionized Bar	&	17 46 57.59	&	-28 31 07.4	&	143x40	&	801	    &	582	&	143x40	&	1950	&	1850 &	\\
Ionized Rim	&	17 46 59.72	&	-28 32 27.6	&	64x114	&	1190	&	908	&	64x114	&	3000	&	2880 &	\\
Sgr B1 Ext1	&	17 47 01.23	&	-28 31 14.8	&	26.0	&	476	    &	394	&	26.0	&	1070	&	1030 &	\\
Sgr B1 E 	&	17 47 04.28	&	-28 33 22.9	&	23.0	&	435	    &	362	&	23.0	&	822	    &	723 &	\\
Sgr B1 I	&	17 47 12.39	&	-28 31 22.9	&	35.0	&	1100	&	938	&	35.0	&	1920	&	1780 &	\\
Sgr B1 J	&	17 47 12.75	&	-28 30 10.0	&	28x80	&	283	    &	207	&	28x80	&	505	    &	410 &	\\	
\enddata
\tablenotetext{\dagger}{From \citet{2011ApJ...736..133A}.}
\tablenotetext{\ast}{Known AGB (OH/IR) star; see \citet{1997ApJ...478..206S}}
\end{deluxetable*}

\begin{deluxetable*}{lccccccccc}
\tabletypesize{\scriptsize}
\tablecolumns{10}
\tablewidth{0pt}
\tablecaption{SOFIA Observational Parameters of Sources in Sgr~B2}
\tablehead{\colhead{  }&
           \colhead{  }&
           \colhead{  }&
           \multicolumn{3}{c}{${\rm 25\mu{m}}$}&
           \multicolumn{3}{c}{${\rm 37\mu{m}}$} \\[-4pt]
           \cmidrule(lr){4-6} \cmidrule(lr){7-9} \\[-12pt]
           \colhead{ Source }&
           \colhead{ R.A.}&
           \colhead{ Decl. }&           
           \colhead{ $R_{\rm int}$ } &
           \colhead{ $F_{\rm int}$ } &
           \colhead{ $F_{\rm int-bg}$ } &
           \colhead{ $R_{\rm int}$ } &
           \colhead{ $F_{\rm int}$ } &
           \colhead{ $F_{\rm int-bg}$ } &
           \colhead{ Aliases } \\[-6pt]
	   \colhead{  } &
          \colhead{(J2000) }&
          \colhead{(J2000) }& 	   
	   \colhead{ ($\arcsec$) } &
	   \colhead{ (Jy) } &
	   \colhead{ (Jy) } &
	   \colhead{ ($\arcsec$) } &
	   \colhead{ (Jy) } &
	   \colhead{ (Jy) } &
          \colhead{   }\\[-12pt]
}
\startdata
\multicolumn{9}{c}{Compact Sources} \\
\hline
Sgr B2 1	&	17 47 12.42	&	-28 24 15.2	&	6.1	    &	4.28	&	0.532	&	6.1	    &	5.06	&	2.85 &	SSTGC772151$^{\dagger}$\\
Sgr B2 2	&	17 47 17.65	&	-28 24 01.9	&	6.9	    &	3.45	&	0.903	&	7.7	    &	6.01	&	4.65 &	\\
Sgr B2 3	&	17 47 18.53	&	-28 24 23.4	&	10.0	&	9.47	&	1.37	&	10.0	&	12.0	&	7.94 &	SSTGC784931$^{\ddagger}$\\
Sgr B2 AA	&	17 47 19.53	&	-28 24 39.5	&	6.9	    &	5.77	&	2.76	&	7.7	    &	11.9	&	11.3 &	SSTGC787884$^{\ddagger}$\\
Sgr B2 4 	&	17 47 19.72	&	-28 22 18.2	&	6.1	    &	$<$0.58	&	UD	    &	8.4	    &	19.3	&	11.9 &	\\
Sgr B2 H	&	17 47 20.38	&	-28 23 42.6	&	8.4	    &	40.3	&	35.0	&	9.2	    &	254	    &	232	 &  South, S \\
Sgr B2 5	&	17 47 20.62	&	-28 23 53.3	&	5.4	    &	4.28	&	2.30	&	5.4	    &	13.3	&	7.10 &	SSTGC790317$^{\dagger}$\\
Sgr B2 6	&	17 47 21.97	&	-28 24 37.1	&	6.1	    &	2.81	&	0.497	&	6.1	    &	1.14	&	1.02 &	SSTGC793536$^{\ddagger}$\\
Sgr B2 BB	&	17 47 22.17	&	-28 22 18.9	&	10.0	&	11.4	&	10.7	&	11.5	&	56.0	&	39.3 &	SSTGC793867$^{\ddagger}$\\
Sgr B2 L	&	17 47 22.45	&	-28 21 55.8	&	7.7	    &	3.40	&	2.62	&	7.7	    &	14.4	&	14.3 &	\\
Sgr B2 7	&	17 47 22.78	&	-28 25 37.0	&	5.4	    &	2.38	&	0.569	&	6.1	    &	3.13	&	2.46 &	SSTGC795418$^{\ddagger}$\\
Sgr B2 O	&	17 47 22.93	&	-28 22 48.0	&	8.4	    &	6.53	&	4.40	&	10.0	&	49.2	&	27.8 &	\\
Sgr B2 8	&	17 47 23.18	&	-28 23 54.8	&	6.1	    &	4.29	&	1.82	&	8.4	    &	15.2	&	9.34 &	SSTGC796410$^{\ddagger}$\\
Sgr B2 9	&	17 47 23.57	&	-28 22 33.4	&	6.9	    &	1.97	&	1.44	&	6.1	    &	7.50	&	2.92 &	SSTGC797252$^{\ddagger}$\\
Sgr B2 P	&	17 47 23.70	&	-28 23 35.6	&	8.4	    &	16.8	&	11.6	&	9.2	    &	44.8	&	35.1 &	SSTGC797384$^{\dagger}$\\
Sgr B2 R	&	17 47 26.12	&	-28 22 04.9	&	11.5	&	47.3	&	38.6	&	12.3	&	132	    &	111	 &  SSTGC803187$^{\dagger}$ \\
Sgr B2 10	&	17 47 26.57	&	-28 24 45.3	&	10.0	&	24.8	&	9.88	&	11.5	&	48.8	&	14.9 &	SSTGC803471$^{\dagger}$\\
Sgr B2 11	&	17 47 27.14	&	-28 27 27.4	&	7.7	    &	10.7	&	3.85	&	11.5	&	17.9	&	12.3 &	SSTGC805200$^{\ddagger}$\\
Sgr B2 12	&	17 47 27.58	&	-28 25 47.5	&	10.7	&	33.3	&	8.13	&	10.7	&	67.1	&	24.1 &	 \\
Sgr B2 13	&	17 47 27.64	&	-28 26 29.0	&	6.9	    &	9.01	&	2.04	&	7.7	    &	16.3	&	5.80 &	SSTGC806191$^{\ddagger,\dagger}$\\
Sgr B2 14	&	17 47 27.98	&	-28 22 00.9	&	10.7	&	25.4	&	12.7	&	12.3	&	72.3	&	30.7 &	\\
Sgr B2 15	&	17 47 28.45	&	-28 25 56.7	&	5.4	    &	8.09	&	1.00	&	5.4	    &	15.4	&	1.81 &	\\
Sgr B2 16	&	17 47 31.08	&	-28 26 41.9	&	8.4	    &	10.5	&	4.59	&	11.5	&	31.1	&	21.4 &	\\
Sgr B2 17	&	17 47 34.06	&	-28 27 17.8	&	10.0	&	25.7	&	11.2	&	11.5	&	59.6	&	33.9 &   \\	
\hline
\multicolumn{9}{c}{Extended Sources} \\
\hline
Sgr B2 V	&	17 47 13.18	&	-28 24 42.1	&	27.0	&	177	&	118	    &	27.0	&	390	    &	353	& \\
Sgr B2 Main	&	17 47 20.43	&	-28 23 01.9	&	27.0	&	139	&	95.4	&	27.0	&	1330	&	1200 & M	\\
Sgr B2 Ext1	&	17 47 29.89	&	-28 26 05.1	&	28x50	&	112	&	69.0	&	28x50	&	269	    &	253	& \\
Sgr B2 Ext2	&	17 47 34.52	&	-28 26 50.9	&	19.0	&	124	&	85.5	&	19.0	&	200	    &	160	& \\
\enddata
\tablecomments{UD means the source is not undetected, and the 5$\sigma$ upper limit on a detection is given. For this source, the $F_{\rm int}$ value is used as an upper limit in the SED modeling.}
\tablenotetext{\dagger}{From \citet{2011ApJ...736..133A}.}
\tablenotetext{\ddagger}{From \citet{2009ApJ...702..178Y}.}
\end{deluxetable*}

\begin{deluxetable*}{lccccccccc}
\tabletypesize{\scriptsize}
\tablecolumns{10}
\tablewidth{0pt}
\tablecaption{SOFIA Observational Parameters of Sources in Sgr~C \label{tb:SgrC_compact}}
\tablehead{\colhead{  }&
           \colhead{  }&
           \colhead{  }&
           \multicolumn{3}{c}{${\rm 25\mu{m}}$}&
           \multicolumn{3}{c}{${\rm 37\mu{m}}$} &
           \colhead{  } \\[-4pt]
           \cmidrule(lr){4-6} \cmidrule(lr){7-9}\\[-12pt]
           \colhead{ Source }&
           \colhead{ R.A.}&
           \colhead{ Decl. }&           
           \colhead{ $R_{\rm int}$ } &
           \colhead{ $F_{\rm int}$ } &
           \colhead{ $F_{\rm int-bg}$ } &
           \colhead{ $R_{\rm int}$ } &
           \colhead{ $F_{\rm int}$ } &
           \colhead{ $F_{\rm int-bg}$ } &
           \colhead{ Aliases } \\[-6pt]
	   \colhead{  } &
          \colhead{(J2000) }&
          \colhead{(J2000) }& 	   
	   \colhead{ ($\arcsec$) } &
	   \colhead{ (Jy) } &
	   \colhead{ (Jy) } &
	   \colhead{ ($\arcsec$) } &
	   \colhead{ (Jy) } &
	   \colhead{ (Jy) } &
           \colhead{  } \\[-12pt]
}
\startdata
\multicolumn{10}{c}{Compact Sources} \\
\hline
Sgr C 1 	&	17 44 31.39	&	-29 27 39.1	&	7.7	    &	6.61	&	3.50	&	10.0	&	25.8	&	14.2 &	SSTGC343554$^{\ddagger,\dagger}$\\
Sgr C 2	    &	17 44 33.15	&	-29 26 59.2	&	12.3	&	38.5	&	25.5	&	14.6	&	85.7	&	49.0 &	SSTGC348392$^{\dagger}$\\
Sgr C 3	    &	17 44 33.39	&	-29 29 00.6	&	9.2	    &	14.9	&	7.53	&	9.2	    &	40.4	&	12.2 &	\\
Sgr C 4	    &	17 44 34.15	&	-29 27 36.1	&	3.8	    &	3.64	&	0.498	&	3.8	    &	11.0	&	1.42 &	\\
Sgr C 5	    &	17 44 34.15	&	-29 28 32.9	&	10.7	&	40.4	&	20.5	&	11.5	&	155	    &	47.7 &	\\
Sgr C 6	    &	17 44 34.56	&	-29 26 07.8	&	8.4	    &	8.80	&	5.44	&	8.4	    &	20.6	&	12.2 &	SSTGC351441$^{\ddagger}$\\
Sgr C 7	    &	17 44 35.86	&	-29 27 44.5	&	9.2	    &	60.9	&	7.94	&	9.2	    &	159	    &	20.9 &	SSTGC354683$^{\dagger}$\\
Sgr C 8	    &	17 44 36.50	&	-29 28 06.8	&	9.2	    &	88.7	&	53.6	&	9.2	    &	194	    &	96.2 &	\\
Sgr C 9	    &	17 44 36.74	&	-29 25 22.4	&	4.6	    &	1.29	&	0.482	&	5.4	    &	4.61	&	1.35 &	\\
Sgr C 10	&	17 44 37.79	&	-29 25 44.7	&	8.4	    &	10.8	&	5.86	&	8.4	    &	28.0	&	8.72 &	SSTGC360055$^{\dagger}$\\
Sgr C 11	&	17 44 38.33	&	-29 29 12.8	&	6.1	    &	8.49	&	6.32	&	6.1	    &	19.9	&	10.1 &	G359.42a\\
Sgr C 12	&	17 44 38.97	&	-29 27 29.9	&	8.4	    &	47.4	&	20.3	&	7.7	    &	91.4	&	39.9 &	\\
Sgr C 13	&	17 44 39.74	&	-29 28 27.5	&	3.1	    &	2.93	&	0.220	&	3.1	    &	7.77	&	0.380 &	\\
Sgr C 14	&	17 44 40.03	&	-29 28 22.9	&	3.1	    &	2.90	&	1.29	&	3.1	    &	7.18	&	1.72 &  \\
Sgr C H3	&	17 44 40.21	&	-29 28 14.5	&	5.4	    &	18.7	&	15.8	&	5.4	    &	62.8	&	52.4 &	G359.44a\\
Sgr C H4	&	17 44 40.56	&	-29 28 15.2	&	3.1	    &	2.38	&	1.99	&	3.1	    &	12.0	&	9.45 &	G359.44b, C103\\
Sgr C H1	&	17 44 41.09	&	-29 27 56.0	&	4.6	    &	2.02	&	1.67	&	3.8	    &	4.29	&	3.72 &	C102\\
Sgr C 15	&	17 44 42.73	&	-29 26 37.7	&	9.2	    &	11.9	&	5.55	&	10.0	&	46.4	&	12.3 &	\\
Sgr C 16	&	17 44 43.62	&	-29 26 18.5	&	7.7	    &	4.24	&	1.01	&	9.2	    &	29.0	&	10.1 &	\\	
\hline
\multicolumn{10}{c}{Extended Sources} \\
\hline
Sgr C HII	&	17 44 36.45	&	-29 27 59.7	&	83.6	&	1630	&	1440	&	83.6	&	4570	&	4540	& \\	
\enddata
\tablenotetext{a}{Source names and aliases: H names are the compact \ion{H}{2} regions found at 1.3\,cm from \citet{2019ApJ...872..171L}; C names from C-band (6\,cm) detections from \citet{2019ApJS..244...35L}; and G names are from the shortened galactic coordinate labels given by \citet{2024arXiv241009253C}.} 
\tablenotetext{\dagger}{From \citet{2011ApJ...736..133A}.}
\tablenotetext{\ddagger}{From \citet{2009ApJ...702..178Y}.}
\end{deluxetable*}

\begin{figure}
\epsscale{1.04}
\plotone{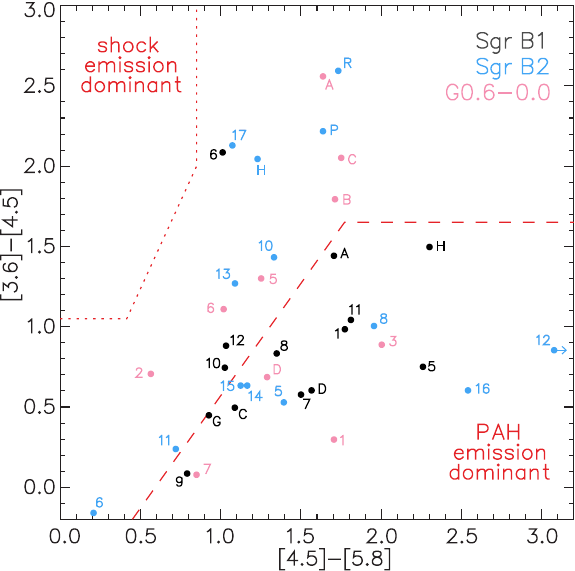}
\caption{\footnotesize A color-color diagram for compact sources in Sgr~B utilizing our background-subtracted Spitzer-IRAC 3.6, 4.5, and 5.8\,$\mu$m source photometry to distinguish ``shocked emission dominant'' and ``PAH emission dominant'' YSO candidates from our list of compact sub-components. Sources marked by black dots and labels are from within Sgr~B1, and blue are from Sgr~B2, and magenta are from G0.0-0.6. Above (up-left) the dotted line indicates shock emission dominant regime. Below (bottom-right) the dashed line indicates PAH dominant regime. We adopt this metric from \citet{2009ApJS..184...18G}. Some sources are not included in this diagram due to non-detection or saturation in the Spitzer-IRAC bands. The arrow for Sgr~B2~12 indicates a high [4.5]-[5.8] value (4.07) which has been cropped off the plot.}\label{fig:Bccd}
\end{figure}

\begin{figure}
\epsscale{1.04}
\plotone{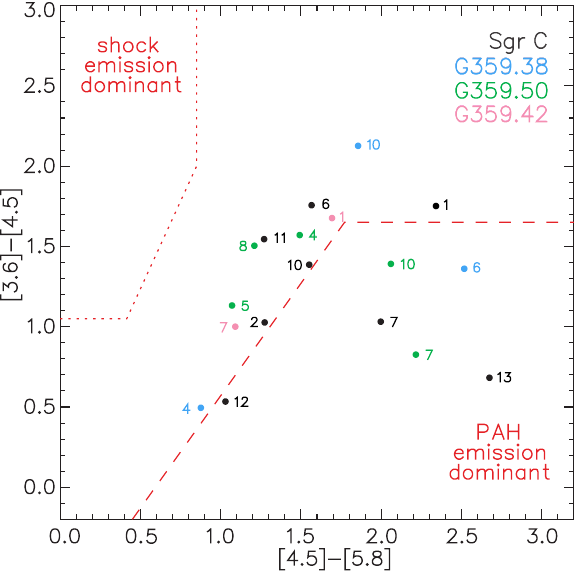}
\caption{\footnotesize A color-color diagram for compact sources in Sgr~C (black), G359.38-0.08 (blue), G359.50-0.09 (green), and G359.42-0.02 (magenta) utilizing our background-subtracted Spitzer-IRAC 3.6, 4.5, and 5.8\,$\mu$m source photometry. See caption in Figure~\ref{fig:Bccd} for more information regarding plot labels.}\label{fig:Cccd}
\end{figure}

\begin{figure*}[tp]
\epsscale{0.98}
\plotone{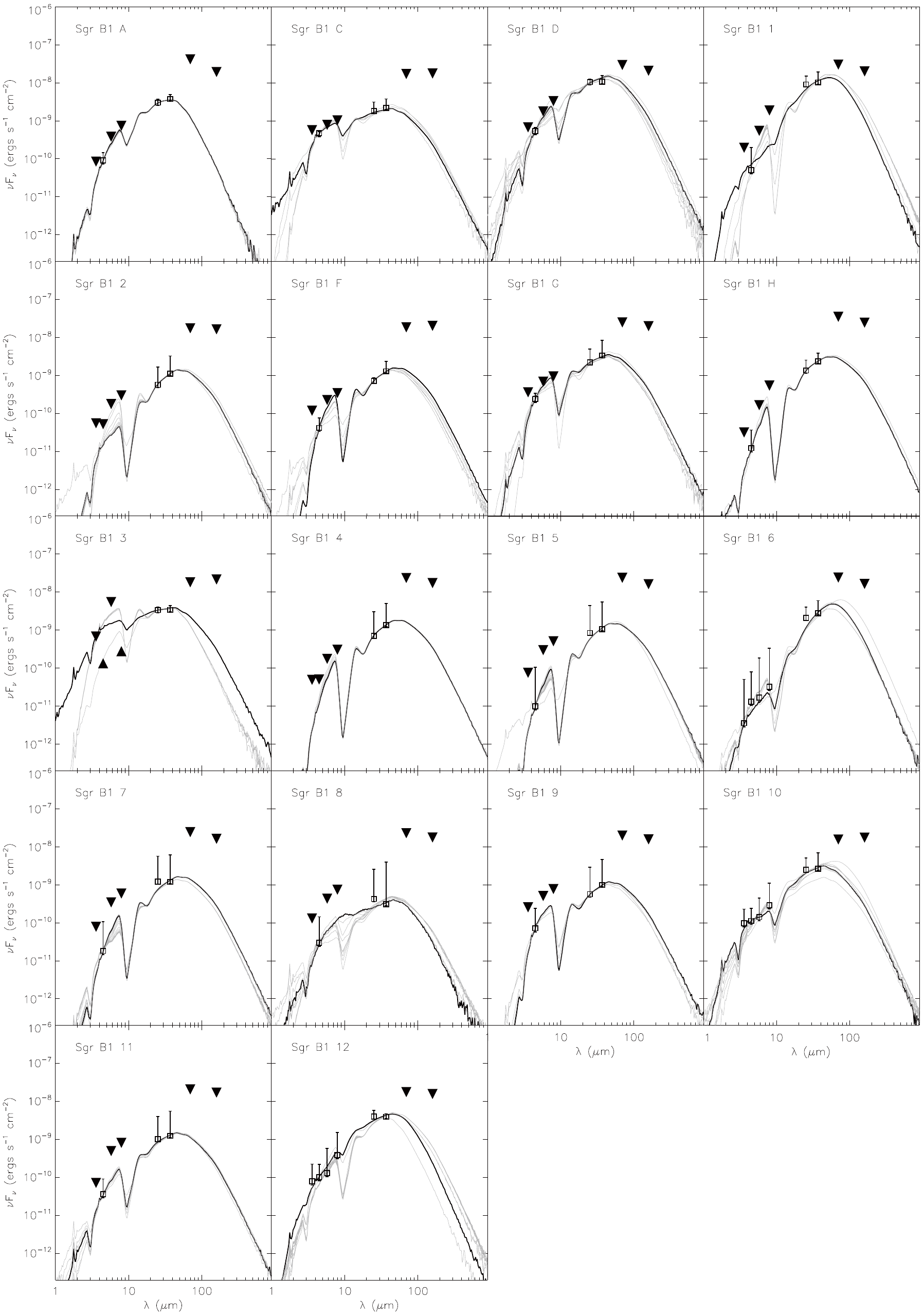}
\caption{SED fitting with ZT model for compact sources in Sgr~B1. Black lines are the best fit model to the SEDs, and the system of gray lines are the remaining fits in the group of best fits (from Table \ref{tb:Bseds}). Upside-down triangles are data that are used as upper limits in the SED fits, and triangles are lower limits.\label{fig:B1SED}}
\end{figure*}

\begin{figure*}[htp]
\epsscale{0.98}
\plotone{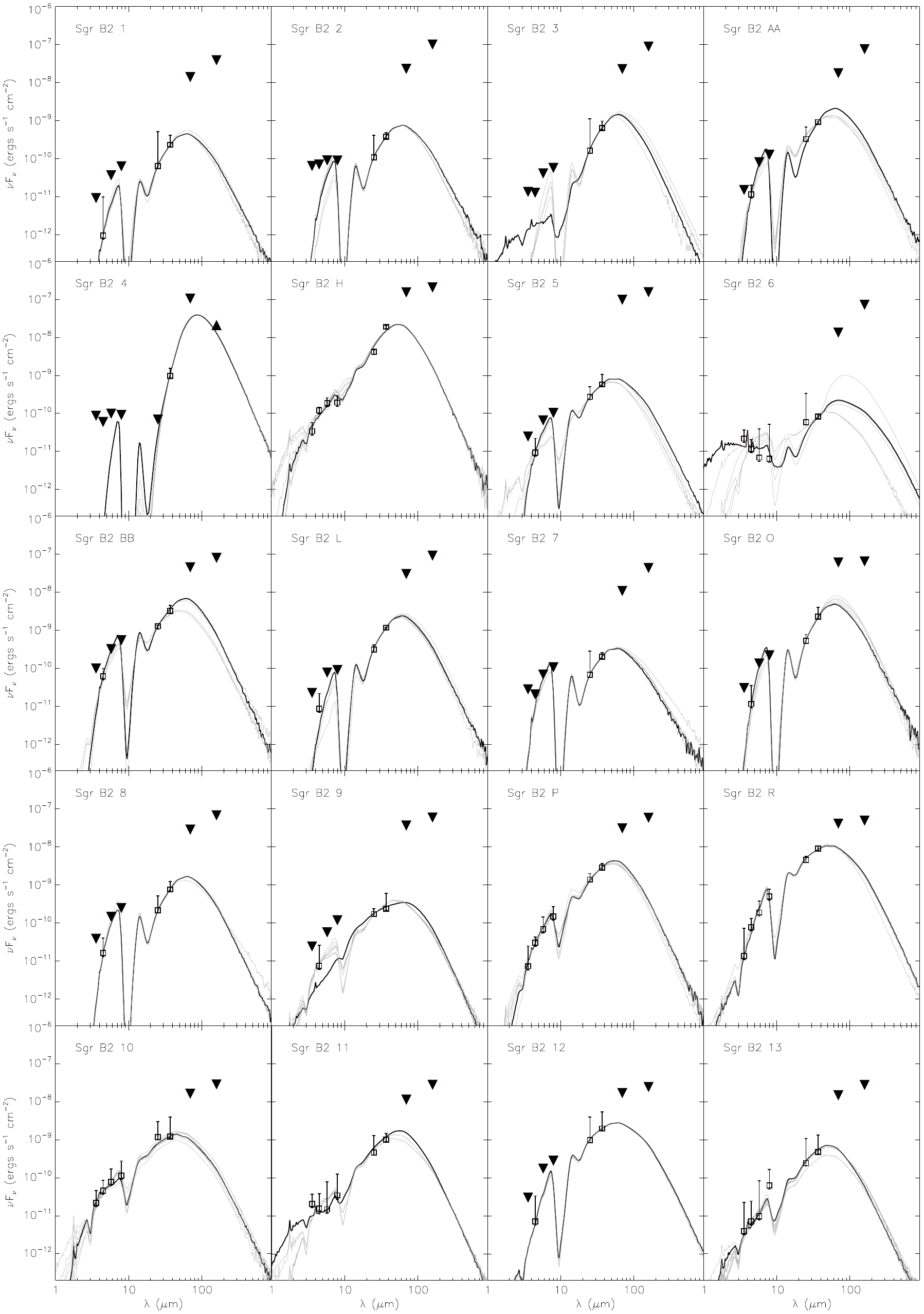}
\caption{SED fitting with ZT model for compact sources in Sgr~B2. Black lines are the best fit model to the SEDs, and the system of gray lines are the remaining fits in the group of best fits (from Table \ref{tb:Bseds}). Upside-down triangles are data that are used as upper limits in the SED fits, and triangles are lower limits.\label{fig:BSED2}}
\end{figure*}

\begin{figure*}[t]
\figurenum{15}
\epsscale{0.98}
\plotone{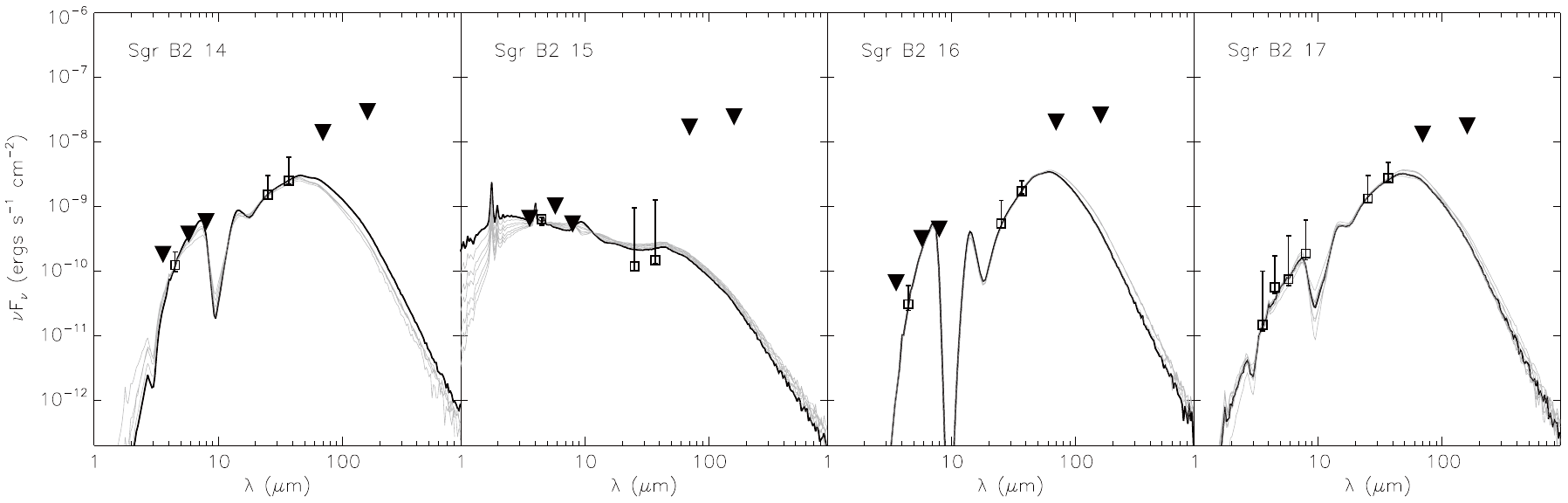}
\caption{\textit{Continued.}}
\end{figure*}

\begin{figure*}[tp]
\epsscale{0.98}
\plotone{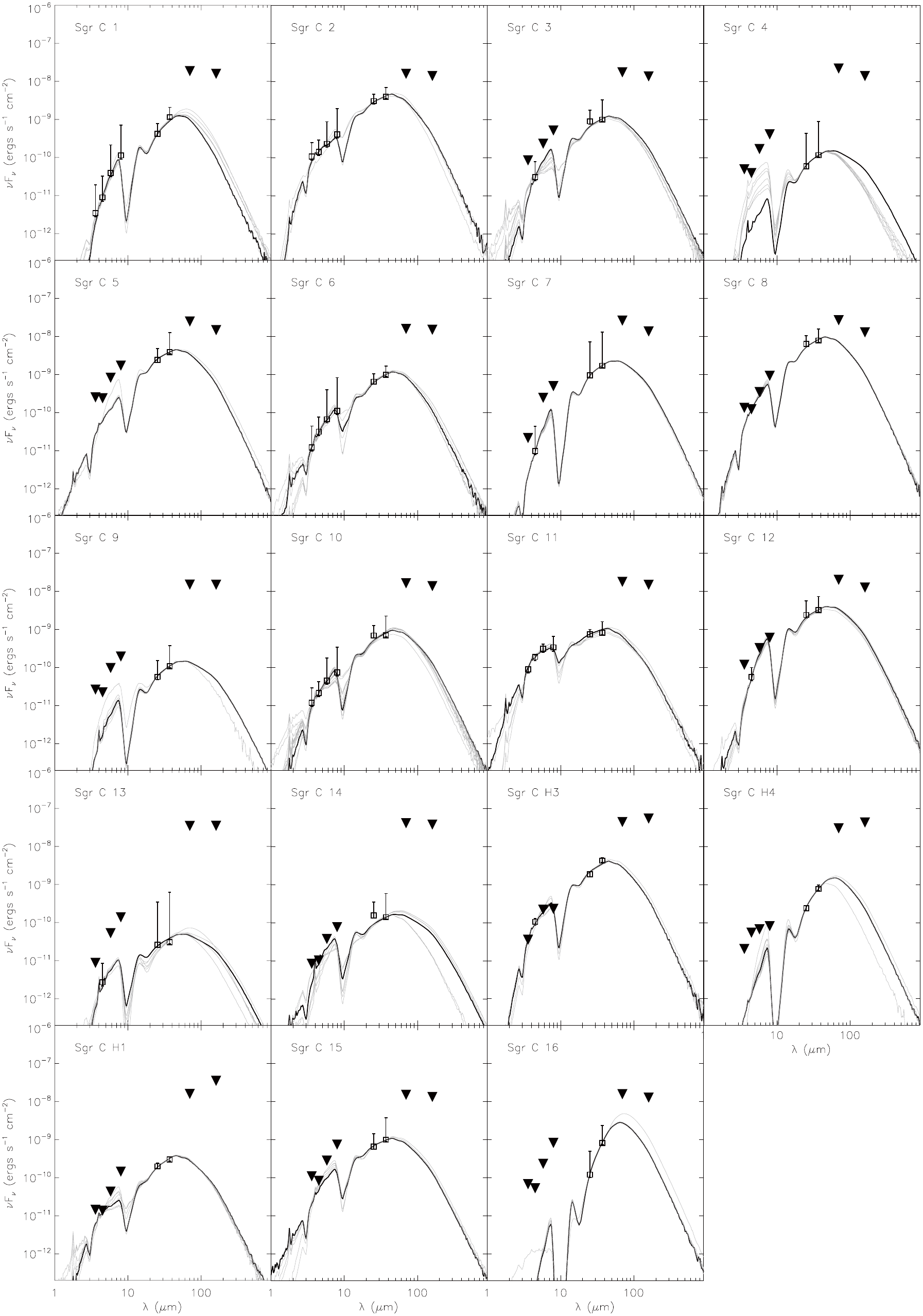}
\caption{SED fitting with ZT model for compact sources in Sgr~C. Black lines are the best fit model to the SEDs, and the system of gray lines are the remaining fits in the group of best fits (from Table \ref{tb:Cseds}). Upside-down triangles are data that are used as upper limits in the SED fits, and triangles are lower limits.\label{fig:CSED}}
\end{figure*}

\section{Data Analysis and Results}\label{sec:data}

As we have done in our other papers in this survey, we classify the infrared emission sources within the G\ion{H}{2} regions of our survey into two groups: compact sources and extended sub-regions. The compact sources are believed to be star-forming cores (typically $\lesssim0.3$\,pc in size), while the extended sub-regions are thought to be larger star-forming molecular clumps. As we have done previously for the compact sources, we will fit SED models to their multi-wavelength photometry to estimate their physical properties and identify potential MYSOs. 

For the extended sub-regions, our ability to follow what we have done in our previous papers breaks down. Previously, we have assessed the evolutionary state of each sub-region by measuring both their luminosity-to-dust mass (L/M) ratio (derived from the infrared SEDs of the radio/mid-infrared sub-regions), as well as the virial parameter of their gas components from $^{13}$CO data. However, as we have mentioned briefly in the discussion of Sgr~B1 and Sgr~C in Section \ref{sec:results1}, while the cm radio continuum and infrared at wavelengths $<$160\,$\mu$m match in morphology, none of these features appear to have cold dust (250-500\,$\mu$m) or definitive molecular CO components. For these reasons, we cannot apply the same L/M and virial analyses to these G\ion{H}{2} regions. In our previous papers, extended sources had their 3-160\,$\mu$m photometry reported along with the descriptions of their evolutionary analyses, however in this paper, since we will not be performing those analyses, we chose to list the extended source photometry at these wavelengths in the same tables as the compact sources (i.e., Tables \ref{tb:SgrB_compact} - \ref{tb:SgrC_compact}). We will discuss this lack of correspondence between the hot and ionized component versus the cold and molecular component in these Galactic Center G\ion{H}{2} regions and the ramifications in Sections \ref{sec:alm} and \ref{sec:genuine}.

\subsection{Physical Properties of Compact Sources: SED Model Fitting and Determining MYSO Candidates}\label{sec:cps}

We define a compact source as one that exhibits a distinct peak which remains consistent in location across different wavelengths and is detected at multiple wavelengths. Therefore, compact source candidates are first identified as resolved sources or peaks in the SOFIA data, and then cross-referenced with Spitzer-IRAC, Herschel-PACS, and cm radio data to check for spatial coincidences. We define compact sources as having physical sizes less than $\sim$0.4~pc, consistent with the typical size of molecular cores, around 0.1~pc \citep[e.g.,][]{2007ARA&A..45..481Z}. We identified 53 compact infrared sources in SOFIA-mapped area containing Sgr~B, 18 of which are associated with the Sgr~B1 G\ion{H}{2} region and 24 with Sgr~B2. In the Sgr~C G\ion{H}{2} region we find 47 compact infrared sources, with 19 being associated directly with the Sgr~C G\ion{H}{2} region. Tables \ref{tb:SgrB_compact} - \ref{tb:SgrC_compact}, for Sgr~B1, Sgr~B2, and Sgr~C respectively, provide details on the compact source positions, radii used for aperture photometry, and background-subtracted flux densities measured at both SOFIA wavelengths (similar information is provided for all other compact sources on the SOFIA fields in Appendix C). We used the same optimal extraction technique as used in our previous studies (see \citetalias{2019ApJ...873...51L}), to determine the best aperture for photometry, and similarly performed background subtraction using background statistics from an annulus outside the optimal extraction radius having the least environmental contamination.   

We conducted additional aperture photometry for all compact sources using archival Spitzer-IRAC data at 3.6, 4.5, 5.8, and 8.0\,$\mu$m, as well as Herschel-PACS data at 70 and 160\,$\mu$m. We applied the same optimal extraction technique to these data as we did to the SOFIA data to obtain the four near-infrared and two far-infrared photometry values. The measured Spitzer and Herschel photometry data for Sgr~B1, Sgr~B2, and Sgr~C are given in the tables in Appendix C. 

To determine how to handle the Spitzer-IRAC photometry data in the construction of our SEDs, we first assessed the potential for flux contamination in the 3.6, 5.8, and 8.0\,$\mu$m bands from polycyclic aromatic hydrocarbons (PAHs) emission and in the 4.5\,$\mu$m band from shock-excited molecular hydrogen emission. As discussed in \citetalias{2019ApJ...873...51L}, a color-color diagram using Spitzer-IRAC data (3.6\,$\mu$m - 4.5\,$\mu$m vs. 4.5\,$\mu$m - 5.8\,$\mu$m) can identify sources with flux densities highly contaminated by shock emission and/or PAH emission. 

For the Sgr~B region, we see from Figure \ref{fig:Bccd} that none of the compact sources are classified as ``shock emission dominated"; however, 21 are ``PAH emission dominated" sources. In particular, 11 sources in Sgr~B1, 6 in Sgr~B2, and 4 in G0.6-0.0 are PAH emission dominated. For the Sgr~C region, we deduce from Figure \ref{fig:Cccd} that there are also no shock-excited sources, but their are 6 PAH emission sources, three in Sgr~C, one in G359.38-0.08, and two in G359.50-0.09. For the PAH emission dominated sources, their 3.6, 5.8, and 8.0\,$\mu$m IRAC fluxes are treated as upper limits in the photometry used for constructing the SEDs. The number of compact sources plotted in Figures \ref{fig:Bccd} and \ref{fig:Cccd} are not the same as those reported in Tables \ref{tb:SgrB_compact} - \ref{tb:SgrC_compact}, because there are compact sources for which there are only IRAC 3.6, 4.6, or 5.8\,$\mu$m upper limits (due to saturation, non-detection, or they are unresolved from other nearby sources). Of the three G\ion{H}{2} regions in the study, this is especially the case for Sgr~C, where more than half of the sources (11 of 19) are missing at least one IRAC photometry value. Consequently, for these compact sources where one or more of the IRAC bands have non-detections or are saturated, the color-color analysis cannot be performed. In these cases, we conservatively assume that the sources are PAH-contaminated (i.e., we only treat the 4.5\,$\mu$m data point as a nominal value, while the rest of the IRAC data points are upper limits).

When constructing the SEDs for our compact sources, we further consider the Herschel-PACS fluxes as upper limits due to the significant and uncertain contamination from the surrounding extended emission, as well as the blending of sources due to poorer angular resolution, which complicates accurate isolation of the 70 and 160\,$\mu$m flux densities for the compact sources. 

Finally, for a couple of compact sources, the Spitzer or Herschel photometry apertures included saturated pixels. In these cases, we use the saturation limit for a point source as a lower limit in the SED fitting.

As for the errors associated with the photometry data, in keeping with our previous methodology (e.g., \citetalias{2019ApJ...873...51L}), we set the upper error bars on our photometry values as the subtracted background flux value, since background subtraction can vary significantly but is never larger than the amount subtracted. The lower error bar values for all sources are based on the average total photometric error at each wavelength, set to 20\%, 15\%, and 10\% for the 4.5, 25, and 37\,$\mu$m bands, respectively. We assume the photometric errors of the Spitzer-IRAC 3.6, 5.8, and 8.0\,$\mu$m fluxes to be 20\% for sources not contaminated by PAH features. Additionally, as in \citetalias{2019ApJ...873...51L}, we assume error bars of 40\% and 30\% for the Herschel 70 and 160\,$\mu$m data points, respectively.

Using the SOFIA, Spitzer, and Herschel photometry data and their uncertainties, we constructed near- to far-infrared SEDs for all compact sources. These SEDs were then fed into an algorithm developed by \citet{2011ApJ...733...55Z} where they were fit with theoretical SED models of MYSOs (referred to as ZT MYSO SED models). Each model fit provides a normalized minimum chi-squared value (so called $\chi_{\rm nonlimit}^2$) as an indication of the goodness-of-fit. As in previous studies, we selected a group of models that show $\chi_{\rm nonlimit}^2$ values similar to the best fit model and distinguishable from the next group of models showing significantly larger $\chi_{\rm nonlimit}^2$ values. Sometimes, the first or first few best fits have significantly lower $\chi_{\rm nonlimit}^2$ values than those that come after, and in such cases we will include those first fits with the first grouping so that we a always have at least 5 best-fit models. 

In Figure \ref{fig:B1SED} for Sgr~B1, Figure \ref{fig:BSED2} for Sgr~B2, and Figure \ref{fig:CSED} for Sgr~C, the ZT MYSO SED model fits are presented as solid lines (black for the best model fit and gray for the other best-fit models) over the measured photometry points and error bars for each individual source (with SED plots for all sources in G0.6-0.0, G359.38-0.08, G359.42-0.02, and G359.50-0.09 are in the Appendix C). Table \ref{tb:Bseds} for Sgr~B1 and Sgr~B2 and Table \ref{tb:Cseds} for Sgr~C list the physical properties of the MYSO SED model fits. For each source, Column 2 presents the observed bolometric luminosity, $L_{\rm obs}$, of the best model, while column 3 shows the true total bolometric luminosity, $L_{\rm tot}$ (corrected for foreground extinction and outflow viewing angle). The extinction and stellar mass of the best model are listed in columns 4 and 5, respectively. Columns 6 and 7 present the ranges of foreground extinction and stellar masses derived from the models in this group. Column 8 indicates the number of models in the group of best-fit models. The rest of the columns provide information related to whether the sources are likely to be MYSO candidate or not, and are described further in Section \ref{sec:MYSOs} below.

\subsubsection{Comparisons to Crowe et al. 2024 Sgr~C SED Modeling Results}

Contemporary to the publication of this paper is one by \citet{2024arXiv241009253C}, in which the authors identify three MYSO candidates from a combination of JWST-NIRCam data as well as the same Spitzer, Herschel, and SOFIA data employed in this paper. They label these sources G359.44a, G359.44b, and G359.42a, which are the same as our sources Sgr~C~H3, Sgr~C~H4, and Sgr~C~11, respectively. Comparing our results to the SED fitting from that work (which also employs the ZT MYSO SED models), we find that we both derive answers consistent with MYSOs for all three sources, but our estimated physical values vary. For one thing, \citet{2024arXiv241009253C} report estimated values for stellar mass and other parameters by averaging over apparently dozens (perhaps hundreds) of models with a much wider range of goodness-of-fit, whereas we report the results of the best fit model (as well as the range of our group of best fits, which all have similar goodness of fit and typically number $\sim$10 models in total; see Section~\ref{sec:cps}.). However, from the figures in \citet{2024arXiv241009253C} one can see they get best-fit stellar masses of 12, 8, and 96\,$M_{\sun}$ for Sgr~C~H3, Sgr~C~H4, and Sgr~C~11, respectively, which one can compare to the 32, 8, and 24\,$M_{\sun}$ that we derive. While the best-fit stellar mass value for Sgr~C~H4 seems to match in this comparison, the other two sources are quite different (especially Sgr~C~11). If we instead compare averaged values, using their model averaging methodology \citet{2024arXiv241009253C} estimate stellar mass values of 20.7$_{-8.4}^{+14.1}$, 20.4$_{-11.0}^{+24.1}$, and 8.5$_{-5.0}^{+11.9}$\,$M_{\sun}$ for Sgr~C~H3, Sgr~C~H4, and Sgr~C~11, respectively. If we average over our groups of approximately a half-dozen best-fit models for each source we get mean stellar masses of 30.4$_{-3.5}^{+1.6}$, 9.3$_{-1.3}^{+3.3}$, and 15.3$_{-6.3}^{+7.1}$\,$M_{\sun}$ for Sgr~C~H3, Sgr~C~H4, and Sgr~C~11, respectively. Thus, their averaged model masses are consistent with both our best fit mass values and our averaged mass values to within their reported errors.

However, the main reason why we don't get exactly the same results using the same data are by-and-large the product of which data we use as nominal estimates of the source flux densities and which we set as upper limits. In \citet{2024arXiv241009253C}, it is assumed that all flux data at wavelengths $<$10\,$\mu$m are upper limits due to the possible presence of PAH contamination. Since this data has the best resolution for resolving sources, we instead choose to test our sources for possible PAH contamination (see Section~\ref{sec:cps}) and include the near-infrared data when we can. This helps pin down the SEDs at the shortest wavelengths. For the three sources under discussion here, this methodology led to us being able to use all four Spitzer-IRAC bands in our fit of Sgr~C~11, and use the 4.5\,$\mu$m data as nominal data in the fits for Sgr~C~H3 \citep[as the 4.5\,$\mu$m IRAC filter does not encompass any bright PAH features; see][]{2006AJ....131.1479R}. We determined that we could not accurately isolate the near-infrared flux of source Sgr~C~H4 because it is too faint and unresolved due to crowding of nearby Sgr~C~H3 (as well as some extended emission), so we set all IRAC fluxes as upper limits. \citet{2024arXiv241009253C} measure IRAC fluxes for this source, though the errors are large and nearly the same value as the reported fluxes. Importantly, however, it should be noted that for all flux densities $\leq$37\,$\mu$m, the reported values from \citet{2024arXiv241009253C} are very similar to ours, except for the 25 and 37\,$\mu$m values for Sgr~C~11 which are a factor of two lower due to the fact that they employed an aperture approximately half the size of the one we used for this source. This is the result of another difference in methodology, i.e. using resolution-dependent apertures vs. fixed apertures for photometry. Since sources can intrinsically have different sizes as a function of wavelength, can be subject to source crowding at certain wavelengths and not others, and because of the large range in data resolutions being used ($2-35\arcsec$), we chose to find the optimal aperture for each source at each wavelength independently. \citet{2024arXiv241009253C} used a method where they found the optimal aperture at 37\,$\mu$m for each source and used that same aperture for all data at wavelengths $\leq$37\,$\mu$m. In the case of Sgr~C~11, there is extended and complicated environmental emission around the source at 37\,$\mu$m that makes the choice of optimal aperture size more subjective, however, the source is free of extended environmental emission at 25\,$\mu$m and a clear background sky level (and hence photometry aperture) can be ascertained. It is therefore evident that source flux is being missed in the smaller aperture of \citet{2024arXiv241009253C} at both SOFIA wavelengths. 

At the longest wavelengths, \citet{2024arXiv241009253C} use the Herschel photometry as source flux estimates, whereas we choose to use these values as upper limits. While one can more confidently use Herschel photometry data in the case of isolated sources, G\ion{H}{2} regions have many MYSOs, often very close to each other, and often embedded within or close to large extended dust substructures. With resolutions of 6, 12, 18, 24, and 35$\arcsec$ at 70, 160, 250, 350, and 500\,$\mu$m with Herschel \citep{2016A&A...591A.149M}, separate MYSOs and larger-scale structures near each other at shorter wavelengths merge into single sources due to lack of resolution. To address this, the method used by \citet{2024arXiv241009253C} involved making decisions about how much flux from the unresolved sources at longer wavelengths should be assigned to which sources based upon their ratio of resolved fluxes at a shorter wavelength. In the case of the close-together sources Sgr~C~H3 and Sgr~C~H4, they chose using the flux ratio between the sources at 37\,$\mu$m (3:1), to apply to their unresolved fluxes measured at all wavelengths $\ge$70\,$\mu$m. This is likely to be a very rough approximation in general as such a ratio likely doesn't hold constant at wavelengths both shorter than and longer than the SED turnover, especially if the real SEDs of the two sources peak at different wavelengths. Furthermore, estimating proper backgrounds at Herschel wavelengths on small scales is also difficult due to large-scale molecular cloud emission, as well as galactic cirrus which can dominate the observed emission, especially at $\lambda\geq250$\,$\mu$m. Where one should choose to select the background is often not at all obvious but has a huge influence on the values reported as a source’s background subtracted flux density (very much akin to the 37$\mu$m photometry issue of Sgr~C~11 just mentioned). Given all of these uncertainties, we choose to use the Herschel fluxes as upper limits in our modeling.

It is clear that, because of all of these differences in methodology, our SED fitting results and those of \citet{2024arXiv241009253C} should not be exactly the same even though the same data and models are being used. While it is encouraging that our results here agree with theirs to within the errors, this is a comparison of only three sources, and in general such similarities may not always be the case.

\subsubsection{Comparisons to Cotera et al. 2024 Source Catalog}

When this paper was in an advanced state, a catalog of point sources derived from the SOFIA-FORCAST data of the Galactic Center was published by \citet{2024ApJ...973..110C}. We cross referenced the point sources found in that work with our list of sources for Sgr~B and Sgr~C produced from the same data. All of the sources we identified, except for source Sgr~B2~11, are also found in the source list of \citet{2024ApJ...973..110C}. However, the \citet{2024ApJ...973..110C} catalog had far more sources than we found: 127 sources total in Sgr~B (compared to our 58), and 84 in Sgr~C (compared to our 47). We checked the observational properties of all \citet{2024ApJ...973..110C} sources not in common with our lists, and the reasons why they were not included in our list were: 1) the source was not detected at shorter or longer wavelengths than SOFIA (i.e., either in the Spitzer 8 and 5.8\,$\mu$m images, the Herschel 70\,$\mu$m images, or the cm radio continuum images); 2) the source was unresolved from another nearby source or from the larger extended emission at multiple wavelengths (in these cases our photometry generally covered the entire emitting region under the assumption that the source was a single elongated compact source, rather than multiple sources); 3) multiple close-together sources were part of what we considered to be a single extended source (and thus we report it as an extended source in our tables); 4) the source had a very broad peak at multiple wavelengths and was embedded in extended emission (and therefore making it difficult to determine if it is actually an independent source or simply a slightly more condensed part of a larger diffuse emission region); 5) the source peak moved around as a function of wavelength (indicating it is not likely to be internally heated).  Related to points 2-4 above, the source selection methodology used by \citet{2024ApJ...973..110C} is likely to give false positives for elongated and extended sources, as the algorithm tends to try to break these structures up into multiple point sources, which is often unlikely to be correct.

In all cases, if there was a peak detected at 70\,$\mu$m at the location of a SOFIA source or peak, it is included in our list, as the object of our source selection was not to find all the peaks in the SOFIA data \citep[which was the object of][]{2024ApJ...973..110C}, but to find compact sources likely to be internally-heated MYSOs. Therefore, we believe that our selection of MYSO candidates in this work is complete to within the detection limits and resolution of the SOFIA data. 

Additionally, we randomly selected several of the sources from our lists and spot-checked their photometry with the values reported in \citet{2024ApJ...973..110C}. In all cases the values appear to be in agreement to within their combined errors.

\subsubsection{Potential Contaminants to the MYSO Candidates}\label{sec:MYSOs}

Unlike many of our G\ion{H}{2} regions previously studied, this study is not the first mid-infrared survey looking for YSOs in the Galactic Center Central Molecular Zone. Most similar to the work presented here (in terms of spatial resolution and wavelength) is that of \citet{2009ApJ...702..178Y}, which leverages Spitzer 24\,$\mu$m data in addition to the four Spitzer-IRAC bands in a search for MYSOs via SED fitting. That work covered not only Sgr~B and Sgr~C, but the entire CMZ. Though plagued with saturation issued, especially in Sgr~B1 and Sgr~C, many of the sources identified in that survey are found in ours (see Tables 2 \& 3). However, we do identify far more MYSO candidates within the three G\ion{H}{2} regions than in that work.

Since the publication of \citet{2009ApJ...702..178Y}, it has been pointed out by several authors that the shapes of near-to-far infrared SEDs of MYSOs created via broadband photometry are not unique, and that other objects can have very similar SEDs. These mostly involve far more evolved stellar objects like: red giant stars and asymptotic giant branch (AGB) stars, post-AGB stars, including proto-planetary nebulae (proto-PNe), planetary nebulae (PNe), and even heavily-extinguished main-sequence stars . In our previous studies of Galactic G\ion{H}{2} regions, we ignored the possible contamination of such evolved sources since they are not commonly found in young massive star formation regions. However, the closeness of Sgr~B1, Sgr~B2, and Sgr~C to the Galactic Center means that these regions are subject to rapid dynamical and environmental changes atypical for G\ion{H}{2} regions farther out in the Galactic plane. As a consequence, more evolved interlopers are far more common in the CMZ (see Section \ref{sec:alm}) and thus possible contamination of these sources in the MYSO counts is possible. We will discuss these potential sources of contamination to our MYSO survey in more detail below.

{\it Main Sequence Stars --} It has been argued that main sequence stars with heavy foreground extinction \citep{2015ApJ...799...53K} may masquerade as YSOs, with their near-to-mid-infrared SEDs appearing similar to YSO SEDs. In particular, the models assume the main sequence stars sit in medium with a typical molecular cloud density range, and we see that (peculiar to these Galactic Center G\ion{H}{2} regions) most of our compact mid-infrared sources do not appear to be within molecular clouds. One advantage our survey has over the \citet{2009ApJ...702..178Y} survey is that we additionally have a 37\,$\mu$m photometry point. The models for non-YSO stars presented by \citet{2015ApJ...799...53K} show flat or decreasing flux from 25 to 37\,$\mu$m, while our sources have increasing fluxes. Moreover, many of our sources are resolved in the SOFIA data (i.e., $\gtrsim$0.2\,pc), indicating a large and extended dust envelope, which is more in line with the sizes of star-forming cores/clumps, and much larger than a main sequence star would appear. It is this dense envelope that is the origin of the 70\,$\mu$m flux seen ubiquitously from MYSOs, as predicted by the SED models for early B and O-type MYSOs. Significant 70\,$\mu$m flux is not expected for a main sequence star (especially one not surrounded by a dense dusty medium which appears to be the general case for the MYSOs in Sgr~B and Sgr~C). While we use the 70\,$\mu$m fluxes as an upper limit in our model fitting, all but a very few of our compact SOFIA sources are undetected at this wavelength (see Tables \ref{tb:BPACS}, \ref{tb:B2PACS}, and \ref{tb:CPACS}). Additionally, some of our sources also have extended green object (EGO) emission, which is almost exclusively associated with YSOs \citep{2008AJ....136.2391C}, and others have methanol maser emission or formaldehyde maser emission both of which are {\it only} associated with MYSOs \citep{2013MNRAS.435..524B,2015ApJS..221...10A}. Overall, we consider the likelihood of contamination by main sequence stars in the regions we are studying to be unlikely.

{\it Red Giant and AGB Stars --} It is argued \citep{2003A&A...405..531S} that YSOs can look the same in infrared colors and SEDs as red giant stars and AGB stars if heavily extinguished ($A_V\sim30$). Additionally, like YSOs, AGB stars can occasionally have OH and water masers \citep[e.g.,][]{2012A&A...547A..40U}, as well as PAH emission \citep[e.g.,][]{2023A&A...670A..97M}. However, while red giant and AGB stars can take on a wide variety of SED shapes, observationally only a small subset appear to mimic YSOs \citep[e.g.,][]{2007AJ....133.2310B,2022A&A...659A.145G,2006A&A...460..555B}, with most having their SEDs turn over at $\lesssim$20\,$\mu$m \citep[e.g.,][]{2000ApJ...530..408V, 1999A&A...352..587S}. Indeed, mid-infrared color-color diagrams using MSX \citep[e.g.,][]{2002AJ....123.2772S} and WISE data \citep[e.g.,][]{2024JKAS...57..123S} have been shown to effectively separate YSO populations from proto-PNe, AGB stars, and post-AGB star populations. More specifically, different color criteria are explored by \citet[][see their Figure 8]{2024JKAS...57..123S}, where they find using the longer wavelengths of WISE (i.e., 12 and 22\,$\mu$m) most clearly separate populations, indicating that the more evolved objects will for the most part appear different in their mid-infrared emission from MYSOs. Additionally, SiO masers are also often found in AGB stars, whereas this species of maser is very rarely seen in YSOs \citep[e.g.,][]{2016ApJ...826..157C}, and thus if present, these masers would indicate a likely AGB star. However, many of our sources have bright cm radio continuum emission, whereas red giant and AGB stars would not show detectable emission within the sensitivity of our radio data \citep[see, for instance,][]{1994ApJ...429L..33K, 2007AJ....133.2291M}. In fact, the presence of bright cm radio continuum in post-AGB stars is defined as the beginning of the PNe phase \citep{2017MNRAS.468.3450C,2022MNRAS.516.2235C}. We have cross-correlated our sample with as many red giant, AGB star, and SiO maser surveys as we could find, and have discovered that two sources are spectroscopically confirmed red giant stars (G359.42-0.02~2 and G359.0-0.09~8) and that three sources are associated with SiO maser emission (Sgr~B1~3, Sgr~B2~15, and G359.50-0.09~9) and thus likely to be AGB stars. In fact, Sgr~B1~3 is a confirmed AGB star \citep[more specifically a OH/IR star;][]{1997ApJ...478..206S}. As predicted, none of these sources have cm radio continuum emission and only G359.0-0.09~8 (and maybe Sgr~B1~3) has an SED like a MYSO.  We consider further contamination of our MYSO candidates by red giant and AGB stars to be the largest contaminant, being more likely than main sequence stars but still minor. 
 
{\it Post-AGB Stars --} Post-AGB stars, including proto-PNe and PNe, are perhaps the most difficult to distinguish from MYSOs. The outer dusty layers of gas ejected during the AGB phase will expand during this phase, often achieving sizes greater than our SOFIA resolution \citep[i.e., $\gtrsim$0.2\,pc;][]{2019A&A...630A.150G} and thus PNe could appear either resolved or unresolved in their infrared emission with SOFIA at the CMZ distance. The PNe nucleus is hot enough to heat and ionize the material from these circumstellar ejecta, so they can be readily seen in cm radio continuum emission \citep{2017MNRAS.468.3450C}, with about 50\% of PNe having detectable cm radio continuum \citep{2011MNRAS.412..223B}. Additionally, post-AGB stars and PNe can display water and hydroxyl masers \citep[e.g.,][]{2009A&A...505..217S,2002AJ....123.2772S}, and can also display PAH emission \citep[e.g.,][]{2002MNRAS.336...66R} -- all similar to YSOs. However, these similarities to YSOs are not the norm. Most importantly, only 10-20\% of PNe have detectable MIR emission with MSX \citep{2003IAUS..209...33C}, which has a 22\,$\mu$m detection limit comparable to the sensitivity of our 25\,$\mu$m SOFIA data. Moreover, only a small subset of all post-AGB stars and PNe that have detectable mid-infrared emission have SEDs like MYSOs, as they display great variability in their SEDs as a class \citep[e.g.,][]{2002ApJ...567..412V, 2000ESASP.456..191H}, with most SED turning over between 10-30\,$\mu$m \citep{2009ApJ...706..252Z}, unlike typical MYSOs. Indeed, as mentioned above, mid-infrared color-color diagrams using MSX data are used to separate star-forming clumps from proto-PNe, and post-AGB stars. Furthermore, unlike AGB stars which are plentiful, proto-PNe and PNe are rare, owing to the short lifetimes of these phases, lasting only a few thousand years for the proto-PNe phase, with the entire PNe lifetime being roughly only 20,000 years \citep{2020Sci...369.1497D}. Being in the turbulent CMZ, any PNe there may disperse even quicker, making this phase very short and thus the occurrence of such sources relatively rare. We believe that post-AGB stars, including proto-PNe and PNe, are unlikely to be a significant contaminant to our MYSO survey.

To avoid some of these issues of misidentifying MYSOs, \citet{2011ApJ...736..133A} used Spitzer IRS ($\sim$5-35\,$\mu$m) spectra to search for signs of the 15.4\,$\mu$m shoulder on the absorption profile of the CO$_2$ ice feature due to the mixing of CO$_2$ ice with methanol ice on grains. This is a signature known only to exist in the spectra of YSOs. However, while sources with this signature are extremely likely to be YSOs, sources without the feature may or may not be YSOs. In addition to variability of chemical abundances and potential environmental effects, MYSOs are known to generate very energetic outflows which clear out the overlying material, and thus at some orientations we would expect to not see such absorption features (i.e. when preferentially looking at pole-on or near-pole-on angles). Furthermore, such observations are extremely dependent upon proper background subtraction, which could erase such spectral signals in legitimate MYSOs. That being said, we believe that the majority of sources without this spectral feature are likely to not be MYSOs. 

In light of all of this, we have defined our sources differently in this paper compared to our previous papers. We have tabulated in Tables \ref{tb:Bseds} and \ref{tb:Cseds} (as well as Table \ref{tb:OTHERseds}) several additional physical properties of our sources taken from the literature and our data. We indicate if the source is resolved or not in the SOFIA data, and if it is well-fit by the MYSO fitter. We further indicate if the source is detected at cm radio wavelengths and/or 70\,$\mu$m. We tabulate which sources are known to have methanol, formaldehyde, hydroxyl, water, and/or SiO masers. We further indicate which sources have EGO emission. Additionally, if a source has been observed spectrally with Spitzer-IRS we indicate if it was found to have a 15.4\,$\mu$m shoulder or not. We also indicate which sources are known red giant and AGB stars from previous studies. Finally, we searched the GAIA Data Release 3 (DR3) catalog for coincidences (with separations $\lesssim$2$\arcsec$) with measured parallactic distances that might indicate a mid-infrared source is a field star unrelated to the CMZ population. 

If a source is well-fit by the MYSO fitter, and has either methanol masers, formaldehyde masers, EGO emission, or was found to have a 15.4\,$\mu$m shoulder in its IRS spectrum, then we definitely mark it as a ``MYSO'' on Tables \ref{tb:Bseds}, \ref{tb:Cseds}, and \ref{tb:OTHERseds}. These types of emissions are {\it only} found associated with YSOs. If it is not well-fit by the MYSO fitter or has model fits with masses $<$8\,$M_{\sun}$, it is not considered a MYSO (though in the latter case it may still be a lower mass YSO). If the source has flat or decreasing flux with wavelength, or was found by GAIA to have a distance indicating it is a field star, it is also not considered to be a MYSO. Additionally, if it has SiO maser emission, it is concluded to be a AGB star and not a MYSO. Of the remaining sources, all are either resolved at SOFIA wavelengths, have cm radio continuum, 70\,$\mu$m, or EGO emission, or have masers, which means none are extinguished main sequence stars. If the source is well-fit by the SED fitter and has cm radio continuum emission it is likely not an AGB star, though there is still a relatively small chance it is a post-AGB star or PNe. We will identify such sources as ``Likely MYSOs''. All other sources will be considered ``Possible MYSOs'', as they, at a minimum, still are well-fit by the SED fitter. 

What is the likelihood of these ``Likely'' (or  ``Possible'') MYSO candidates being actual MYSOs? Unfortunately, there are no comprehensive direct measurements of the stellar density of AGB stars in the CMZ to understand better our most likely contaminant. According to \citet{2011ApJ...736..133A}, they find that about half of the previously identified YSO candidates that they observed do not display the 15.4\,$\mu$m shoulder, and thus are contaminants to the MYSO population. However, many of the claims for these YSOs were based upon less data than we present here, or come from surveys that covered fields throughout the CMZ and not just in the suspected star formation regions, so 50\% should be considered an extremely conservative upper limit on contaminants.

\begin{deluxetable*}{lccccccccccc}[ht]
\tabletypesize{\scriptsize}
\tablewidth{0pt}
\tablecaption{SED Fitting Parameters of Selected Compact Infrared Sources in Sgr~B1 and Sgr~B2 \label{tb:Bseds}}
\tablehead{\colhead{Source   }  &
           \colhead{  $L_{\rm obs}$   } &
           \colhead{  $L_{\rm tot}$   } &
           \colhead{ $A_v$ } &
           \colhead{  $M_{\rm star}$  } &
           \colhead{$A_v$ Range}&
           \colhead{$M_{\rm star}$ Range}&
           \colhead{ Best }&
           \colhead{ Well }&
           \colhead{ Reso- }&
           \colhead{  }&
           \colhead{  }\\[-6pt]
	   \colhead{        } &
	   \colhead{ ($\times 10^3 L_{\sun}$) } &
	   \colhead{ ($\times 10^3 L_{\sun}$) } &
	   \colhead{ (mag.) } &
	   \colhead{ ($M_{\sun}$) } &
       \colhead{(mag.)}&
       \colhead{($M_{\sun}$)}&
       \colhead{  Models   }&
       \colhead{ Fit? }&
           \colhead{ lved? }&
           \colhead{ Features }&
           \colhead{ MYSO? }
}
\startdata
\multicolumn{12}{c}{Sgr B1} \\
\hline
Sgr B1 A &      9.36 &     102 &       27 &     16 &    27 - 27 &  16 - 16 &  5 &Y      &Y &ice,cm,70   &Yes   \\ 
Sgr B1 C &      7.93 &    16.1 &      1.7 &     12 &   1.7 - 55 &   8 - 48 &  5 &Y      &Y &cm,70       &Likely    \\ 
Sgr B1 D &      42.9 &     153 &       12 &     32 &   3.4 - 27 &  24 - 48 & 10 &Y      &Y &cm,70       &Likely    \\ 
Sgr B1 1 &      35.6 &     113 &       27 &     16 &    27 - 80 &  16 - 24 & 11 &Y      &Y &cm,70       &Likely    \\ 
Sgr B1 2 &      3.64 &    16.1 &       39 &     12 &   2.6 - 67 &   8 - 16 & 11 &Y$^a$  &Y &cm,70       &Likely    \\ 
Sgr B1 F &      4.62 &    9.67 &       51 &      8 &   7.9 - 51 &   8 - 24 &  6 &Y      &Y &cm,70,W,H   &Likely    \\ 
Sgr B1 G &      10.8 &    31.5 &       22 &     16 &    12 - 81 &  12 - 48 &  7 &Y      &Y &cm,70       &Likely    \\ 
Sgr B1 H &      8.65 &    10.5 &      0.8 &      8 &   0.8 - 6.7&   8 -  8 &  5 &Y      &Y &ice,cm,70   &Yes   \\ 
Sgr B1 3 &      15.2 &    74.7 &      2.5 &     24 &   2.5 - 80 &  16 - 24 &  5 &Y$^a$  &Y &S,70        &No   \\ 
Sgr B1 4 &      4.96 &    9.95 &       42 &      8 &    40 - 50 &   8 -  8 &  5 &Y$^a$  &Y &cm,70       &Likely    \\ 
Sgr B1 5 &      4.02 &    9.67 &       45 &      8 &    17 - 53 &   8 -  8 & 13 &Y      &Y &cm,70       &Likely    \\ 
Sgr B1 6 &      10.8 &    49.1 &       53 &     12 &    19 - 67 &   8 - 16 &  9 &Y      &Y &cm,70       &Likely    \\ 
Sgr B1 7 &      4.52 &    9.67 &       31 &      8 &    14 - 37 &   8 -  8 & 13 &Y      &Y &cm,70?      &Likely    \\ 
Sgr B1 8 &      1.34 &     158 &      8.4 &     32 &   2.6 - 34 &  12 - 32 &  9 &Y      &Y &ice         &Yes      \\ 
Sgr B1 9 &      3.39 &    28.8 &       63 &     16 &    32 - 63 &   8 - 24 &  5 &Y      &Y &70          &Possible   \\ 
Sgr B1 10 &     8.57 &    74.7 &      9.2 &     24 &   3.4 - 31 &   8 - 32 & 10 &Y      &Y &cm,70,      &Likely    \\ 
Sgr B1 11 &     4.23 &    10.2 &      8.4 &      8 &   1.7 - 27 &   8 - 16 &  7 &Y      &Y &cm,70       &Likely    \\ 
Sgr B1 12 &     13.6 &    49.4 &       27 &     12 &   5.9 - 39 &  12 - 32 & 12 &Y$^b$  &Y &no\,ice,cm,70   &No?$^g$  \\ 
\hline
\multicolumn{12}{c}{Sgr B2} \\
\hline
Sgr B2 1 &      1.00 &     19.6 &    140 &     12 &   130 - 196 &   8 - 24 &  5 &Y      &Y &ice,70      &Yes    \\ 
Sgr B2 2 &      1.72 &       81 &    204 &     24 &   160 - 243 &  16 - 24 &  6 &Y$^a$  &Y &70          &Possible      \\ 
Sgr B2 3 &      2.88 &     11.8 &    8.4 &      8 &   8.4 - 212 &   8 - 48 &  5 &Y$^a$  &Y &70,M,F,H    &Yes  \\ 
Sgr B2 AA &     4.55 &     88.4 &    223 &     24 &    68 - 223 &   8 - 24 &  5 &Y      &Y &cm,70       &Likely   \\ 
Sgr B2 4 &      67.8 &      294 &    556 &     24 &   503 - 556 &  24 - 24 & 15 &Y$^c$  &Y &70,M,F,W,H  &Yes  \\ 
Sgr B2 H &      47.6 &      457 &     25 &     48 &   8.4 -  36 &  32 - 64 &  8 &Y      &Y &cm,70,M,F,H &Yes   \\ 
Sgr B2 5 &      2.14 &     9.48 &     81 &      8 &   5.3 -  81 &   8 - 12 &  6 &Y      &Y &ice?        &Possible    \\ 
Sgr B2 6 &      0.67 &     0.57 &    3.4 &      2 &   3.4 -  77 &   2 - 16 &  6 &Y$^d$  &N &70          &No$^{e,h}$       \\  
Sgr B2 BB &     15.0 &      213 &    168 &     32 &    53 - 180 &  24 - 64 &  5 &Y      &Y &cm,70       &Likely    \\ 
Sgr B2 L &      5.07 &     88.4 &    151 &     24 &    34 - 151 &   8 - 32 &  5 &Y      &Y &cm,70       &Likely     \\ 
Sgr B2 7 &      0.84 &      158 &    235 &     32 &   201 - 236 &  16 - 32 &  6 &Y$^a$  &Y &70          &Possible     \\ 
Sgr B2 O &      10.1 &      617 &    244 &     64 &   185 - 260 &  12 - 64 &  7 &Y      &Y &cm,70       &Likely    \\ 
Sgr B2 8 &      3.70 &     26.6 &    231 &     12 &   220 - 262 &   8 - 48 &  5 &Y      &Y &70          &Possible?$^{h}$    \\ 
Sgr B2 9 &      1.07 &     2.59 &     27 &      2 &    19 -  46 &   2 - 16 &  8 &Y      &Y &70?         &Possible?$^{e,f}$      \\ 
Sgr B2 P &      9.99 &      196 &     59 &     32 &    29 -  80 &  12 - 32 &  7 &Y      &Y &ice,cm,70   &Yes   \\ 
Sgr B2 R &      29.9 &     38.6 &    4.2 &     16 &   0.8 -  34 &  16 - 24 &  7 &Y      &Y &ice,cm,70,E &Yes  \\ 
Sgr B2 10 &     3.80 &     28.8 &     18 &     16 &   5.3 -  29 &   8 - 24 &  9 &Y      &Y &ice,70      &Yes        \\ 
Sgr B2 11 &     4.20 &     11.8 &    8.4 &      8 &   2.6 -  17 &   8 - 16 &  8 &Y      &Y &70          &Possible    \\ 
Sgr B2 12 &     7.71 &     8.83 &    7.5 &      8 &   5.0 -  24 &   8 -  8 &  5 &Y      &Y &70          &Possible      \\ 
Sgr B2 13 &     1.68 &     11.2 &     17 &      8 &   8.4 -  34 &   8 - 16 & 10 &Y      &Y &ice?,70     &Possible       \\ 
Sgr B2 14 &     8.72 &     31.5 &     48 &     16 &    18 -  50 &  16 - 48 &  5 &Y      &Y &70?         &Possible     \\ 
Sgr B2 15 &     4.79 &     24.7 &    1.7 &     16 &   1.7 -  14 &  16 - 16 &  7 &Y$^d$  &Y &S           &No       \\ 
Sgr B2 16 &     7.43 &      119 &    243 &     24 &   233 - 252 &  16 - 32 &  5 &Y      &Y &70          &Possible      \\ 
Sgr B2 17 &     7.97 &      158 &     24 &     32 &    19 -  34 &  16 - 32 &  8 &Y      &Y &70          &Possible       
\enddata
\tablecomments{\footnotesize Abbreviations in the ``Features'' column are: 
cm = cm radio continuum emission; 
70 = 70\,$\mu$m emission; 
70? = 70\,$\mu$m emission present but unresolved from nearby sources; 
M=class II methanol maser \citep{2022AA...666A..59N};
F=formaldehyde maser \citep{1994ApJ...434..237M}; 
W=water maser \citep{2014MNRAS.442.2240W,1993ApJ...402L..69M};
H=hydroxyl maser \citep{2015MNRAS.451...74G, 1993ApJ...402L..69M}; 
S=SiO maser \citep{2002AA...393..115M,1997ApJ...478..206S};
E=EGO emission \citep{2009ApJ...702..178Y}; 
and from \citet{2011ApJ...736..133A}: 
ice=15.4\,$\mu$m ice feature, 
ice?= maybe 15.4\,$\mu$m ice feature, 
no ice= no 15.4\,$\mu$m ice feature.}
\tablenotetext{a}{Only two nominal data points used in the SED fitting, thus the results of the modeling are less reliable.} 
\tablenotetext{b}{The extended emission from this source lies partially off-field at both 25 and 37\,$\mu$m which may affect the accuracy of the resultant parameters from the SED fits.}
\tablenotetext{c}{Only one nominal data point used in the SED fitting. However,  since all resultant fits are poor (they all violate lower limit set at 160\,$\mu$m), the results from the SED fitting are unreliable.}
\tablenotetext{d}{This source has a flat or decreasing flux with wavelength and is thus likely not a MYSO.}
\tablenotetext{e}{Has SED model fits less than 8\,$M_{\sun}$. If well fit by SED fitter, it may be a low-to-intermediate mass YSO.}
\tablenotetext{f}{Most fits imply a MYSO.}
\tablenotetext{g}{Has no 15.4\,$\mu$m ice feature, but no other indicators point to it not being a MYSO.}
\tablenotetext{h}{Coincident with a GAIA source to within 2$\arcsec$ with a measured parallactic distance placing it in the foreground.}
\end{deluxetable*}

\begin{deluxetable*}{lccccccclccc}[ht]
\tabletypesize{\scriptsize}
\tablecolumns{1}
\tablewidth{0pt}
\tablecaption{SED Fitting Parameters of Selected Compact Infrared Sources in Sgr~C \label{tb:Cseds}}
\tablehead{\colhead{   Source   }  &
           \colhead{  $L_{\rm obs}$   } &
           \colhead{  $L_{\rm tot}$   } &
           \colhead{ $A_v$ } &
           \colhead{  $M_{\rm star}$  } &
           \colhead{$A_v$ Range}&
           \colhead{$M_{\rm star}$ Range}&
           \colhead{ Best }&
           \colhead{ Well }&
           \colhead{  }&
           \colhead{  }&
           \colhead{  }\\[-6pt]
	   \colhead{        } &
	   \colhead{ ($\times 10^3 L_{\sun}$) } &
	   \colhead{ ($\times 10^3 L_{\sun}$) } &
	   \colhead{ (mag.) } &
	   \colhead{ ($M_{\sun}$) } &
       \colhead{(mag.)}&
       \colhead{($M_{\sun}$)}&
       \colhead{  Models   }&
       \colhead{ Fit? }&
           \colhead{ Resolved? }&
           \colhead{ Features }&
           \colhead{ MYSO? }
}
\startdata
\multicolumn{12}{c}{Sgr C} \\
\hline
Sgr C 1  &      3.29 &      40.7 &       84 &     12 &    37 -  84   &   8 - 12 &  7 &Y      &Y &no\,ice,cm,70   &No?$^g$  \\
Sgr C 2  &      14.2 &       147 &       23 &     32 &    22 -  27   &  12 - 32 &  5 &Y      &Y &no\,ice         &No?$^i$   \\
Sgr C 3  &      3.98 &      10.2 &       23 &      8 &   2.6 -  307  &   8 - 24 & 11 &Y      &Y &70              &Possible    \\ 
Sgr C 4  &      0.52 &      0.79 &      3.4 &      4 &   3.4 - 109   &   4 -  8 &  7 &Y$^a$  &N &                &No$^{e}$   \\ 
Sgr C 5  &      14.0 &      33.1 &      3.4 &     16 &   2.5 -   5.0 &  12 - 16 &  6 &Y      &Y &cm,70           &Likely    \\
Sgr C 6  &      3.56 &      80.5 &      13  &     24 &   7.9 -  25   &   8 - 24 &  8 &Y      &Y &70              &Possible    \\ 
Sgr C 7  &      7.23 &      9.45 &       10 &      8 &   4.2 -  14   &   8 -  8 &  8 &Y      &Y &no\,ice,cm,70   &No?$^g$  \\
Sgr C 8  &      29.1 &      82.0 &       16 &     24 &    16 -  22   &  24 - 24 &  7 &Y      &Y &cm,70           &Likely    \\
Sgr C 9  &      0.51 &      0.79 &       17 &      4 &    10 -  64   &   4 - 4  &  5 &Y$^a$  &N &70              &No$^{e}$   \\ 
Sgr C 10 &      3.05 &      9.48 &       17 &      8 &   5.3 -  40   &   8 - 24 & 15 &Y      &Y &no\,ice,cm,70?  &No?$^g$  \\
Sgr C 11 &      4.02 &      71.4 &      6.7 &     24 &   0.8 -  59   &   8 - 24 &  6 &Y      &Y &70,W            &Possible    \\ 
Sgr C 12 &      13.3 &      19.9 &       15 &     12 &   5.9 -  39   &  12 - 16 &  7 &Y      &Y &70              &Possible    \\ 
Sgr C 13 &      0.19 &      0.24 &      9.2 &      2 &   9.2 - 134   &   2 -  4 &  7 &Y      &N &70?             &No$^{e}$   \\ 
Sgr C 14 &      0.65 &      0.79 &       11 &      4 &   4.2 -  27   &   4 -  8 &  7 &N$^a$  &Y &70?             &No$^{e}$   \\ 
Sgr C H3 &      12.2 &       147 &       44 &     32 &    41 -  53   &  24 - 32 &  5 &Y      &Y &cm,70,M,F,E     &Yes   \\
Sgr C H4 &      3.72 &      11.7 &       66 &      8 &    56 - 212   &   8 - 16 &  6 &Y$^a$  &Y &70?,M,F         &Yes   \\
Sgr C H1 &      1.16 &      13.6 &       27 &     12 &   2.6 -  34   &  12 - 16 &  8 &Y$^a$  &Y &cm,70,M         &Yes   \\
Sgr C 15 &      3.66 &      9.48 &      0.8 &      8 &   0.8 -  42   &   8 - 16 &  6 &Y$^a$  &Y &cm,70           &Likely    \\
Sgr C 16 &      6.15 &      33.8 &      186 &     12 &    25 - 193   &   8 - 16 &  7 &Y$^a$  &Y &70              &Possible  
\enddata
\tablecomments{Symbols same as for Table \ref{tb:Bseds}. W here is the water maser detection from \citet{2019ApJ...872..171L}.}
\tablenotetext{i}{Spectroscopically determined to be a K or M red giant by \citet{2022ApJ...930...16J}.}
\end{deluxetable*}

\subsubsection{Identifying MYSO Candidates Among the Compact Infrared Sources}

Based upon the information compiled in Tables~\ref{tb:Bseds} and \ref{tb:Cseds} we will discuss below the properties of the compact mid-infrared sources contained within the confines of Sgr~B1, Sgr~B2, and Sgr~C individually. In particular, we will identify those sources thought to be MYSO or MYSO candidates and discuss estimates to the present MYSO stellar densities in each G\ion{H}{2} region to compare to the values we have previously derived for other Milky Way G\ion{H}{2} regions. As mentioned in our previous papers, given the angular resolution limitations of FORCAST ($\sim$3$\arcsec$) and the distance to these G\ion{H}{2} regions ($\sim$8.0\,kpc), we can only resolve structures as small as $\sim$0.11\,pc. Therefore, it is likely that in many cases, the infrared sources discussed here contain protobinaries or even protoclusters. While the assumption of a single MYSO is reasonable when the core contains a dominant primary MYSO, we cannot be certain that this would be the case in general. However, these CMZ G\ion{H}{2} regions are all at a distance similar to the average distance of the regions we have studied so far (6.6\,kpc), and thus we are making comparisons using approximately the same level of spatial information. With that caveat in mind, as we have done for our previously studied G\ion{H}{2} regions, in this section we will also discuss the brightest SOFIA source in each region, as our past observations have shown that the sources that are brightest at the wavelengths of our SOFIA data generally trace the most massive stars in the present MYSO population. We will additionally use the results from the SED fitting and give the derived mass estimates of these highest mass sources in each G\ion{H}{2} region. Since the stellar mass distribution function appears to be more or less universal from modest to the most massive star-forming clusters \citep{2002Sci...295...68P}, this would imply that the more massive the highest mass member, the larger the underlying stellar cluster total mass in general. Therefore, we have previously used the masses of the most massive MYSOs as an additional rough proxy for understanding the relative sizes of the presently-forming stellar populations within each G\ion{H}{2} region we have studied. However, it is unknown if the stellar mass functions that seem universal in stellar clusters in the Galactic plane should even apply to those in the CMZ given the unique environment and rapid dynamical evolution. With that caveat, and in keeping with our prior studies, we will report the highest mass MYSO for each CMZ G\ion{H}{2} region as well as their MYSO densities below and compare and contrast them to our results from other G\ion{H}{2} regions. 

\paragraph{Sgr~B1 Compact Sources}\label{sec:sgrb1}

\hspace{5pt} For the compact mid-infrared sources identified by SOFIA within the Sgr~B1 G\ion{H}{2} region, we conclude from the information complied in Table\,\ref{tb:Bseds} that there are 3 MYSOs (17\%) and 12 Likely MYSOs (67\%). One source falls into the Possible MYSO category, while two are found not be MYSOs. This means that 83\% of the compact mid-infrared sources in Sgr~B1 are MYSOs or Likely MYSOs. The results in Table\,\ref{tb:Bseds} show that the absolute best model fits for all the mid-infrared detected YSO candidates in all of Sgr~B1 yield protostellar masses in the range  $m_*$\,=\,8--32\,$M_{\sun}$, which is approximately equivalent to a range of ZAMS spectral type B1--O7 stars \citep{2000AJ....119.1860B}.

Two sources, Sgr~B1~D and Sgr~B1~8, are tied for the highest best-fit mass of 32\,$M_{\sun}$. To determine which one is most likely the most massive, we calculated a non-weighted average mass from the group of best fits for each source, and find Sgr~B1~D has by far the highest of all sources at 36.8\,$M_{\sun}$ (Sgr~B1~8 was 19.5\,$M_{\sun}$). Though Sgr~B1~D is categorized as a Likely MYSO, the values for it's best fit luminosity and average luminosity from all fits (1.5$\times$10$^{5}$ and 2.5$\times$10$^{5}$\,$L_{\sun}$, respectively) are very large. This close to (or over, in the case of the average luminosity) the largest theoretical limit of for AGB luminosity of $\sim$1.5$\times$10$^{5}$\,$L_{\sun}$, for the most massive AGB stars (ranging from metal-rich to metal-poor, i.e. $0.001 < Z < 0.04$; see \citealt{2014MNRAS.439..977V,2020A&A...641A.103V}). If this source is indeed at the distance of the CMZ it could not be an AGB star, adding further confidence to it being a legitimate MYSO.

For the two sources that are labeled as not being MYSOs, the emission of Sgr~B1~3 is coincident the location of the OH/IR star OH~0.548-0.059 and has SiO masers \citep{1997ApJ...478..206S} typical of AGB stars. The infrared SED of this source is flat, with saturated fluxes at 4.5 and 8.0\,$\mu$m, something not expected for MYSOs in highly extinguished environments. The other source, Sgr~B1~12 was found to not have a 15.4\,$\mu$m shoulder on the absorption profile of the CO$_2$ ice feature, however its infrared SED looks very similar to a MYSO.

Based upon the number of MYSOs and Likely MYSOs combined, we calculate a MYSO density of 0.13~MYSOs/pc$^2$. This would mean that Sgr~B1 has a MYSO density slightly smaller but similar to both W49A (\citetalias{2021ApJ...923..198D}) and W51A:G49.4-0.3 (\citetalias{2019ApJ...873...51L}) of 0.15~MYSOs/pc$^2$, but below the average of all G\ion{H}{2} regions we have studied thus far (0.18~MYSOs/pc$^2$). Sgr~B1 also has a total MYSO content (15) similar to both NGC~3603 (14; \citetalias{2022ApJ...933...60D}) and W49A (22), which are the first and second most-luminous G\ion{H}{2} regions in the Galaxy. If we used our extremely conservative lower limit estimate that half of the unconfirmed MYSOs (i.e., the Likely plus Possible MYSOs) are true MYSO and add that number (6.5) to the 3 confirmed MYSOs, the MYSO density value for Sgr~B1 would drop to 0.08 MYSOs/pc$^2$, which would place it on the lower end range of G\ion{H}{2} stellar densities, but still higher than the least dense Galactic plane G\ion{H}{2} region we have studied (i.e., 0.05 MYSOs/pc$^2$ in NGC~3603). At the upper limit, if we add in the one Possible MYSO with the confirmed and Likely MYSOs, this value basically stays the same at 0.13~MYSOs/pc$^2$.

While the more extended regions like the Ionized Rim and Sgr~B1~E are the sources with the highest integrated intensity at 6\,cm, the brightest compact source at 6\,cm is Sgr~B1~D \citep[120 mJy;][]{1992ApJ...401..168M}, which, as we mentioned above, is the most massive source in the region, and which is also the brightest MIR source at both 25 (88\,Jy) and 37\,$\mu$m (131\,Jy).

Besides the lettered sources in Table \ref{tb:Bseds}, which were defined via their radio continuum emission at cm wavelengths, there are additionally several of the MIR-defined compact sources (i.e., the numbered sources which do not already have radio labels) in Sgr~B1 that are coincident with radio continuum peaks or sources. In total, 9 of the 12 sources identified in the mid-infrared are associated with 6\,cm and/or 20\,cm radio continuum peaks (see Table \ref{tb:Bseds}), with the definite exceptions being Sgr~B1~3, Sgr~B1~8, and Sgr~B1~9. Sgr~B1~3, as we have said, is an AGB star, but Sgr~B1~8 is a confirmed MYSO. In our previous studies of the MYSO populations of G\ion{H}{2} regions, we have cataloged many sources that appear to be MYSOs but do not have radio continuum emission, and contend that these may be sources that are at a more youthful phase of stellar evolution, prior to the onset of compact \ion{H}{2} regions. Sgr~B1~8 is likely a similar object. Though we include Sgr~B1~2 and Sgr~B1~10 in the list of cm continuum emitters, we caution that they both have potential issues. Sgr~B1~2 lies close to a brighter radio sources and appears as tongue of emission extending out from that bright source, so its radio peak is not well-defined. The radio emission for Sgr~B1~10 is weak and consistent with high frequency noise artifacts seen across the radio image, however we tentatively assert that this emission is real due to its coincidence in location with the MIR source and the fact that it has a very similar shape and extent. 

Sgr B1 E are the sources with the highest integrated intensity at 6\,cm, they are not considered compact sources in our study. The brightest compact source at 6\,cm is, Sgr~B1~D \citep[120 mJy;][]{1992ApJ...401..168M}, and this is also the brightest MIR source at both 25 (88\,Jy) and 37\,$\mu$m (131\,Jy). 

\paragraph{Sgr~B2 Compact Sources} \label{sec:sgrb2}

\hspace{5pt} For the Sgr~B2 G\ion{H}{2} region, of the 24 total mid-infrared compact sources found, we conclude (Table\,\ref{tb:Bseds}) that it contains 7 MYSOs (29\%) and 4 Likely MYSOs (17\%). The larger number of confirmed MYSOs compared to Sgr~B1 is a product of the larger number of sources with methanol and formaldehyde masers, as well as EGO-emitting sources. However, there are a large number of Possible MYSOs (10, or 42\%) which require further confirming evidence to assess their true nature. The results in Table\,\ref{tb:Bseds} show that the absolute best model fits for all the mid-infrared detected YSO candidates in all of Sgr~B2 yield protostellar masses in the range $m_*$\,=\,2--64\,$M_{\sun}$, which is approximately equivalent to a range of ZAMS spectral type A2--O4 stars.

According to the best-fit SED models, the most massive source in Sgr~B2 is Sgr~B2~O with a stellar mass of 64\,$M_{\sun}$, or the equivalent of a spectral type O4 ZAMS star. However, Sgr~B2~O is tied with Sgr~B2~H (a.k.a. Sgr~B2~South, or S) and Sgr~B2~BB for having the highest top mass given by its mass range from the group of best fits at 64\,$M_{\sun}$. Though Sgr~B2~H has the second highest best-fit mass in Sgr~B2, all 8 of the fits from the group of best fits lie within the range of 32-64\,$M_{\sun}$, and this source has the highest non-weighted average model fit (50.0\,$M_{\sun}$) of all compact sources identified and is thus likely to be the highest mass source.

Perhaps consistent with this, Sgr~B2~H has by far the brightest flux at 37\,$\mu$m among all compact sources, and is also the brightest peak after Sgr~B2~Main at 70\,$\mu$m in the Herschel data. Indeed, though the peak of Sgr~B2~Main is the brightest peak in the mid-infrared at the SOFIA and Herschel wavelengths, this source is extended in our images and is known to be comprised of multiple compact and ultracompact \ion{H}{2} regions. Interestingly, the third brightest peak at 70\,$\mu$m is radio source Sgr~B2~K (a.k.a. Sgr~B2~North, or N), which we do not see in the SOFIA data, likely due to extremely high extinction. 

For the two Sgr~B2 mid-infrared sources that are labeled as not being MYSOs in Table\,\ref{tb:Bseds}, source Sgr~B2~15 has an SED that is generally decreasing with wavelength from near to mid-infrared, atypical of a MYSO. We also found it to be coincident (to within 2$\arcsec$) of SiO maser emission, which means that this source is most likely an AGB star. Similarly, source Sgr~B2~6 has a strange but generally flat SED with wavelength, and most of the fits from the SED fitter imply a mass $\lesssim$4\,$M_{\sun}$ assuming a CMZ distance. However, this source is coincident with a GAIA object (Gaia DR3 4057530371326037248, with a separation of 0.77$\arcsec$, or a single FORCAST pixel) which has a measured parallactic distance of 5078\,pc, and thus is likely a source in the foreground of the CMZ. Similarly, source Sgr~B2~8 is also only 1.72$\arcsec$ from a GAIA source (Gaia DR3 4057530405685287424) that has a measured parallactic distance of only 588\,pc. However, this source separation is at the limit of our expected astrometric error, and given the high stellar density of field stars toward the Galactic Center it may be a chance alignment with a foreground star, especially given that the the SED fitter yielded good results and the SED is typical of a MYSO (as opposed to Sgr~B2~6). We therefore label this source as a Possible MYSO, but with a question mark in Table\,\ref{tb:Bseds}. Finally, like Sgr~B2~6, Sgr~B2~9 has a best fit of 2\,$M_{\sun}$, however unlike Sgr~B2~6, Sgr~B2~9 has an SED more typical of YSOs. Also, all fits for Sgr~B2~9 except the best fit SED, are $\gtrsim$12\,$M_{\sun}$, implying that it's likely a MYSO, with an average mass from all fits of 12.9\,$M_{\sun}$. However, given the higher uncertainty we also label this source as a Possible MYSO, but with a question mark in Table\,\ref{tb:Bseds}.

Besides the two sources we identify as not being MYSOs (8\%) in Sgr~B2, there are significantly more Possible MYSOs than we tabulated for Sgr~B1 with 11 (46\%), mostly due to a higher portion of sources with no detected cm radio continuum emission. As we have stated before, our previous surveys have revealed a great number of MYSOs that are at such a youthful stage that they are being detected in the mid-infrared prior to the onset of their UC\ion{H}{2} regions. This might point to Sgr~B2 as having a more youthful population of MYSOs than Sgr~B1, as further evidenced by the prolific number of observed UC\ion{H}{2} regions \citep{1995ApJ...449..663G, 2022A&A...666A..31M}, masers \citep{1997ApJ...474..346M, 2010MNRAS.404.1029C, 2004ApJS..155..577M}, and mm cores \citep{2018ApJ...853..171G} in Sgr~B2\footnote{Of all the mm cores identified by \citet{2018ApJ...853..171G}, Sgr~B2~2 is the only compact mid-infrared-identified sources coincident with a small cluster of these cores. Unresolved and extended mid-infrared emission is also present in and around the cluster of cores in Sgr~B2~Main.}.

Relatedly, several of the radio emitting sources in Sgr~B2 are seen in the SOFIA data at one or both wavelengths (e.g., Figure \ref{fig:SgrB2radio}). Radio sources that we detect clearly in the mid-infrared are AA, BB, H, L, O, P, and R (as well as V, but again, this source may not be related to Sgr~B2 given its anomalous velocity). These sources are all well-fit as MYSOs from the SED fitting of their infrared emission, with three being confirmed MYSOs via other indicators. The other four of these radio continuum emitters are categorized as Likely MYSOs. Given their proximity to the molecular clump centered on Sgr~B2~Main, their likelihood of being true MYSOs is further enhanced. Interestingly, we do not detect mid-infrared emission with SOFIA from radio sources K, Q, W, Y, or Z. Radio sources K, W, Y, and Z all lie north or west of the peak of Sgr~B2~Main which is thought to be the region of highest extinction, thus the reason for their non-detections (but these are all likely to house MYSOs). Source Q is perhaps barely detected at 37\,$\mu$m the the couple sigma level, but the emission is point-like. Since it is not observed at any shorter wavelengths we did not try to fit it with our SED models. Radio source K is a diffuse patch of extended radio continuum, and we detect slightly extended mid-infrared emission from Sgr~B2~4 on the southwest edge of its emission in the SOFIA data, but only at 37\,$\mu$m. It is unclear if this is emission leaking out of a less-extinguished part of radio source K, or is in fact a true MYSO \citep[it is not coincident with any compact radio or mm cores, of which there are many in this general area;][]{2022A&A...666A..31M,2018ApJ...853..171G}. Under the assumption that Sgr~B2~4 is an independent source, it is tenuously found to be a MYSO from the SED fitting because said fits are based upon a single nominal data point at 37\,$\mu$m (but it has highly constraining upper and lower limits at the other wavelengths), however the source is coincident to with 2$\arcsec$ of methanol, formaldehyde, water and hydroxyl masers, so it's categorization as such a MYSO seems more certain. 

In total, Sgr~B2 has 11 sources identified as either an MYSO or a Likely MYSO (46\%). Based upon this number, we calculate a MYSO density of 0.07~MYSOs/pc$^2$, which is consistent with the value for NGC~3603 of 0.05~MYSOs/pc$^2$, the least mid-infrared MYSO-dense Galactic plane G\ion{H}{2} region we have studied. This is also almost half what we calculated for Sgr~B1. This total number of confirmed and Likely MYSOs (11) is comparable to the MYSO content of the less-prolific G\ion{H}{2} regions W51A:G49.4-0.3 (10) and K3-50 (8). However, using our worst-case lower limit scenario (as we discuss for Sgr~B1 above), we derive a slightly higher value 0.09~MYSOs/pc$^2$, and similar to the lower limit we calculated for Sgr~B1. For the upper limit, if we include the Possible MYSOs with the confirmed and Likely MYSOs (22 total), the value for Sgr~B2 would rise to 0.14~MYSOs/pc$^2$, comparable to the Sgr~B1 upper limit and just below the average for Galactic Plane G\ion{H}{2} regions. Also at 22 MYSOs, the total number of MYSOs would be the same as one of the most prolific G\ion{H}{2} regions we have studied, W49A. 

\paragraph{Sgr~C Compact Sources}

\hspace{5pt} For the Sgr~C G\ion{H}{2} region we find, based upon the results reported in Table \ref{tb:Cseds}, that of the 19 infrared compact sources identified, 3 are confirmed to be MYSOs (16\%) and another 3 are Likely MYSOs (16\%). All three of the confirmed MYSOs are sources found within the IRDC located here, and all have methanol maser emission (and/or formaldehyde masers and EGO emission) which are tracers of current massive star formation. Sgr~C also has 5 Possible MYSOs (26\%). We do caution that 1 Likely MYSO (Sgr~C~15) and 1 Possible MYSO (Sgr~C~16) are well-fit with MYSO SED models but only two nominal data points used in the SED fitting, so their confidence is less robust than the others in their categories.

The absolute best model fits for the mid-infrared detected sources in all of the Sgr~C G\ion{H}{2} region yield protostellar masses in the range $m_*$\,=\,2--32\,$M_{\sun}$ (Table \ref{tb:Cseds}), which is approximately equivalent to a range of ZAMS spectral type A2--O7 stars. Sgr~C~H3 and Sgr~C~2 are tied for being the highest mass sources at 32\,$M_{\sun}$ for their absolute best-fit mass, however, if we calculate the non-weighted average mass from their groups of best fit models, Sgr~C~H3 is clearly more massive (30.4\,$M_{\sun}$) than the rest of the sources in the region. Consistent with this, the brightest peak in the mid-infrared, is Sgr~C~H3 (572 mJy/pix at 25\,$\mu$m and 1305 mJy/pix at 37\,$\mu$m), however in terms of integrated brightness, Sgr~C~8 is the brightest source in the mid-infrared. Sgr~C~8 is also the brightest peak at 6\,cm, which may not be surprising as it lies near the center of the extended Sgr~C \ion{H}{2} region.

If we count just the confirmed and Likely MYSOs (6), we measure a MYSO density of 0.08 MYSOs/pc$^2$, which is comparable to the value for confirmed and Likely MYSOs in Sgr~B2. For further comparison, 6 confirmed and Likely MYSOs is comparable to the MYSO content of the less-prolific G\ion{H}{2} regions M17 (7) and K3-50 (8). Deriving our worst-case lower limit MYSOs scenario (as we discuss for Sgr~B1 above), we derive a value of 0.09~MYSOs/pc$^2$, very similar to the lower limit values we derived for Sgr~B2 (0.09~MYSOs/pc$^2$) and Sgr~B1 (0.08~MYSOs/pc$^2$), but, again, still higher than the value of 0.05 MYSOs/pc$^2$ in NGC~3603 which we derived in the least dense Galactic plane G\ion{H}{2} region we have studied. In our upper limit case where all Likely and Possible MYSOs (8) are counted along with the 3 confirmed MYSOs we get a value of 0.14~MYSOs/pc$^2$, which again is similar to the upper limit cases for both Sgr~B2 (0.14~MYSOs/pc$^2$) and Sgr~B1 (0.13~MYSOs/pc$^2$) and a little below the average of all G\ion{H}{2} regions we have studied thus far (0.18~MYSOs/pc$^2$). 

Despite these similarities, Sgr~C is different than Sgr~B1 and Sgr~B2 in that it has a high incidence of mid-infrared sources that are suspected to not be MYSOs (8, compared to 2 for both Sgr~B1 and Sgr~B2). The two reasons for this are that half of these sources (Sgr~C~1, Sgr~C~2, Sgr~C~7, and Sgr~C~10) have been observed spectroscopically by \citet{2011ApJ...736..133A} and shown not to have the 15\,$\mu$m ice-shoulder indicative of a MYSO, and the remaining four sources (Sgr~C~4, Sgr~C~9, Sgr~C~13, and Sgr~C~14) not only have SED fits less than 8\,$M_{\sun}$, but they all have an average mass from all fits of less than 8\,$M_{\sun}$ (with Sgr~C~9 and Sgr~C~13 both having maximum masses of only 4\,$M_{\sun}$ in their range of best fits). These four sources with lower mass fits are all well-fit by the YSO fitting algorithm, and thus may be legitimate low-mass YSOs. However, this higher incidence of mid-infrared sources not being MYSOs may mean Sgr~C has an intrinsically different stellar population than Sgr~B1 and Sgr~B2, with potentially more interlopers and/or low-mass YSOs present. 

Of the 11 MYSO candidates (i.e., confirmed, Likely, and Possible combined) in Sgr~C, 5 (45\%) are associated with obvious 6\,cm and/or 20\,cm continuum peaks: Sgr~C~5, Sgr~C~8, Sgr~C~H3, Sgr~C~H1, and Sgr~C~15 (Table \ref{tb:Cseds}). This, again, provides confirming evidence of the MYSO nature of these sources. However, Sgr~C~1, Sgr~C~7, and Sgr~C~10 also have cm radio continuum but do not have detected 15\,$\mu$m ice-shoulder indicative of YSOs \citep{2011ApJ...736..133A}. Again, there are likely occasional exceptions to the rule, and a small subset of legitimate YSOs may lack this spectral ice feature, but it is unlikely that this is the case for all three sources. It would also be surprising if all sources were (proto)PNe given their rarity. These sources, therefore would be interesting to follow-up to determine their true nature.

There are only a few radio-defined sources in Sgr~C (see Figure \ref{fig:SgrCradio}), the main Sgr~C \ion{H}{2} region, and the UC\ion{H}{2} regions G359.436-0.102 (our source Sgr~C~H3) and G359.425-0.111 \citep{2000ApJ...530..371F}. Of these, the G359.425-0.111 UC\ion{H}{2} region is the only source without a detected mid-infrared component in the SOFIA data. Sgr~C~6 is interesting in that there is no radio continuum emission detected at its mid-infrared peak, however this source lies near (at least in projection) the non-thermal radio filament and there appears to be a ring of radio continuum emission around the source at both 6 and 20\,cm (Figure \ref{fig:SgrCradio}). However, Sgr~C~15 is also coincident with the non-thermal filament and has peaks at both 6 and 20\,cm, but the third mid-infrared source in/near the non-thermal filament is Sgr~C~16 which has no obvious radio emission.

When this manuscript was in an advanced state, a study by \citet{2024arXiv241009253C} that was based upon SED modeling using a combination of this SOFIA-FORCAST data and JWST-NIRCam data, claim that Sgr~C~H3 (their G359.44a), Sgr~C~H4 (their G359.44b), and Sgr~C~11 (their G359.42a) are MYSOs. In the work presented here, the first two are confirmed MYSOs, and the third is considered a Possible MYSO.

\subsection{Physical Properties of Extended Sub-Regions: Evolutionary Analysis and Status}\label{sec:es}

In our previous studies, we used the radio continuum maps to help us determine where the different extended sub-regions were that make up the larger G\ion{H}{2} region. These sub-regions were believed to be the star-forming molecular clumps within the larger molecular cloud hosting the G\ion{H}{2} because they not only had similar morphologies in their mid-infrared emission, but in their CO emission (e.g., $^{13}$CO $J=1-0$) and cold dust (far-infrared continuum) emission as well. The CO and mid- to far-infrared data allowed us to derive kinematic and evolutionary information for each sub-region. The relative ages of these regions were then used to infer something about the origin of the G\ion{H}{2} region and/or how it has evolved to its present appearance.

However, we see that the radio continuum regions within Sgr~B1, Sgr~B2, and Sgr~C have little to no correspondence to the clumps and structures seen in CO emission data, nor in the far-infrared/sub-mm continuum. In particular, we find that there is no CO or far-infrared emission corresponding directly with any large radio structure in the G\ion{H}{2} regions of Sgr~B1 or Sgr~C. Sgr~C does have CO and far-infrared emission corresponding to the IRDC, but this structure itself has no radio continuum emission and, more importantly, no mid-infrared emission for us to perform our normal analyses. 

We identified two molecular clumps in Sgr~B2, Sgr~B2~Main and Sgr~B2~V, to apply evolutionary analysis similar to our G\ion{H}{2} regions. Notably, Sgr~B2~V is likely unassociated with Sgr~B2 (\S~\ref{sec:sgrb}), leaving Sgr~B2~Main as the sole clump for analysis. Our methodology involved deriving clump masses through pixel-by-pixel graybody fits to Herschel data and calculating bolometric luminosities with two-temperature fits across various bands. Background fluxes were adjusted per resolution and environmental emission, with virial properties calculated from $^{13}$CO(1-0) data \citep{1998ApJS..118..455O}. For further details of methodologies, refer to previous papers (e.g., \citetalias{2019ApJ...873...51L}).

Sgr~B2~Main has an unusually low virial parameter ($\alpha_{\rm vir} \approx 0.23$), suggesting it may be among the youngest molecular clumps in our sample of all G\ion{H}{2} regions. It also has extreme extinction and a high mass surface density of approximately 3.5~g/cm², which aligns with its early evolutionary state near a IRDC. Notably, this mass surface density corresponds to an approximate extinction of $\sim 787 \, \mathrm{mag}$ in $A_V$ (adopting $1 \, \mathrm{g \, cm^{-2}} = A_V / 224.8 \, \mathrm{mag}$; \citealt{kt13,2016ApJ...829L..19L}). This extinction value aligns well with the high extinction ($A_V > 500 \, \mathrm{mag}$) derived for a nearby point source Sgr~B2~4 from YSO fitting results (see Table 4). However, Sgr~B2~Main’s high $L/M$ value ($\approx 580$) contrasts with other low $\alpha_{\rm vir}$ clumps, such as G49.5-0.4~b of W51A (where $\alpha_{\rm vir} \approx 0.18$ and $L/M \approx 26$), and is likely due to contamination from photodissociation regions (PDRs). The bright mid-infrared and 20\,cm emission in Sgr~B2~Main support this interpretation, as PDR influence can artificially raise near and mid-infrared fluxes, increasing the bolometric luminosity -- a trend also noted in our previous G\ion{H}{2} studies (e.g., \citetalias{2020ApJ...888...98L}).

The evolutionary histories of Sgr~B and Sgr~C are much harder to understand from the present observations than our previous studies of G\ion{H}{2} regions far from the Galactic Center and in the spiral arms of the Galactic disk. This is because, for regions further out in the Galactic plane, young stars do not stray from their birthplaces within the host giant molecular cloud. However for Sgr~B and Sgr~C, which are only 100 and 90\,pc, respectively, from the Galactic Center \citepalias{2022ApJ...933...60D}, this does not seem to be the case. It only takes $\sim$5\,Myr for an object at the locations of Sgr~B and Sgr~C to rotate around the Galactic Center, assuming both regions have a circular velocity of $\sim$110\,km/s \citep{2022MNRAS.514L...1S}, which is fraction of a stellar lifetime.  At these Galactic Center distances, stars will stray much quicker from their birth places; even bound star clusters could dissolve in $\sim$30\,Myr \citep{2001ApJ...546L.101P}. Furthermore, the turbulent crossing time of typical molecular clouds in the CMZ is $\sim$0.3\,Myr \citep{2017A&A...603A..89K}, and therefore the feedback from star formation will change them over timescales longer than this \citep{2019ApJ...872..171L}. This indicates that, although many of the MYSOs we are detecting are likely young and may still be in a phase before the onset of radio continuum emission -- making them too youthful to have strayed far from their birthplaces -- their natal environments may have already undergone significant changes. This could potentially be one reason behind the lack of correlation of molecular material and the mid-infrared and radio continuum.

\subsubsection{The Nature of the Ionized Emission in Sgr~B1, Sgr~B2, and Sgr~C}\label{sec:alm}

In our previous studies, the G\ion{H}{2} regions were ionized and heated primarily by revealed or nascent massive stars that were born in situ within the molecular clouds hosting the G\ion{H}{2} regions. However, it appears that in Sgr~B there are several ionizing Wolf-Rayet (WR) and O supergiant stars that are sufficiently evolved that they have lived long enough to have at least encircled the Galactic Center once. Likewise, \citet{2021A&A...649A..43C} find that Sgr~C contains two highly ionizing WR stars. 

In just Sgr~B1, \citet{2010ApJ...710..706M} and \citet{2021A&A...649A..43C} find a total of 3 WR and one O supergiant star, which are visible as point sources in the IRAC data. One is situated in the center of the the shell of Source I, and the other is situated west of the Ionized Rim (Figure \ref{fig:SgrB1radio}). Given their locations on the sky, it may be that these two evolved but powerful objects are ionizing Source I and the Ionized Rim. \citet{2021ApJ...910...59S} state that the four evolved stars in Sgr~B1 account for a combined Lyman continuum photon rate of $1.4\times10^{50}\, s^{-1}$. They state that this is half the amount needed to power the entire Sgr~B1 G\ion{H}{2} region, which they claim has a total of $3\times10^{50}\,s^{-1}$. However, we calculate, based on radio continuum observations, an observed Lyman continuum photon rate for Sgr~B1 of $N^{\prime}_{LyC}$ = 2.0$\times$10$^{50}\,s^{-1}$ in \citetalias{2022ApJ...933...60D}, meaning the WR stars would account for almost three quarters of the total ionizing photons in the Sgr~B1 G\ion{H}{2} region. Though observations by \citet{2022NatAs...6.1178N} suggest that Sgr~B1 may have an excess of stars $<$60\,Myr old, \citet{2018ApJ...867L..13S} speculate that Sgr~B1 is not ionized by a central cluster, and instead claim that these interlopers are now presently ionizing the molecular material and may be triggering present star formation. These more evolved stars likely were formed in a different cluster elsewhere in the CMZ and already orbited the GC and are now recent interlopers to the Sgr~B1 region. 

In Sgr~C, one of the two WR stars (2MASS J17443734-2927557) is fairly centrally located within the main Sgr~C \ion{H}{2} region and visible in the Spitzer IRAC images (Figure \ref{fig:SgrCradio}). This star is believed to be a WC8-9 star, and these stars typically have luminosities of $2.5-4.0\times10^5\,L_{\sun}$, which alone is similar to the luminosity of the entire Sgr~C~\ion{H}{2} region ($L_{HII}=4\times10^5\,L_{\sun}$; \citealt{1995ApJS...98..259L}). The second WR star is similarly thought to be a WC8-9 star, and lies 75$\arcsec$ northeast of the center of the Sgr~C~\ion{H}{2} region (i.e., just outside its $\sim70\arcsec$ radio continuum radius). Though \citet{2024A&A...681L..21N} find the Sgr~C region has a significant population of young stars around 20~Myr old, like Sgr B1, Sgr C may not be predominantly ionized by a stellar cluster, but rather the two evolved stars alone could be responsible for the vast majority of the ionization within the Sgr~C G\ion{H}{2} region. Given that the WC stars discussed have ages of about 3-8\,Myr \citep{1995A&A...298..767M}, these two stars, as is the case for Sgr~B1, likely did not form in-situ to the clumps they are now ionizing.  The relationship between the tip finger-shaped IRDC and the \ion{H}{2} region is not well understood, but it may be that the WC interloper may be externally heating and ionizing this IRDC finger, creating the \ion{H}{2} region we see, and perhaps even triggering the star formation in the IRDC. 

As we have just suggested for Sgr~C, there is also some speculation \citep{2010ApJ...710..706M, 2021A&A...649A..43C} that perhaps these interloper stars in Sgr~B1 have triggered the collapse of that region's clouds and clumps on their journeys through the Sgr~B1 region. Though all of Sgr~B1 and the ionized regions of Sgr~C appear to lack dense molecular material and cold dust, they may have had widespread but diffuse material that was held apart by high turbulence which were then triggered to collapse by these passing interloper stars. Additionally, these interlopers are massive enough to ionize these regions as they move through them, and as we mentioned, this may be causing much of the more wide-spread radio continuum emission we see. 

The situation is different for Sgr~B2. There is one evolved O supergiant star located almost equidistant between Sgr~B2~7, Sgr~B2~10, and Sgr~B2~12 (Figure \ref{fig:SgrB2radio}). However, this star is located $\sim$2.5$\arcmin$ southeast of Sgr~B2~Main and is coincident with an extended infrared emission region, but no substantial radio continuum emission. It therefore likely contributes little to the radio-defined Lyman continuum photon rate of Sgr~B2, which is dominated by the contributions from the many compact \ion{H}{2} regions. An additional WR star is also found in G0.6-0.0. 

Finally, though G\ion{H}{2} regions are dominated by thermal emission, there is increasing evidence that integrated radio flux of \ion{H}{2} regions in general may have a substantial non-thermal component \citep[e.g.,][]{1996ApJ...459..193G,2002ApJ...571..366M}. Cosmic ray ionization is potentially one of the leading sources of non-thermal emission \citep{2019A&A...630A..72P}, and may be particularly enhanced in CMZ \ion{H}{2} regions \citep{2019A&A...630A..73M,2024arXiv241210983B}. Indeed, the CMZ on large scales has been measured to have typically 2 orders of magnitude higher cosmic ray ionization rates than in the Solar galactic environment \citep{2023ASPC..534...83H}. However, this cosmic ray flux becomes too attenuated when penetrating the embedded star-forming environments of molecular clouds, and instead it is believed that relativistically-accelerated electrons within \ion{H}{2} region shocks may be a dominant source of locally-generated non-thermal emission \citep{2019A&A...630A..72P}. Since the strength of this non-thermal emission is dependent upon magnetic field strength, it has been conjectured that the generally stronger magnetic fields of the CMZ \citep[$\sim$1\,mG vs. $1-100$\,$\mu$G in the Solar neighborhood;][]{2023ASPC..534...83H} may lead to an enhanced non-thermal component in CMZ G\ion{H}{2} regions \citep{2024arXiv241210983B}. Evidence supporting this idea comes from \citet{2019A&A...630A..73M}, who measured a 60\% non-thermal component to the radio flux of Sgr~B2~DS, and \citet{2024arXiv241210983B} who find that the Sgr C G\ion{H}{2} region seems to be riddled with small-scale (sub-parsec) non-thermal filaments. Therefore, in addition to the ionizing emission contributed by evolved interloper stars, another difference in the total measured ionized flux in CMZ G\ion{H}{2} regions versus Galactic plane G\ion{H}{2} regions is the potentially much larger level of non-thermal emission that comes from cosmic rays accelerated locally.

\subsubsection{Are Sgr~B1, Sgr~B2, and Sgr~C bonafide \texorpdfstring{G\ion{H}{2}} regions?}\label{sec:genuine}

The most basic definition of a G\ion{H}{2} region is a source that exceeds a Lyman continuum photon rate of $N_{LyC} = 10^{50}$ photons/sec. However, the naive assumption is that when a region has this Lyman continuum photon flux, it is due to the presence of a cluster of massive ionizing stars either recently formed from or is presently forming within a host giant molecular cloud. Thus, the moniker ``G\ion{H}{2} regions'' is generally taken to be synonymous with young \ion{H}{2} regions hosting the most powerful sites of massive star formation in a galaxy \citep[e.g.,][]{1978AA....66...65S, 2004MNRAS.355..899C}.  

With their lack of molecular material and scarcity of cold dust, both Sgr~B1 and Sgr~C are peculiar compared to G\ion{H}{2} regions that are located in the Galaxy's spiral arms. As we have mentioned, for all seven G\ion{H}{2} regions in our prior studies, emission from the material of the host giant molecular cloud (as traced by Herschel 160, 250, 350, and 500~$\mu$m dust emission as well as emission lines of molecular species) was intermingled with the radio continuum and/or mid-infrared continuum emission, and for all G\ion{H}{2} regions even had very similar morphology. A lack of molecular gas and cold dust is usually associated with older and evolved massive star clusters. Indeed, \citet{2004MNRAS.355..899C} rejected some bright radio sources from their final G\ion{H}{2} census due to them being much older regions that lacked molecular gas (i.e., Westerlund 1). Given their high levels of present star formation and the rapidity in which these CMZ regions evolve, it would be hard to classify Sgr~B1 and Sgr~C as old G\ion{H}{2} regions. Therefore, one could make an argument that these sources may represent a new class of G\ion{H}{2} region. Conversely, however, if we combine the lack of cold dust/molecular material with the fact that the origin of the high Lyman continuum photon rate for both of these regions is due to evolved interloper stars, Sgr~B1 and Sgr~C do not fit the classical picture of G\ion{H}{2} regions, nor our previous observations of G\ion{H}{2} regions. Therefore, even though Sgr~B1 and Sgr~C contain MYSOs at stellar densities similar to previously studied G\ion{H}{2} regions, based upon these other non-conformant properties, one could reasonably argue that Sgr~B1 and Sgr~C are not legitimate G\ion{H}{2} regions.

Sgr~B2, on the other hand, which wasn't in our prior G\ion{H}{2} region census, is rich in molecular material and surrounded by cold dust, and presently undergoing very rigorous massive star formation. However, though it is extremely bright in its radio continuum emission and has a very high Lyman continuum photon rate ($log(N_{LyC})$ = 51.04\,$s^{-1}$), it appears to have relatively little mid-infrared emission. This could be due to the fact that there is an extremely high level of local and galactic extinction towards this region, or that it is a very youthful starburst region in the process of generating its first-generation massive star cluster out of an infrared dark cloud (or a combination of both). In any case, this lack of bright near-to-mid-infrared emission was the original reason for it being excluded from our G\ion{H}{2} source list. Given all of observable characteristics of Sgr~B2 other than it's overall level of mid-infrared emission (i.e., very high $N_{LyC}$; typical MYSO stellar density; coincidence of mid-infrared, far-infrared, cm radio continuum, and molecular material) it certainly seems to better fit the classical definition of a G\ion{H}{2} region than Sgr~B1 or Sgr~C. Or perhaps given the fact that there is so little extended radio continuum emission, and given it's apparent extreme youth, one might instead consider Sgr~B2 to be a proto-G\ion{H}{2} region. 

Since massive star forming regions are typically quite bright in the mid-infrared, one would assume that identifying G\ion{H}{2} regions based in part on their 12-100\,$\mu$m fluxes, as was done by \citet{2004MNRAS.355..899C}, would seem reasonable. However, Sgr~B1, Sgr~B2, and Sgr~C show the limitations of using this method when trying to determine whether a source fits this classical idea of a G\ion{H}{2} region. These idiosyncrasies of Sgr~B1, Sgr~B2, and Sgr~C compared to our previously studied G\ion{H}{2} regions may be simply be due in large part to the bizarre environment of the Galactic Central region. However, it may also be reasonable to conclude that any surveys searching for G\ion{H}{2} regions (either within our Galaxy or in external galaxies) should cross-correlate the radio continuum data with molecular (e.g., CO) and/or cold dust (e.g., Herschel 250\,$\mu$m) maps, which appear to be better indicators of host giant molecular clouds (and thus, ongoing massive star forming regions) than the mid-infrared.

\section{Summary}\label{sec:sum}

In this, our seventh paper from our mid-infrared imaging survey of Milky Way G\ion{H}{2} regions, we used SOFIA-FORCAST 25 and 37\,$\mu$m images with $\lesssim$3$\arcsec$ spatial resolution that covered the entire mid-infrared and radio continuum emitting areas of three G\ion{H}{2} regions in the Milky Way's Central Molecular Zone: Sgr~B1, Sgr~B2, and Sgr~C. We compared these SOFIA-FORCAST images with previous multi-wavelength observations from various ground- and space-based telescopes in order to discern the morphological and physical properties of these regions. Our previous studies of have shown that, at our SOFIA wavelengths, the majority of the compact mid-infrared sources we find are harboring massive young stars that are presently in the act of forming. However, the closeness of these particular G\ion{H}{2} regions to the Galactic Center means that these regions are subject to rapid dynamical and environmental changes atypical for G\ion{H}{2} regions farther out in the Galactic plane. The practical upshot of this is that Sgr~B and Sgr~C may contain a much higher incidence of evolved interloper stars masquerading (in terms of their near-to-far infrared SEDs) as MYSOs. We argue that while the uncertainty of any particular source as and MYSO candidate in the Galactic Center G\ion{H}{2} regions is higher (in the absence of evidence beyond merely having a MYSO-like SED), overall we still expect the majority of our sources to indeed be MYSOs and, indeed, we have evidence for several sources indicating that they are in fact MYSOs. Additionally, we compared the more global properties of these three CMZ G\ion{H}{2} regions to the G\ion{H}{2} regions we previously studied farther out in the Galactic plane. In many ways, the CMZ G\ion{H}{2} regions are similar to those in the Galactic plane, however Sgr~B1 and Sgr~C are fundamentally different in several respects. Below we describe these results in more detail, and itemize the other major take-aways from this study.

1) In the whole of Sgr~B, which includes both Sgr~B1 and Sgr~B2 G\ion{H}{2} regions, and their nearby surrounding areas, a total of 53 compact mid-infrared sources are found, of which 11 are considered to be confirmed MYSOs via SED fitting of their infrared photometry and the presence of definitive MYSO tracers, such as methanol masers. We categorize a further 20 sources as Likely MYSOs based upon SED fitting of their infrared photometry and the presence of cm continuum radio emission, with an additional 17 categorized as Possible MYSOs based solely upon SED fitting alone. In the whole of Sgr~C, which includes the Sgr~C G\ion{H}{2} region and its nearby surroundings covered by the SOFIA data, a total of 47 compact mid-infrared sources are found, of which only 4 are considered to be confirmed MYSOs, with 7 categorized as Likely MYSOs, and 18 Possible MYSOs. In total, within all of the fields covered in our SOFIA-FORCAST data, we have identified 77 CMZ MYSO candidates. 

2) Within the confines of just the Sgr~B1 G\ion{H}{2} region, we find 18 compact mid-infrared sources for which we believe 16 (89\%) to be either MYSOs (3) or MYSO candidates (12 Likely MYSOs and 1 Possible MYSO). Source Sgr~B1~D has the highest mass at 32\,$M_{\sun}$, as derived from the best fit MYSO model, and highest average mass (37\,$M_{\sun}$) from the group of best fit models. It is also the brightest compact source at cm radio continuum wavelengths, as well as the brightest source at 25 and 37\,$\mu$m. The extended radio continuum emission comprising the G\ion{H}{2} region and the extended mid-infrared continuum emission as seen by SOFIA are fairly well-matched across all of Sgr~B1.

3) Within the confines of just the Sgr~B2 G\ion{H}{2} region, we find 24 compact mid-infrared sources for which we believe 16 (92\%) to be either MYSOs (7) or MYSO candidates (4 Likely MYSOs and 11 Possible MYSO). Sgr~B2~H (a.k.a. Sgr~B2~South) has highest average mass (50\,$M_{\sun}$) from the group of best fit models, and is believed to be the most massive mid-infrared source in Sgr~B2. Consistent with this claim, Sgr~B2~H is the brightest peak at 37\,$\mu$m as well as 70\,$\mu$m (after Sgr~B2~Main, which is not considered a compact source). Extinction appears to be extremely high to the west of Sgr~B2~Main as we do not see any mid-infrared emission from prominent radio continuum sources K, Q, W, Y, and Z even at 37\,$\mu$m. Furthermore, we have many more Possible MYSOs in Sgr~B2 (46\%) than in Sgr~B1 (6\%), mostly due to a higher portion of compact mid-infrared sources with no detected cm radio continuum emission. This may mean that Sgr~B2 has a larger percentage of MYSOs at a more youthful stage prior to the onset of a ultracompact \ion{H}{2} region than Sgr~B1, which is perhaps in keeping with the idea that Sgr~B2 is a more youthful star-forming environment overall. 

4) Within the confines of just the Sgr~C G\ion{H}{2} region, we find 19 compact mid-infrared sources for which we believe 11 (58\%) to be either MYSOs (3) or MYSO candidates (3 Likely MYSOs and 5 Possible MYSO). Sources Sgr~C~2 and Sgr~C~H3 (a.k.a. G359.44a) are tied with the highest mass at 32\,$M_{\sun}$ (as derived from the best fit MYSO model). Sgr~C~H3 , however, has the highest average mass fit (30\,$M_{\sun}$) and also has also highest peak brightness in the SOFIA data at 25 and 37\,$\mu$m. Sgr~C differs from Sgr~B1 and Sgr~B2 in a much lower percentage of combined MYSO and MYSO candidates as well as a much higher percentage of mid-infrared sources that are confirmed as not being MYSOs (46\%, compared to 11\% for Sgr~B1 and 8\% for Sgr~B2). Half (4) of the non-MYSO sources lack the 15.4\,$\mu$m spectral ice feature indicative of a MYSO, and other half (4) have SED fits that indicate they are low-mass YSOs (or perhaps red giant or AGB interlopers/foreground objects). Furthermore, the only mid-infrared sources confirmed to be MYSOs are all located within the tip of the IRDC present here.  

5) Despite several distinct differences between the three regions, they are have very similar lower and upper limits for their MYSO stellar density estimates. Our worst-case scenario estimates yield MYSO stellar density values of 0.08, 0.09, and 0.09\,MYSOs/pc$^{2}$ for Sgr~B1, Sgr~B2, and Sgr~C, respectively. Our most optimistic upper limit estimates are 0.13, 0.14, and 0.14\,MYSOs/pc$^{2}$. Both limits are in the range (0.05--0.29\,MYSOs/pc$^{2}$) of the G\ion{H}{2} regions we have studied previously, but both are both also less than the average (0.18\,MYSOs/pc$^{2}$). This result is consistent with the perception that CMZ star forming regions do not appear to be as prolific as one might expect, however they are indeed presently producing MYSOs at a rate consistent with G\ion{H}{2} regions further out in the Galactic plane, albeit at the rate below the average. Additionally, the fact that MYSO stellar density values are comparable across all three CMZ G\ion{H}{2} regions is perhaps surprising given that Sgr~B2 is thought to be a much more prolific star formation environment than the others, and differences in extinction between the regions is surely a factor. 

6) The MYSO with the highest best fit mass of 64\,$M_{\sun}$ for Sgr~B2 is consistent with several of the other G\ion{H}{2} regions we have studied farther out in the Galactic plane. In particular W51A:G49.4-0.3, M17, and NGC~3603 all top out with MYSOs having 64\,$M_{\sun}$ for their best fit masses. However, the most massive MYSOs found in both Sgr~B1 and Sgr~C are only 32\,$M_{\sun}$, which would place them at the bottom of the rankings above only DR~7, whose status as a G\ion{H}{2} region is questionable. Since the most massive member of a cluster informs us as to the likely underlying cluster mass function, this is another indication that the present star formation in Sgr~B1 and Sgr~C may be less prolific than Sgr~B2. It may also indicate that while the CMZ G\ion{H}{2} regions appear similar in many ways to Galactic G\ion{H}{2} regions, it may be more difficult to form the highest mass O stars here.

7) For all of our previously studied G\ion{H}{2} regions, their morphologies were similar in hot and cold dust emission (as traced by the SOFIA-FORCAST and Herschel-PACS/SPIRE data) and in their ionized gas emission (traced by cm radio continuum), as well as in their molecular gas emission (in maps like $^{13}$CO $J=1-0$). However, in Sgr~B1 and Sgr~C, as well as (but to a lesser degree) Sgr~B2, we have good morphological correspondence between cm radio continuum emission and 25--70\,$\mu$m emission, but sources and features seen at these these wavelengths do not display any emission from cold dust ($\gtrsim$160\,$\mu$m) or molecular material. Sgr~B2 and Sgr~C do have molecular gas and cold dust reservoirs (in the form of IRDCs) for continued star formation, however Sgr~B1 does not.      

8) Sgr~B1 contains 3 Wolf-Rayet and one evolved O supergiant, and Sgr~C contains two Wolf-Rayet stars. In both Sgr~B1 and Sgr~C it appears that the dominant contribution to the overall Lyman continuum photon rate to the G\ion{H}{2} regions is by the ionization provided by these evolved interloper stars that were very likely not even formed from the same material as the presently forming stars in each region. It could be that the Sgr~B1 G\ion{H}{2} region, and most of the Sgr~C G\ion{H}{2} region, were interspersed by diffuse material held apart by high turbulence and were triggered to collapse into the present mid-infrared structures and MYSOs by the passing of these interloper stars. This could account for the lack of dense molecular material and cold dust directly associated with these features and MYSOs. Additionally, these interlopers are ionizing their immediate surroundings as they move through these regions, and may be causing much of the more wide-spread radio continuum emission we see there.

9) For the Sgr~B1 and Sgr~C G\ion{H}{2} regions, given the anti-correlation between the location of the overwhelming majority of their hot dust/ionized emission and the location of cold dust/molecular material, these two sources may represent a new category of G\ion{H}{2} region. Further evidence of their nonconforming character comes from the fact that they are most likely predominantly ionized by stars not of their own making. Both characteristics may be in large part due to the fact that everything is moving swiftly around the Galactic Center at the small Galactic radii of the Sgr~B1 and Sgr~C orbits, as well as the generally highly turbulent environment of the CMZ. On the other hand, since the naive assumption is that G\ion{H}{2} regions are the result of clusters of massive stars forming from, and still residing within, their natal giant molecular clouds, one could argue that instead of being a new category of G\ion{H}{2} region, that Sgr~B1 and Sgr~C are not legitimate G\ion{H}{2} regions at all. 

\begin{acknowledgments}
We would like to thank the valuable input from an anonymous referee, which helped improve the quality of the  manuscript. This research is based on archival data from the NASA/DLR Stratospheric Observatory for Infrared Astronomy (SOFIA). SOFIA was jointly operated by the Universities Space Research Association, Inc. (USRA), under a contract with NASA, and the Deutsches SOFIA Institut (DSI), under a contract from DLR to the University of Stuttgart. This work is also based in part on archival data obtained with the Spitzer Space Telescope, which was operated by the Jet Propulsion Laboratory, California Institute of Technology under a contract with NASA. This work is also based in part on archival data obtained with Herschel, an European Space Agency (ESA) space observatory with science instruments provided by European-led Principal Investigator consortia and with important participation from NASA. This research has made use of \textit{Aladin Sky Atlas}, CDS, Strasbourg Astronomical Observatory, France.  

The lead author wishes to acknowledge NASA funding via an ADAP Award (80NSSC24K0640) which made this work possible.

\facilities{SOFIA(FORCAST), Spitzer, Herschel}
\software{sofia\_redux \citep{https://doi.org/10.5281/zenodo.8219569}}
\end{acknowledgments}

\bibliography{GCreferences}{}

\begin{thebibliography}{}
\expandafter\ifx\csname natexlab\endcsname\relax\def\natexlab#1{#1}\fi
\providecommand{\url}[1]{\href{#1}{#1}}
\providecommand{\dodoi}[1]{doi:~\href{http://doi.org/#1}{\nolinkurl{#1}}}
\providecommand{\doeprint}[1]{\href{http://ascl.net/#1}{\nolinkurl{http://ascl.net/#1}}}
\providecommand{\doarXiv}[1]{\href{https://arxiv.org/abs/#1}{\nolinkurl{https://arxiv.org/abs/#1}}}

\bibitem[{{An} {et~al.}(2011){An}, {Ram{\'\i}rez}, {Sellgren}, {Arendt}, {Adwin Boogert}, {Robitaille}, {Schultheis}, {Cotera}, {Smith}, \& {Stolovy}}]{2011ApJ...736..133A}
{An}, D., {Ram{\'\i}rez}, S.~V., {Sellgren}, K., {et~al.} 2011, \apj, 736, 133, \dodoi{10.1088/0004-637X/736/2/133}

\bibitem[{{Ao} {et~al.}(2013){Ao}, {Henkel}, {Menten}, {Requena-Torres}, {Stanke}, {Mauersberger}, {Aalto}, {M{\"u}hle}, \& {Mangum}}]{2013A&A...550A.135A}
{Ao}, Y., {Henkel}, C., {Menten}, K.~M., {et~al.} 2013, \aap, 550, A135, \dodoi{10.1051/0004-6361/201220096}

\bibitem[{{Araya} {et~al.}(2015){Araya}, {Olmi}, {Morales Ortiz}, {Brown}, {Hofner}, {Kurtz}, {Linz}, \& {Creech-Eakman}}]{2015ApJS..221...10A}
{Araya}, E.~D., {Olmi}, L., {Morales Ortiz}, J., {et~al.} 2015, \apjs, 221, 10, \dodoi{10.1088/0067-0049/221/1/10}

\bibitem[{{Bally} {et~al.}(2024){Bally}, {Crowe}, {Fedriani}, {Ginsburg}, {Sch{\"o}del}, {Andersen}, {Tan}, {Li}, {Nogueras-Lara}, {Cheng}, {Law}, {Wang}, {Zhang}, \& {Zhang}}]{2024arXiv241210983B}
{Bally}, J., {Crowe}, S., {Fedriani}, R., {et~al.} 2024, arXiv e-prints, arXiv:2412.10983, \dodoi{10.48550/arXiv.2412.10983}

\bibitem[{{Balser} {et~al.}(2011){Balser}, {Rood}, {Bania}, \& {Anderson}}]{2011ApJ...738...27B}
{Balser}, D.~S., {Rood}, R.~T., {Bania}, T.~M., \& {Anderson}, L.~D. 2011, \apj, 738, 27, \dodoi{10.1088/0004-637X/738/1/27}

\bibitem[{{Benson} \& {Johnston}(1984)}]{1984ApJ...277..181B}
{Benson}, J.~M., \& {Johnston}, K.~J. 1984, \apj, 277, 181, \dodoi{10.1086/161680}

\bibitem[{{Bieging} {et~al.}(1980){Bieging}, {Downes}, {Wilson}, {Martin}, \& {Guesten}}]{1980A&AS...42..163B}
{Bieging}, J., {Downes}, D., {Wilson}, T.~L., {Martin}, A.~H.~M., \& {Guesten}, R. 1980, \aaps, 42, 163

\bibitem[{{Blommaert} {et~al.}(2006){Blommaert}, {Groenewegen}, {Okumura}, {Ganesh}, {Omont}, {Cami}, {Glass}, {Habing}, {Schultheis}, {Simon}, \& {van Loon}}]{2006A&A...460..555B}
{Blommaert}, J.~A.~D.~L., {Groenewegen}, M.~A.~T., {Okumura}, K., {et~al.} 2006, \aap, 460, 555, \dodoi{10.1051/0004-6361:20066145}

\bibitem[{{Blum} {et~al.}(2000){Blum}, {Conti}, \& {Damineli}}]{2000AJ....119.1860B}
{Blum}, R.~D., {Conti}, P.~S., \& {Damineli}, A. 2000, \aj, 119, 1860, \dodoi{10.1086/301317}

\bibitem[{{Boji{\v{c}}i{\'c}} {et~al.}(2011){Boji{\v{c}}i{\'c}}, {Parker}, {Filipovi{\'c}}, \& {Frew}}]{2011MNRAS.412..223B}
{Boji{\v{c}}i{\'c}}, I.~S., {Parker}, Q.~A., {Filipovi{\'c}}, M.~D., \& {Frew}, D.~J. 2011, \mnras, 412, 223, \dodoi{10.1111/j.1365-2966.2010.17900.x}

\bibitem[{{Breen} {et~al.}(2013){Breen}, {Ellingsen}, {Contreras}, {Green}, {Caswell}, {Stevens}, {Dawson}, \& {Voronkov}}]{2013MNRAS.435..524B}
{Breen}, S.~L., {Ellingsen}, S.~P., {Contreras}, Y., {et~al.} 2013, \mnras, 435, 524, \dodoi{10.1093/mnras/stt1315}

\bibitem[{{Busso} {et~al.}(2007){Busso}, {Guandalini}, {Persi}, {Corcione}, \& {Ferrari-Toniolo}}]{2007AJ....133.2310B}
{Busso}, M., {Guandalini}, R., {Persi}, P., {Corcione}, L., \& {Ferrari-Toniolo}, M. 2007, \aj, 133, 2310, \dodoi{10.1086/512612}

\bibitem[{{Cala} {et~al.}(2022){Cala}, {G{\'o}mez}, {Miranda}, {Uscanga}, {Breen}, {Dawson}, {de Gregorio-Monsalvo}, {Imai}, {Qiao}, \& {Su{\'a}rez}}]{2022MNRAS.516.2235C}
{Cala}, R.~A., {G{\'o}mez}, J.~F., {Miranda}, L.~F., {et~al.} 2022, \mnras, 516, 2235, \dodoi{10.1093/mnras/stac2341}

\bibitem[{{Caswell} \& {Haynes}(1987)}]{1987A&A...171..261C}
{Caswell}, J.~L., \& {Haynes}, R.~F. 1987, \aap, 171, 261

\bibitem[{{Caswell} {et~al.}(2010){Caswell}, {Fuller}, {Green}, {Avison}, {Breen}, {Brooks}, {Burton}, {Chrysostomou}, {Cox}, {Diamond}, {Ellingsen}, {Gray}, {Hoare}, {Masheder}, {McClure-Griffiths}, {Pestalozzi}, {Phillips}, {Quinn}, {Thompson}, {Voronkov}, {Walsh}, {Ward-Thompson}, {Wong-McSweeney}, {Yates}, \& {Cohen}}]{2010MNRAS.404.1029C}
{Caswell}, J.~L., {Fuller}, G.~A., {Green}, J.~A., {et~al.} 2010, \mnras, 404, 1029, \dodoi{10.1111/j.1365-2966.2010.16339.x}

\bibitem[{{Cerrigone} {et~al.}(2017){Cerrigone}, {Umana}, {Trigilio}, {Leto}, {Buemi}, \& {Ingallinera}}]{2017MNRAS.468.3450C}
{Cerrigone}, L., {Umana}, G., {Trigilio}, C., {et~al.} 2017, \mnras, 468, 3450, \dodoi{10.1093/mnras/stx690}

\bibitem[{{Cho} {et~al.}(2016){Cho}, {Yun}, {Kim}, {Liu}, {Kim}, \& {Choi}}]{2016ApJ...826..157C}
{Cho}, S.-H., {Yun}, Y., {Kim}, J., {et~al.} 2016, \apj, 826, 157, \dodoi{10.3847/0004-637X/826/2/157}

\bibitem[{{Clark} {et~al.}(2021){Clark}, {Patrick}, {Najarro}, {Evans}, \& {Lohr}}]{2021A&A...649A..43C}
{Clark}, J.~S., {Patrick}, L.~R., {Najarro}, F., {Evans}, C.~J., \& {Lohr}, M. 2021, \aap, 649, A43, \dodoi{10.1051/0004-6361/202039205}

\bibitem[{Clarke \& Vander~Vliet(2023)}]{https://doi.org/10.5281/zenodo.8219569}
Clarke, M., \& Vander~Vliet, R. 2023, SOFIA-USRA/sofia\_redux: v1.3.3,  Zenodo, \dodoi{10.5281/ZENODO.8219569}

\bibitem[{{Cohen} \& {Parker}(2003)}]{2003IAUS..209...33C}
{Cohen}, M., \& {Parker}, Q.~A. 2003, in IAU Symposium, Vol. 209, Planetary Nebulae: Their Evolution and Role in the Universe, ed. S.~{Kwok}, M.~{Dopita}, \& R.~{Sutherland}, 33

\bibitem[{{Conti} \& {Crowther}(2004)}]{2004MNRAS.355..899C}
{Conti}, P.~S., \& {Crowther}, P.~A. 2004, \mnras, 355, 899, \dodoi{10.1111/j.1365-2966.2004.08367.x}

\bibitem[{{Conti} \& {McCray}(1980)}]{1980Sci...208....9C}
{Conti}, P.~S., \& {McCray}, R. 1980, Science, 208, 9, \dodoi{10.1126/science.208.4439.9}

\bibitem[{{Cotera} {et~al.}(2024){Cotera}, {Hankins}, {Bally}, {Barnes}, {Battersby}, {Hatchfield}, {Herter}, {Lau}, {Longmore}, {Mills}, {Morris}, {Radomski}, {Simpson}, {Stephens}, \& {Walker}}]{2024ApJ...973..110C}
{Cotera}, A.~S., {Hankins}, M.~J., {Bally}, J., {et~al.} 2024, \apj, 973, 110, \dodoi{10.3847/1538-4357/ad55f2}

\bibitem[{{Cox} \& {Laureijs}(1989)}]{1989IAUS..136..121C}
{Cox}, P., \& {Laureijs}, R. 1989, in IAU Symposium, Vol. 136, The Center of the Galaxy, ed. M.~{Morris}, 121

\bibitem[{{Crowe} {et~al.}(2024){Crowe}, {Fedriani}, {Tan}, {Kinman}, {Zhang}, {Andersen}, {Bravo Ferres}, {Nogueras-Lara}, {Sch{\"o}del}, {Bally}, {Ginsburg}, {Cheng}, {Yang}, {Kendrew}, {Law}, {Armstrong}, \& {Li}}]{2024arXiv241009253C}
{Crowe}, S., {Fedriani}, R., {Tan}, J.~C., {et~al.} 2024, arXiv e-prints, arXiv:2410.09253, \dodoi{10.48550/arXiv.2410.09253}

\bibitem[{{Cyganowski} {et~al.}(2011){Cyganowski}, {Brogan}, {Hunter}, \& {Churchwell}}]{2011ApJ...743...56C}
{Cyganowski}, C.~J., {Brogan}, C.~L., {Hunter}, T.~R., \& {Churchwell}, E. 2011, \apj, 743, 56, \dodoi{10.1088/0004-637X/743/1/56}

\bibitem[{{Cyganowski} {et~al.}(2008){Cyganowski}, {Whitney}, {Holden}, {Braden}, {Brogan}, {Churchwell}, {Indebetouw}, {Watson}, {Babler}, {Benjamin}, {Gomez}, {Meade}, {Povich}, {Robitaille}, \& {Watson}}]{2008AJ....136.2391C}
{Cyganowski}, C.~J., {Whitney}, B.~A., {Holden}, E., {et~al.} 2008, \aj, 136, 2391, \dodoi{10.1088/0004-6256/136/6/2391}

\bibitem[{{De Buizer} {et~al.}(2024){De Buizer}, {Lim}, {Karnath}, \& {Radomski}}]{2024ApJ...963...55D}
{De Buizer}, J.~M., {Lim}, W., {Karnath}, N., \& {Radomski}, J.~T. 2024, \apj, 963, 55, \dodoi{10.3847/1538-4357/ad19d1}

\bibitem[{{De Buizer} {et~al.}(2022){De Buizer}, {Lim}, {Karnath}, {Radomski}, \& {Bonne}}]{2022ApJ...933...60D}
{De Buizer}, J.~M., {Lim}, W., {Karnath}, N., {Radomski}, J.~T., \& {Bonne}, L. 2022, \apj, 933, 60, \dodoi{10.3847/1538-4357/ac6fd8}

\bibitem[{{De Buizer} {et~al.}(2021){De Buizer}, {Lim}, {Liu}, {Karnath}, \& {Radomski}}]{2021ApJ...923..198D}
{De Buizer}, J.~M., {Lim}, W., {Liu}, M., {Karnath}, N., \& {Radomski}, J.~T. 2021, \apj, 923, 198, \dodoi{10.3847/1538-4357/ac2d25}

\bibitem[{{De Buizer} {et~al.}(2023){De Buizer}, {Lim}, {Radomski}, \& {Liu}}]{2023ApJ...949...82D}
{De Buizer}, J.~M., {Lim}, W., {Radomski}, J.~T., \& {Liu}, M. 2023, \apj, 949, 82, \dodoi{10.3847/1538-4357/acc9c6}

\bibitem[{{De Buizer} \& {Vacca}(2010)}]{2010AJ....140..196D}
{De Buizer}, J.~M., \& {Vacca}, W.~D. 2010, \aj, 140, 196, \dodoi{10.1088/0004-6256/140/1/196}

\bibitem[{{Decin} {et~al.}(2020){Decin}, {Montarg{\`e}s}, {Richards}, {Gottlieb}, {Homan}, {McDonald}, {El Mellah}, {Danilovich}, {Wallstr{\"o}m}, {Zijlstra}, {Baudry}, {Bolte}, {Cannon}, {De Beck}, {De Ceuster}, {de Koter}, {De Ridder}, {Etoka}, {Gobrecht}, {Gray}, {Herpin}, {Jeste}, {Lagadec}, {Kervella}, {Khouri}, {Menten}, {Millar}, {M{\"u}ller}, {Plane}, {Sahai}, {Sana}, {Van de Sande}, {Waters}, {Wong}, \& {Yates}}]{2020Sci...369.1497D}
{Decin}, L., {Montarg{\`e}s}, M., {Richards}, A.~M.~S., {et~al.} 2020, Science, 369, 1497, \dodoi{10.1126/science.abb1229}

\bibitem[{{Downes} {et~al.}(1979){Downes}, {Goss}, {Schwarz}, \& {Wouterloot}}]{1979A&AS...35....1D}
{Downes}, D., {Goss}, W.~M., {Schwarz}, U.~J., \& {Wouterloot}, J.~G.~A. 1979, \aaps, 35, 1

\bibitem[{{Downes} {et~al.}(1980){Downes}, {Wilson}, {Bieging}, \& {Wink}}]{1980A&AS...40..379D}
{Downes}, D., {Wilson}, T.~L., {Bieging}, J., \& {Wink}, J. 1980, \aaps, 40, 379

\bibitem[{{Ferri{\`e}re}(2009)}]{2009A&A...505.1183F}
{Ferri{\`e}re}, K. 2009, \aap, 505, 1183, \dodoi{10.1051/0004-6361/200912617}

\bibitem[{{Forster} \& {Caswell}(2000)}]{2000ApJ...530..371F}
{Forster}, J.~R., \& {Caswell}, J.~L. 2000, \apj, 530, 371, \dodoi{10.1086/308347}

\bibitem[{{Fujii} {et~al.}(2006){Fujii}, {Deguchi}, {Ita}, {Izumiura}, {Kameya}, {Miyazaki}, \& {Nakada}}]{2006PASJ...58..529F}
{Fujii}, T., {Deguchi}, S., {Ita}, Y., {et~al.} 2006, \pasj, 58, 529, \dodoi{10.1093/pasj/58.3.529}

\bibitem[{{Garay} {et~al.}(1996){Garay}, {Ramirez}, {Rodriguez}, {Curiel}, \& {Torrelles}}]{1996ApJ...459..193G}
{Garay}, G., {Ramirez}, S., {Rodriguez}, L.~F., {Curiel}, S., \& {Torrelles}, J.~M. 1996, \apj, 459, 193, \dodoi{10.1086/176882}

\bibitem[{{Gatley} {et~al.}(1978){Gatley}, {Becklin}, {Werner}, \& {Harper}}]{1978ApJ...220..822G}
{Gatley}, I., {Becklin}, E.~E., {Werner}, M.~W., \& {Harper}, D.~A. 1978, \apj, 220, 822, \dodoi{10.1086/155971}

\bibitem[{{Gaume} \& {Claussen}(1990)}]{1990ApJ...351..538G}
{Gaume}, R.~A., \& {Claussen}, M.~J. 1990, \apj, 351, 538, \dodoi{10.1086/168492}

\bibitem[{{Gaume} {et~al.}(1995){Gaume}, {Claussen}, {de Pree}, {Goss}, \& {Mehringer}}]{1995ApJ...449..663G}
{Gaume}, R.~A., {Claussen}, M.~J., {de Pree}, C.~G., {Goss}, W.~M., \& {Mehringer}, D.~M. 1995, \apj, 449, 663, \dodoi{10.1086/176087}

\bibitem[{{Gautier} {et~al.}(1984){Gautier}, {Hauser}, {Beichman}, {Low}, {Neugebauer}, {Rowan-Robinson}, {Aumann}, {Boggess}, {Emerson}, {Harris}, {Houck}, {Jennings}, \& {Marsden}}]{1984ApJ...278L..57G}
{Gautier}, T.~N., {Hauser}, M.~G., {Beichman}, C.~A., {et~al.} 1984, \apjl, 278, L57, \dodoi{10.1086/184222}

\bibitem[{{Ginsburg} {et~al.}(2016){Ginsburg}, {Henkel}, {Ao}, {Riquelme}, {Kauffmann}, {Pillai}, {Mills}, {Requena-Torres}, {Immer}, {Testi}, {Ott}, {Bally}, {Battersby}, {Darling}, {Aalto}, {Stanke}, {Kendrew}, {Kruijssen}, {Longmore}, {Dale}, {Guesten}, \& {Menten}}]{2016A&A...586A..50G}
{Ginsburg}, A., {Henkel}, C., {Ao}, Y., {et~al.} 2016, \aap, 586, A50, \dodoi{10.1051/0004-6361/201526100}

\bibitem[{{Ginsburg} {et~al.}(2018){Ginsburg}, {Bally}, {Barnes}, {Bastian}, {Battersby}, {Beuther}, {Brogan}, {Contreras}, {Corby}, {Darling}, {De Pree}, {Galv{\'a}n-Madrid}, {Garay}, {Henshaw}, {Hunter}, {Kruijssen}, {Longmore}, {Lu}, {Meng}, {Mills}, {Ott}, {Pineda}, {S{\'a}nchez-Monge}, {Schilke}, {Schmiedeke}, {Walker}, \& {Wilner}}]{2018ApJ...853..171G}
{Ginsburg}, A., {Bally}, J., {Barnes}, A., {et~al.} 2018, \apj, 853, 171, \dodoi{10.3847/1538-4357/aaa6d4}

\bibitem[{{Gonz{\'a}lez-Santamar{\'\i}a} {et~al.}(2019){Gonz{\'a}lez-Santamar{\'\i}a}, {Manteiga}, {Manchado}, {Ulla}, \& {Dafonte}}]{2019A&A...630A.150G}
{Gonz{\'a}lez-Santamar{\'\i}a}, I., {Manteiga}, M., {Manchado}, A., {Ulla}, A., \& {Dafonte}, C. 2019, \aap, 630, A150, \dodoi{10.1051/0004-6361/201936162}

\bibitem[{{Green} {et~al.}(2015){Green}, {Caswell}, \& {McClure-Griffiths}}]{2015MNRAS.451...74G}
{Green}, J.~A., {Caswell}, J.~L., \& {McClure-Griffiths}, N.~M. 2015, \mnras, 451, 74, \dodoi{10.1093/mnras/stv936}

\bibitem[{{Groenewegen}(2022)}]{2022A&A...659A.145G}
{Groenewegen}, M.~A.~T. 2022, \aap, 659, A145, \dodoi{10.1051/0004-6361/202142648}

\bibitem[{{Gutermuth} {et~al.}(2009){Gutermuth}, {Megeath}, {Myers}, {Allen}, {Pipher}, \& {Fazio}}]{2009ApJS..184...18G}
{Gutermuth}, R.~A., {Megeath}, S.~T., {Myers}, P.~C., {et~al.} 2009, \apjs, 184, 18, \dodoi{10.1088/0067-0049/184/1/18}

\bibitem[{{Hankins} {et~al.}(2020){Hankins}, {Lau}, {Radomski}, {Cotera}, {Morris}, {Mills}, {Walker}, {Barnes}, {Simpson}, {Herter}, {Longmore}, {Bally}, {Kasliwal}, {Sabha}, \& {Garc{\'\i}a-Mar{\'\i}n}}]{2020ApJ...894...55H}
{Hankins}, M.~J., {Lau}, R.~M., {Radomski}, J.~T., {et~al.} 2020, \apj, 894, 55, \dodoi{10.3847/1538-4357/ab7c5d}

\bibitem[{{Harris} {et~al.}(2021){Harris}, {G{\"u}sten}, {Requena-Torres}, {Riquelme}, {Morris}, {Stacey}, {Mart{\`\i}n-Pintado}, {Stutzki}, {Simon}, {Higgins}, \& {Risacher}}]{2021ApJ...921...33H}
{Harris}, A.~I., {G{\"u}sten}, R., {Requena-Torres}, M.~A., {et~al.} 2021, \apj, 921, 33, \dodoi{10.3847/1538-4357/ac1863}

\bibitem[{{Harvey} {et~al.}(1977){Harvey}, {Campbell}, \& {Hoffmann}}]{1977ApJ...211..786H}
{Harvey}, P.~M., {Campbell}, M.~F., \& {Hoffmann}, W.~F. 1977, \apj, 211, 786, \dodoi{10.1086/154988}

\bibitem[{{Henshaw} {et~al.}(2023){Henshaw}, {Barnes}, {Battersby}, {Ginsburg}, {Sormani}, \& {Walker}}]{2023ASPC..534...83H}
{Henshaw}, J.~D., {Barnes}, A.~T., {Battersby}, C., {et~al.} 2023, in Astronomical Society of the Pacific Conference Series, Vol. 534, Protostars and Planets VII, ed. S.~{Inutsuka}, Y.~{Aikawa}, T.~{Muto}, K.~{Tomida}, \& M.~{Tamura}, 83, \dodoi{10.48550/arXiv.2203.11223}

\bibitem[{{Hoffmann} {et~al.}(1971){Hoffmann}, {Fredrick}, \& {Emery}}]{1971ApJ...170L..89H}
{Hoffmann}, W.~F., {Fredrick}, C.~L., \& {Emery}, R.~J. 1971, \apjl, 170, L89, \dodoi{10.1086/180847}

\bibitem[{{Hrivnak}(2000)}]{2000ESASP.456..191H}
{Hrivnak}, B.~J. 2000, in ISO Beyond the Peaks: The 2nd ISO Workshop on Analytical Spectroscopy, ed. A.~{Salama}, M.~F. {Kessler}, K.~{Leech}, \& B.~{Schulz}, 191

\bibitem[{{Israel}(1980)}]{1980A&A....90..246I}
{Israel}, F.~P. 1980, \aap, 90, 246

\bibitem[{{Jang} {et~al.}(2022){Jang}, {An}, {Sellgren}, {Ram{\'\i}rez}, {Boogert}, \& {Schultheis}}]{2022ApJ...930...16J}
{Jang}, D., {An}, D., {Sellgren}, K., {et~al.} 2022, \apj, 930, 16, \dodoi{10.3847/1538-4357/ac5d51}

\bibitem[{{Jones} {et~al.}(2011){Jones}, {Crocker}, {Ott}, {Protheroe}, \& {Ekers}}]{2011AJ....141...82J}
{Jones}, D.~I., {Crocker}, R.~M., {Ott}, J., {Protheroe}, R.~J., \& {Ekers}, R.~D. 2011, \aj, 141, 82, \dodoi{10.1088/0004-6256/141/3/82}

\bibitem[{{Jones} {et~al.}(2013){Jones}, {Burton}, {Cunningham}, {Tothill}, \& {Walsh}}]{2013MNRAS.433..221J}
{Jones}, P.~A., {Burton}, M.~G., {Cunningham}, M.~R., {Tothill}, N.~F.~H., \& {Walsh}, A.~J. 2013, \mnras, 433, 221, \dodoi{10.1093/mnras/stt717}

\bibitem[{{Kainulainen} \& {Tan}(2013)}]{kt13}
{Kainulainen}, J., \& {Tan}, J.~C. 2013, \aap, 549, A53, \dodoi{10.1051/0004-6361/201219526}

\bibitem[{{Kauffmann} {et~al.}(2017){Kauffmann}, {Pillai}, {Zhang}, {Menten}, {Goldsmith}, {Lu}, \& {Guzm{\'a}n}}]{2017A&A...603A..89K}
{Kauffmann}, J., {Pillai}, T., {Zhang}, Q., {et~al.} 2017, \aap, 603, A89, \dodoi{10.1051/0004-6361/201628088}

\bibitem[{{Kendrew} {et~al.}(2013){Kendrew}, {Ginsburg}, {Johnston}, {Beuther}, {Bally}, {Cyganowski}, \& {Battersby}}]{2013ApJ...775L..50K}
{Kendrew}, S., {Ginsburg}, A., {Johnston}, K., {et~al.} 2013, \apjl, 775, L50, \dodoi{10.1088/2041-8205/775/2/L50}

\bibitem[{{Knapp} {et~al.}(1994){Knapp}, {Bowers}, {Young}, \& {Phillips}}]{1994ApJ...429L..33K}
{Knapp}, G.~R., {Bowers}, P.~F., {Young}, K., \& {Phillips}, T.~G. 1994, \apjl, 429, L33, \dodoi{10.1086/187406}

\bibitem[{{Koepferl} {et~al.}(2015){Koepferl}, {Robitaille}, {Morales}, \& {Johnston}}]{2015ApJ...799...53K}
{Koepferl}, C.~M., {Robitaille}, T.~P., {Morales}, E. F.~E., \& {Johnston}, K.~G. 2015, \apj, 799, 53, \dodoi{10.1088/0004-637X/799/1/53}

\bibitem[{{Kramer} {et~al.}(1998){Kramer}, {Staguhn}, {Ungerechts}, \& {Sievers}}]{1998IAUS..184..173K}
{Kramer}, C., {Staguhn}, J., {Ungerechts}, H., \& {Sievers}, A. 1998, in IAU Symposium, Vol. 184, The Central Regions of the Galaxy and Galaxies, ed. Y.~{Sofue}, 173

\bibitem[{{Kruijssen} {et~al.}(2015){Kruijssen}, {Dale}, \& {Longmore}}]{2015MNRAS.447.1059K}
{Kruijssen}, J.~M.~D., {Dale}, J.~E., \& {Longmore}, S.~N. 2015, \mnras, 447, 1059, \dodoi{10.1093/mnras/stu2526}

\bibitem[{{Kuchar} \& {Clark}(1997)}]{1997ApJ...488..224K}
{Kuchar}, T.~A., \& {Clark}, F.~O. 1997, \apj, 488, 224, \dodoi{10.1086/304697}

\bibitem[{{Lee} {et~al.}(2013){Lee}, {Giavalisco}, {Williams}, {Guo}, {Lotz}, {Van der Wel}, {Ferguson}, {Faber}, {Koekemoer}, {Grogin}, {Kocevski}, {Conselice}, {Wuyts}, {Dekel}, {Kartaltepe}, \& {Bell}}]{2013ApJ...774...47L}
{Lee}, B., {Giavalisco}, M., {Williams}, C.~C., {et~al.} 2013, \apj, 774, 47, \dodoi{10.1088/0004-637X/774/1/47}

\bibitem[{Lim(2019)}]{DVN/POYMK5_2019}
Lim, W. 2019, {Surveying the Giant HII Regions of the Milky Way with SOFIA}, V5,  Harvard Dataverse, \dodoi{10.7910/DVN/POYMK5}

\bibitem[{{Lim} \& {De Buizer}(2019)}]{2019ApJ...873...51L}
{Lim}, W., \& {De Buizer}, J.~M. 2019, \apj, 873, 51, \dodoi{10.3847/1538-4357/ab0288}

\bibitem[{{Lim} {et~al.}(2020){Lim}, {De Buizer}, \& {Radomski}}]{2020ApJ...888...98L}
{Lim}, W., {De Buizer}, J.~M., \& {Radomski}, J.~T. 2020, \apj, 888, 98, \dodoi{10.3847/1538-4357/ab5fd0}

\bibitem[{{Lim} {et~al.}(2016){Lim}, {Tan}, {Kainulainen}, {Ma}, \& {Butler}}]{2016ApJ...829L..19L}
{Lim}, W., {Tan}, J.~C., {Kainulainen}, J., {Ma}, B., \& {Butler}, M.~J. 2016, \apjl, 829, L19, \dodoi{10.3847/2041-8205/829/1/L19}

\bibitem[{{Lis} \& {Carlstrom}(1994)}]{1994ApJ...424..189L}
{Lis}, D.~C., \& {Carlstrom}, J.~E. 1994, \apj, 424, 189, \dodoi{10.1086/173882}

\bibitem[{{Lis} {et~al.}(1991){Lis}, {Carlstrom}, \& {Keene}}]{1991ApJ...380..429L}
{Lis}, D.~C., {Carlstrom}, J.~E., \& {Keene}, J. 1991, \apj, 380, 429, \dodoi{10.1086/170601}

\bibitem[{{Liszt} \& {Spiker}(1995)}]{1995ApJS...98..259L}
{Liszt}, H.~S., \& {Spiker}, R.~W. 1995, \apjs, 98, 259, \dodoi{10.1086/192160}

\bibitem[{{Little} \& {Price}(1985)}]{1985AJ.....90.1812L}
{Little}, S.~J., \& {Price}, S.~D. 1985, \aj, 90, 1812, \dodoi{10.1086/113882}

\bibitem[{{Lu} {et~al.}(2020){Lu}, {Cheng}, {Ginsburg}, {Longmore}, {Kruijssen}, {Battersby}, {Zhang}, \& {Walker}}]{2020ApJ...894L..14L}
{Lu}, X., {Cheng}, Y., {Ginsburg}, A., {et~al.} 2020, \apjl, 894, L14, \dodoi{10.3847/2041-8213/ab8b65}

\bibitem[{{Lu} {et~al.}(2019{\natexlab{a}}){Lu}, {Mills}, {Ginsburg}, {Walker}, {Barnes}, {Butterfield}, {Henshaw}, {Battersby}, {Kruijssen}, {Longmore}, {Zhang}, {Bally}, {Kauffmann}, {Ott}, {Rickert}, \& {Wang}}]{2019ApJS..244...35L}
{Lu}, X., {Mills}, E. A.~C., {Ginsburg}, A., {et~al.} 2019{\natexlab{a}}, \apjs, 244, 35, \dodoi{10.3847/1538-4365/ab4258}

\bibitem[{{Lu} {et~al.}(2019{\natexlab{b}}){Lu}, {Zhang}, {Kauffmann}, {Pillai}, {Ginsburg}, {Mills}, {Kruijssen}, {Longmore}, {Battersby}, {Liu}, \& {Gu}}]{2019ApJ...872..171L}
{Lu}, X., {Zhang}, Q., {Kauffmann}, J., {et~al.} 2019{\natexlab{b}}, \apj, 872, 171, \dodoi{10.3847/1538-4357/ab017d}

\bibitem[{{Lu} {et~al.}(2021){Lu}, {Li}, {Ginsburg}, {Longmore}, {Kruijssen}, {Walker}, {Feng}, {Zhang}, {Battersby}, {Pillai}, {Mills}, {Kauffmann}, {Cheng}, \& {Inutsuka}}]{2021ApJ...909..177L}
{Lu}, X., {Li}, S., {Ginsburg}, A., {et~al.} 2021, \apj, 909, 177, \dodoi{10.3847/1538-4357/abde3c}

\bibitem[{{Mangilli} {et~al.}(2019){Mangilli}, {Aumont}, {Bernard}, {Buzzelli}, {de Gasperis}, {Durrive}, {Ferriere}, {Fo{\"e}nard}, {Hughes}, {Lacourt}, {Misawa}, {Montier}, {Mot}, {Ristorcelli}, {Roussel}, {Ade}, {Alina}, {de Bernardis}, {de Gouveia Dal Pino}, {Dubois}, {Engel}, {Guillet}, {Hargrave}, {Laureijs}, {Longval}, {Maffei}, {Magalhaes}, {Marty}, {Masi}, {Montel}, {Pajot}, {P{\'e}rot}, {Rodriguez}, {Salatino}, {Saccoccio}, {Savini}, {Stever}, {Tauber}, {Tibbs}, \& {Tucker}}]{2019A&A...630A..74M}
{Mangilli}, A., {Aumont}, J., {Bernard}, J.~P., {et~al.} 2019, \aap, 630, A74, \dodoi{10.1051/0004-6361/201935072}

\bibitem[{{Marini} {et~al.}(2023){Marini}, {Dell'Agli}, {Kamath}, {Ventura}, {Mattsson}, {Marchetti}, {Garc{\'\i}a-Hern{\'a}ndez}, {Carini}, {Fabrizio}, \& {Tosi}}]{2023A&A...670A..97M}
{Marini}, E., {Dell'Agli}, F., {Kamath}, D., {et~al.} 2023, \aap, 670, A97, \dodoi{10.1051/0004-6361/202245501}

\bibitem[{{Matthews} \& {Reid}(2007)}]{2007AJ....133.2291M}
{Matthews}, L.~D., \& {Reid}, M.~J. 2007, \aj, 133, 2291, \dodoi{10.1086/512613}

\bibitem[{{Mauerhan} {et~al.}(2010){Mauerhan}, {Muno}, {Morris}, {Stolovy}, \& {Cotera}}]{2010ApJ...710..706M}
{Mauerhan}, J.~C., {Muno}, M.~P., {Morris}, M.~R., {Stolovy}, S.~R., \& {Cotera}, A. 2010, \apj, 710, 706, \dodoi{10.1088/0004-637X/710/1/706}

\bibitem[{{McGrath} {et~al.}(2004){McGrath}, {Goss}, \& {De Pree}}]{2004ApJS..155..577M}
{McGrath}, E.~J., {Goss}, W.~M., \& {De Pree}, C.~G. 2004, \apjs, 155, 577, \dodoi{10.1086/424486}

\bibitem[{{Mehringer} {et~al.}(1994){Mehringer}, {Goss}, \& {Palmer}}]{1994ApJ...434..237M}
{Mehringer}, D.~M., {Goss}, W.~M., \& {Palmer}, P. 1994, \apj, 434, 237, \dodoi{10.1086/174721}

\bibitem[{{Mehringer} \& {Menten}(1997)}]{1997ApJ...474..346M}
{Mehringer}, D.~M., \& {Menten}, K.~M. 1997, \apj, 474, 346, \dodoi{10.1086/303454}

\bibitem[{{Mehringer} {et~al.}(1993{\natexlab{a}}){Mehringer}, {Palmer}, \& {Goss}}]{1993ApJ...402L..69M}
{Mehringer}, D.~M., {Palmer}, P., \& {Goss}, W.~M. 1993{\natexlab{a}}, \apjl, 402, L69, \dodoi{10.1086/186702}

\bibitem[{{Mehringer} {et~al.}(1995){Mehringer}, {Palmer}, \& {Goss}}]{1995ApJS...97..497M}
---. 1995, \apjs, 97, 497, \dodoi{10.1086/192148}

\bibitem[{{Mehringer} {et~al.}(1993{\natexlab{b}}){Mehringer}, {Palmer}, {Goss}, \& {Yusef-Zadeh}}]{1993ApJ...412..684M}
{Mehringer}, D.~M., {Palmer}, P., {Goss}, W.~M., \& {Yusef-Zadeh}, F. 1993{\natexlab{b}}, \apj, 412, 684, \dodoi{10.1086/172954}

\bibitem[{{Mehringer} {et~al.}(1992){Mehringer}, {Yusef-Zadeh}, {Palmer}, \& {Goss}}]{1992ApJ...401..168M}
{Mehringer}, D.~M., {Yusef-Zadeh}, F., {Palmer}, P., \& {Goss}, W.~M. 1992, \apj, 401, 168, \dodoi{10.1086/172050}

\bibitem[{{Meng} {et~al.}(2019){Meng}, {S{\'a}nchez-Monge}, {Schilke}, {Padovani}, {Marcowith}, {Ginsburg}, {Schmiedeke}, {Schw{\"o}rer}, {DePree}, {Veena}, \& {M{\"o}ller}}]{2019A&A...630A..73M}
{Meng}, F., {S{\'a}nchez-Monge}, {\'A}., {Schilke}, P., {et~al.} 2019, \aap, 630, A73, \dodoi{10.1051/0004-6361/201935920}

\bibitem[{{Meng} {et~al.}(2022){Meng}, {S{\'a}nchez-Monge}, {Schilke}, {Ginsburg}, {DePree}, {Budaiev}, {Jeff}, {Schmiedeke}, {Schw{\"o}rer}, {Veena}, \& {M{\"o}ller}}]{2022A&A...666A..31M}
---. 2022, \aap, 666, A31, \dodoi{10.1051/0004-6361/202243674}

\bibitem[{{Messineo} {et~al.}(2002){Messineo}, {Habing}, {Sjouwerman}, {Omont}, \& {Menten}}]{2002AA...393..115M}
{Messineo}, M., {Habing}, H.~J., {Sjouwerman}, L.~O., {Omont}, A., \& {Menten}, K.~M. 2002, \aap, 393, 115, \dodoi{10.1051/0004-6361:20021017}

\bibitem[{{Meynet}(1995)}]{1995A&A...298..767M}
{Meynet}, G. 1995, \aap, 298, 767

\bibitem[{{Mills} {et~al.}(2018){Mills}, {Ginsburg}, {Immer}, {Barnes}, {Wiesenfeld}, {Faure}, {Morris}, \& {Requena-Torres}}]{2018ApJ...868....7M}
{Mills}, E.~A.~C., {Ginsburg}, A., {Immer}, K., {et~al.} 2018, \apj, 868, 7, \dodoi{10.3847/1538-4357/aae581}

\bibitem[{{Molinari} {et~al.}(2016){Molinari}, {Schisano}, {Elia}, {Pestalozzi}, {Traficante}, {Pezzuto}, {Swinyard}, {Noriega-Crespo}, {Bally}, {Moore}, {Plume}, {Zavagno}, {di Giorgio}, {Liu}, {Pilbratt}, {Mottram}, {Russeil}, {Piazzo}, {Veneziani}, {Benedettini}, {Calzoletti}, {Faustini}, {Natoli}, {Piacentini}, {Merello}, {Palmese}, {Del Grande}, {Polychroni}, {Rygl}, {Polenta}, {Barlow}, {Bernard}, {Martin}, {Testi}, {Ali}, {Andr{\'e}}, {Beltr{\'a}n}, {Billot}, {Carey}, {Cesaroni}, {Compi{\`e}gne}, {Eden}, {Fukui}, {Garcia-Lario}, {Hoare}, {Huang}, {Joncas}, {Lim}, {Lord}, {Martinavarro-Armengol}, {Motte}, {Paladini}, {Paradis}, {Peretto}, {Robitaille}, {Schilke}, {Schneider}, {Schulz}, {Sibthorpe}, {Strafella}, {Thompson}, {Umana}, {Ward-Thompson}, \& {Wyrowski}}]{2016A&A...591A.149M}
{Molinari}, S., {Schisano}, E., {Elia}, D., {et~al.} 2016, \aap, 591, A149, \dodoi{10.1051/0004-6361/201526380}

\bibitem[{{Morris} \& {Serabyn}(1996)}]{1996ARA&A..34..645M}
{Morris}, M., \& {Serabyn}, E. 1996, \araa, 34, 645, \dodoi{10.1146/annurev.astro.34.1.645}

\bibitem[{{M{\"u}cke} {et~al.}(2002){M{\"u}cke}, {Koribalski}, {Moffat}, {Corcoran}, \& {Stevens}}]{2002ApJ...571..366M}
{M{\"u}cke}, A., {Koribalski}, B.~S., {Moffat}, A.~F.~J., {Corcoran}, M.~F., \& {Stevens}, I.~R. 2002, \apj, 571, 366, \dodoi{10.1086/339843}

\bibitem[{{Nguyen} {et~al.}(2022){Nguyen}, {Rugel}, {Murugeshan}, {Menten}, {Brunthaler}, {Urquhart}, {Dokara}, {Dzib}, {Gong}, {Khan}, {Medina}, {Ortiz-Le{\'o}n}, {Reich}, {Wyrowski}, {Yang}, {Beuther}, {Cotton}, \& {Pandian}}]{2022AA...666A..59N}
{Nguyen}, H., {Rugel}, M.~R., {Murugeshan}, C., {et~al.} 2022, \aap, 666, A59, \dodoi{10.1051/0004-6361/202244115}

\bibitem[{{Nogueras-Lara}(2024)}]{2024A&A...681L..21N}
{Nogueras-Lara}, F. 2024, \aap, 681, L21, \dodoi{10.1051/0004-6361/202348712}

\bibitem[{{Nogueras-Lara} {et~al.}(2022){Nogueras-Lara}, {Sch{\"o}del}, \& {Neumayer}}]{2022NatAs...6.1178N}
{Nogueras-Lara}, F., {Sch{\"o}del}, R., \& {Neumayer}, N. 2022, Nature Astronomy, 6, 1178, \dodoi{10.1038/s41550-022-01755-3}

\bibitem[{{Oka} \& {Geballe}(2022)}]{2022ApJ...927...97O}
{Oka}, T., \& {Geballe}, T.~R. 2022, \apj, 927, 97, \dodoi{10.3847/1538-4357/ac3912}

\bibitem[{{Oka} {et~al.}(1998){Oka}, {Hasegawa}, {Sato}, {Tsuboi}, \& {Miyazaki}}]{1998ApJS..118..455O}
{Oka}, T., {Hasegawa}, T., {Sato}, F., {Tsuboi}, M., \& {Miyazaki}, A. 1998, \apjs, 118, 455, \dodoi{10.1086/313138}

\bibitem[{{Padovani} {et~al.}(2019){Padovani}, {Marcowith}, {S{\'a}nchez-Monge}, {Meng}, \& {Schilke}}]{2019A&A...630A..72P}
{Padovani}, M., {Marcowith}, A., {S{\'a}nchez-Monge}, {\'A}., {Meng}, F., \& {Schilke}, P. 2019, \aap, 630, A72, \dodoi{10.1051/0004-6361/201935919}

\bibitem[{{Peeters} {et~al.}(2002){Peeters}, {Mart{\'\i}n-Hern{\'a}ndez}, {Damour}, {Cox}, {Roelfsema}, {Baluteau}, {Tielens}, {Churchwell}, {Kessler}, {Mathis}, {Morisset}, \& {Schaerer}}]{2002A&A...381..571P}
{Peeters}, E., {Mart{\'\i}n-Hern{\'a}ndez}, N.~L., {Damour}, F., {et~al.} 2002, \aap, 381, 571, \dodoi{10.1051/0004-6361:20011516}

\bibitem[{{Portegies Zwart} {et~al.}(2001){Portegies Zwart}, {Makino}, {McMillan}, \& {Hut}}]{2001ApJ...546L.101P}
{Portegies Zwart}, S.~F., {Makino}, J., {McMillan}, S. L.~W., \& {Hut}, P. 2001, \apjl, 546, L101, \dodoi{10.1086/318869}

\bibitem[{{Pudritz}(2002)}]{2002Sci...295...68P}
{Pudritz}, R.~E. 2002, Science, 295, 68, \dodoi{10.1126/science.1068298}

\bibitem[{{Reach} {et~al.}(2006){Reach}, {Rho}, {Tappe}, {Pannuti}, {Brogan}, {Churchwell}, {Meade}, {Babler}, {Indebetouw}, \& {Whitney}}]{2006AJ....131.1479R}
{Reach}, W.~T., {Rho}, J., {Tappe}, A., {et~al.} 2006, \aj, 131, 1479, \dodoi{10.1086/499306}

\bibitem[{{Reid} {et~al.}(2009){Reid}, {Menten}, {Zheng}, {Brunthaler}, \& {Xu}}]{2009ApJ...705.1548R}
{Reid}, M.~J., {Menten}, K.~M., {Zheng}, X.~W., {Brunthaler}, A., \& {Xu}, Y. 2009, \apj, 705, 1548, \dodoi{10.1088/0004-637X/705/2/1548}

\bibitem[{{Reid} {et~al.}(2014){Reid}, {Menten}, {Brunthaler}, {Zheng}, {Dame}, {Xu}, {Wu}, {Zhang}, {Sanna}, {Sato}, {Hachisuka}, {Choi}, {Immer}, {Moscadelli}, {Rygl}, \& {Bartkiewicz}}]{2014ApJ...783..130R}
{Reid}, M.~J., {Menten}, K.~M., {Brunthaler}, A., {et~al.} 2014, \apj, 783, 130, \dodoi{10.1088/0004-637X/783/2/130}

\bibitem[{{Ridley} {et~al.}(2017){Ridley}, {Sormani}, {Tre{\ss}}, {Magorrian}, \& {Klessen}}]{2017MNRAS.469.2251R}
{Ridley}, M. G.~L., {Sormani}, M.~C., {Tre{\ss}}, R.~G., {Magorrian}, J., \& {Klessen}, R.~S. 2017, \mnras, 469, 2251, \dodoi{10.1093/mnras/stx944}

\bibitem[{{Rinehart} {et~al.}(2002){Rinehart}, {Houck}, {Smith}, \& {Wilson}}]{2002MNRAS.336...66R}
{Rinehart}, S.~A., {Houck}, J.~R., {Smith}, J.~D., \& {Wilson}, J.~C. 2002, \mnras, 336, 66, \dodoi{10.1046/j.1365-8711.2002.05707.x}

\bibitem[{{Schultheis} {et~al.}(2003){Schultheis}, {Lan{\c{c}}on}, {Omont}, {Schuller}, \& {Ojha}}]{2003A&A...405..531S}
{Schultheis}, M., {Lan{\c{c}}on}, A., {Omont}, A., {Schuller}, F., \& {Ojha}, D.~K. 2003, \aap, 405, 531, \dodoi{10.1051/0004-6361:20030459}

\bibitem[{{Scoville} {et~al.}(1975){Scoville}, {Solomon}, \& {Penzias}}]{1975ApJ...201..352S}
{Scoville}, N.~Z., {Solomon}, P.~M., \& {Penzias}, A.~A. 1975, \apj, 201, 352, \dodoi{10.1086/153892}

\bibitem[{{Sevenster}(2002)}]{2002AJ....123.2772S}
{Sevenster}, M.~N. 2002, \aj, 123, 2772, \dodoi{10.1086/339827}

\bibitem[{{Shiki} {et~al.}(1997){Shiki}, {Ohishi}, \& {Deguchi}}]{1997ApJ...478..206S}
{Shiki}, S., {Ohishi}, M., \& {Deguchi}, S. 1997, \apj, 478, 206, \dodoi{10.1086/303796}

\bibitem[{{Simpson} {et~al.}(2018){Simpson}, {Colgan}, {Cotera}, {Kaufman}, \& {Stolovy}}]{2018ApJ...867L..13S}
{Simpson}, J.~P., {Colgan}, S. W.~J., {Cotera}, A.~S., {Kaufman}, M.~J., \& {Stolovy}, S.~R. 2018, \apjl, 867, L13, \dodoi{10.3847/2041-8213/aae8e4}

\bibitem[{{Simpson} {et~al.}(2021){Simpson}, {Colgan}, {Cotera}, {Kaufman}, \& {Stolovy}}]{2021ApJ...910...59S}
---. 2021, \apj, 910, 59, \dodoi{10.3847/1538-4357/abe636}

\bibitem[{{Smith} {et~al.}(1978){Smith}, {Biermann}, \& {Mezger}}]{1978AA....66...65S}
{Smith}, L.~F., {Biermann}, P., \& {Mezger}, P.~G. 1978, \aap, 66, 65

\bibitem[{{Sofue}(2024)}]{2024MNRAS.532.4187S}
{Sofue}, Y. 2024, \mnras, 532, 4187, \dodoi{10.1093/mnras/stae1724}

\bibitem[{{Sormani} {et~al.}(2022){Sormani}, {Gerhard}, {Portail}, {Vasiliev}, \& {Clarke}}]{2022MNRAS.514L...1S}
{Sormani}, M.~C., {Gerhard}, O., {Portail}, M., {Vasiliev}, E., \& {Clarke}, J. 2022, \mnras, 514, L1, \dodoi{10.1093/mnrasl/slac046}

\bibitem[{{Staguhn} {et~al.}(2004){Staguhn}, {Benford}, {Morris}, \& {Uchida}}]{2004dimg.conf..277S}
{Staguhn}, J., {Benford}, D., {Morris}, M., \& {Uchida}, K. 2004, in The Dense Interstellar Medium in Galaxies, ed. S.~{Pfalzner}, C.~{Kramer}, C.~{Staubmeier}, \& A.~{Heithausen}, Vol.~91, 277, \dodoi{10.1007/978-3-642-18902-9_50}

\bibitem[{{Su{\'a}rez} {et~al.}(2009){Su{\'a}rez}, {G{\'o}mez}, {Miranda}, {Torrelles}, {G{\'o}mez}, {Anglada}, \& {Morata}}]{2009A&A...505..217S}
{Su{\'a}rez}, O., {G{\'o}mez}, J.~F., {Miranda}, L.~F., {et~al.} 2009, \aap, 505, 217, \dodoi{10.1051/0004-6361/200911777}

\bibitem[{{Suh}(2024)}]{2024JKAS...57..123S}
{Suh}, K.-W. 2024, Journal of Korean Astronomical Society, 57, 123, \dodoi{10.5303/JKAS.2024.57.2.123}

\bibitem[{{Sylvester} {et~al.}(1999){Sylvester}, {Kemper}, {Barlow}, {de Jong}, {Waters}, {Tielens}, \& {Omont}}]{1999A&A...352..587S}
{Sylvester}, R.~J., {Kemper}, F., {Barlow}, M.~J., {et~al.} 1999, \aap, 352, 587, \dodoi{10.48550/arXiv.astro-ph/9910368}

\bibitem[{{Uscanga} {et~al.}(2012){Uscanga}, {G{\'o}mez}, {Su{\'a}rez}, \& {Miranda}}]{2012A&A...547A..40U}
{Uscanga}, L., {G{\'o}mez}, J.~F., {Su{\'a}rez}, O., \& {Miranda}, L.~F. 2012, \aap, 547, A40, \dodoi{10.1051/0004-6361/201219760}

\bibitem[{{Ventura} {et~al.}(2020){Ventura}, {Dell'Agli}, {Lugaro}, {Romano}, {Tailo}, \& {Yag{\"u}e}}]{2020A&A...641A.103V}
{Ventura}, P., {Dell'Agli}, F., {Lugaro}, M., {et~al.} 2020, \aap, 641, A103, \dodoi{10.1051/0004-6361/202038289}

\bibitem[{{Ventura} {et~al.}(2014){Ventura}, {Dell'Agli}, {Schneider}, {Di Criscienzo}, {Rossi}, {La Franca}, {Gallerani}, \& {Valiante}}]{2014MNRAS.439..977V}
{Ventura}, P., {Dell'Agli}, F., {Schneider}, R., {et~al.} 2014, \mnras, 439, 977, \dodoi{10.1093/mnras/stu028}

\bibitem[{{Volk} {et~al.}(2002){Volk}, {Kwok}, {Hrivnak}, \& {Szczerba}}]{2002ApJ...567..412V}
{Volk}, K., {Kwok}, S., {Hrivnak}, B.~J., \& {Szczerba}, R. 2002, \apj, 567, 412, \dodoi{10.1086/337992}

\bibitem[{{Volk} {et~al.}(2000){Volk}, {Xiong}, \& {Kwok}}]{2000ApJ...530..408V}
{Volk}, K., {Xiong}, G.-Z., \& {Kwok}, S. 2000, \apj, 530, 408, \dodoi{10.1086/308355}

\bibitem[{{Walsh} {et~al.}(2014){Walsh}, {Purcell}, {Longmore}, {Breen}, {Green}, {Harvey-Smith}, {Jordan}, \& {Macpherson}}]{2014MNRAS.442.2240W}
{Walsh}, A.~J., {Purcell}, C.~R., {Longmore}, S.~N., {et~al.} 2014, \mnras, 442, 2240, \dodoi{10.1093/mnras/stu989}

\bibitem[{{Westbrook} {et~al.}(1976){Westbrook}, {Werner}, {Elias}, {Gezari}, {Hauser}, {Lo}, \& {Neugebauer}}]{1976ApJ...209...94W}
{Westbrook}, W.~E., {Werner}, M.~W., {Elias}, J.~H., {et~al.} 1976, \apj, 209, 94, \dodoi{10.1086/154695}

\bibitem[{{Whitworth} {et~al.}(1994){Whitworth}, {Bhattal}, {Chapman}, {Disney}, \& {Turner}}]{1994A&A...290..421W}
{Whitworth}, A.~P., {Bhattal}, A.~S., {Chapman}, S.~J., {Disney}, M.~J., \& {Turner}, J.~A. 1994, \aap, 290, 421

\bibitem[{{Yusef-Zadeh} {et~al.}(2022){Yusef-Zadeh}, {Arendt}, {Wardle}, {Heywood}, \& {Cotton}}]{2022MNRAS.517..294Y}
{Yusef-Zadeh}, F., {Arendt}, R.~G., {Wardle}, M., {Heywood}, I., \& {Cotton}, W. 2022, \mnras, 517, 294, \dodoi{10.1093/mnras/stac2415}

\bibitem[{{Yusef-Zadeh} {et~al.}(2009){Yusef-Zadeh}, {Hewitt}, {Arendt}, {Whitney}, {Rieke}, {Wardle}, {Hinz}, {Stolovy}, {Lang}, {Burton}, \& {Ramirez}}]{2009ApJ...702..178Y}
{Yusef-Zadeh}, F., {Hewitt}, J.~W., {Arendt}, R.~G., {et~al.} 2009, \apj, 702, 178, \dodoi{10.1088/0004-637X/702/1/178}

\bibitem[{{Yusef-Zadeh} {et~al.}(2024){Yusef-Zadeh}, {Wardle}, {Arendt}, {Hewitt}, {Hu}, {Lazarian}, {Kassim}, {Hyman}, \& {Heywood}}]{2024MNRAS.527.1275Y}
{Yusef-Zadeh}, F., {Wardle}, M., {Arendt}, R., {et~al.} 2024, \mnras, 527, 1275, \dodoi{10.1093/mnras/stad3203}

\bibitem[{{Zhang} \& {Kwok}(2009)}]{2009ApJ...706..252Z}
{Zhang}, Y., \& {Kwok}, S. 2009, \apj, 706, 252, \dodoi{10.1088/0004-637X/706/1/252}

\bibitem[{{Zhang} \& {Tan}(2011)}]{2011ApJ...733...55Z}
{Zhang}, Y., \& {Tan}, J.~C. 2011, \apj, 733, 55, \dodoi{10.1088/0004-637X/733/1/55}

\bibitem[{{Zinnecker} \& {Yorke}(2007)}]{2007ARA&A..45..481Z}
{Zinnecker}, H., \& {Yorke}, H.~W. 2007, \araa, 45, 481, \dodoi{10.1146/annurev.astro.44.051905.092549}

\end{thebibliography}
\bibliographystyle{aasjournal}


\appendix

\vspace{-6mm}

\begin{deluxetable*}{lrrrrrrrrrrrr}
\tabletypesize{\scriptsize}
\tablecaption{Spitzer-IRAC Observational Parameters of Sources in Sgr~B1 \label{tb:BIRAC}}
\tablehead{\colhead{  }&
           \multicolumn{3}{c}{${\rm 3.6\mu{m}}$}&
           \multicolumn{3}{c}{${\rm 4.5\mu{m}}$}&
           \multicolumn{3}{c}{${\rm 5.8\mu{m}}$}&
           \multicolumn{3}{c}{${\rm 8.0\mu{m}}$}\\[-4pt]
           \cmidrule(lr){2-4} \cmidrule(lr){5-7} \cmidrule(lr){8-10} \cmidrule(lr){11-13}\\[-12pt]
           \colhead{ Source }&
           \colhead{ $R_{\rm int}$ } &
           \colhead{ $F_{\rm int}$ } &
           \colhead{ $F_{\rm int-bg}$ } &
                      \colhead{ $R_{\rm int}$ } &
           \colhead{ $F_{\rm int}$ } &
           \colhead{ $F_{\rm int-bg}$ } &
                      \colhead{ $R_{\rm int}$ } &
           \colhead{ $F_{\rm int}$ } &
           \colhead{ $F_{\rm int-bg}$ } &
                      \colhead{ $R_{\rm int}$ } &
           \colhead{ $F_{\rm int}$ } &
           \colhead{ $F_{\rm int-bg}$ } \\[-6pt]
	   \colhead{  } &
	   \colhead{ ($\arcsec$) } &
	   \colhead{ (mJy) } &
	   \colhead{ (mJy) } &
	   \colhead{ ($\arcsec$) } &
	   \colhead{ (mJy) } &
	   \colhead{ (mJy) } &
	   \colhead{ ($\arcsec$) } &
	   \colhead{ (mJy) } &
	   \colhead{ (mJy) } &
	   \colhead{ ($\arcsec$) } &
	   \colhead{ (mJy) } &
	   \colhead{ (mJy) } \\[-12pt]
}
\startdata
\multicolumn{13}{c}{Compact Sources} \\
\hline
Sgr B1 A	&	4.8	&	101	    &	51.0	&	6.0	&	218	    &	135	    &	6.0	&	736	    &	412	    &	6.0	&	1960	&	1270	\\
Sgr B1 C	&	6.0	&	676	    &	603	    &	6.0	&	780	    &	704	    &	6.0	&	1510	&	1230	&	7.2	&	2730	&	1860	\\
Sgr B1 D	&	8.4	&	803	    &	626	    &	8.4	&	989	    &	800	    &	9.6	&	3400	&	2160	&	9.6	&	8760	&	5740	\\
Sgr B1 1	&	8.4	&	234	    &	42.3	&	8.4	&	296	    &	75.3	&	8.4	&	1050	&	247	    &	8.4	&	4970	&	1190	\\
Sgr B1 2	&	4.8	&	$<$65.5	&	UR	    &	4.8	&	$<$77.5	&	UR	    &	4.8	&	340	    &	90.6	&	4.8	&	793	    &	190	\\
Sgr B1 F	&	4.8	&	$<$141	&	UR	    &	4.8	&	114	    &	62.3	&	4.8	&	435	    &	206	    &	4.8	&	912	    &	405	\\
Sgr B1 G	&	7.2	&	431	    &	322	    &	7.2	&	507	    &	361	    &	8.4	&	1320	&	542	    &	8.4	&	2520	&	798	\\
Sgr B1 H	&	4.8	&	37.9	&	6.63	&	4.8	&	54.6	&	18.4	&	6.0	&	319	    &	98.0    &	8.4	&	1440	&	466	\\
Sgr B1 3	&	6.0	&	808	    &	746	    &	sat	&	sat	    &	sat	    &	7.2	&	10500	&	9920	&	sat	&	sat	    &	sat	\\
Sgr B1 4	&	4.8	&	58.7	&	UD	    &	4.8	&	$<$75.4	&	UD	    &	4.8	&	336	    &	44.9	&	4.8	&	808	    &	20.0	\\
Sgr B1 5	&	6.0	&	89.9	&	10.1	&	6.0	&	157	    &	14.8	&	6.0	&	572	    &	75.9	&	6.0	&	1350	&	277	\\
Sgr B1 6	&	4.8	&	60.5	&	4.28	&	4.8	&	119	    &	19.7	&	4.8	&	356	    &	32.0	&	4.8	&	875	    &	85.5	\\
Sgr B1 7	&	4.8	&	93.4	&	21.9	&	4.8	&	159	    &	27.3	&	6.0	&	659	    &	69.3	&	6.0	&	1560	&	241	\\
Sgr B1 8	&	7.2	&	155	    &	28.6	&	7.2	&	213	    &	44.6	&	7.2	&	827	    &	98.7	&	7.2	&	1970	&	229	\\
Sgr B1 9	&	7.2	&	304	    &	130	    &	7.2	&	354	    &	107	    &	7.2	&	970	    &	141	    &	7.2	&	2050	&	244	\\
Sgr B1 10	&	7.2	&	273	    &	116	    &	7.2	&	357	    &	167	    &	7.2	&	862	    &	274	    &	9.6	&	2890	&	768	\\
Sgr B1 11	&	4.8	&	85.6	&	29.0	&	4.8	&	135	    &	54.4	&	7.2	&	948	    &	184	    &	7.2	&	2150	&	441	\\
Sgr B1 12	&	8.4	&	268	    &	93.3	&	8.4	&	330	    &	151	    &	8.4	&	1120	&	250	    &	10.8&	4060	&	1030	\\
\hline
\multicolumn{13}{c}{Extended Sources} \\
\hline
Ionized Bar	&	143x40	&	$<$8.15	&	UD	&	143x40	&	$<$8.01	&	con	    &	143x40	&	$<$24.2	&	con	    &	143x40	&	51000	&	41200	\\
Ionized Rim	&	64x114	&	$<$7.97	&	con	&	64x114	&	$<$8.80	&	con	    &	64x114	&	$<$28.7	&	con	    &  	64x114	&	65200	&	54500	\\
Sgr B1 Ext1	&	26.0	&	$<$2.27	&	con	&	26.0	&	$<$2.54	&	con	    &	26.0	&	8940	&	6610	&	26.0	&	21000	&	16300	\\
Sgr B1 E 	&	23.0	&	$<$3.18	&	con	&	23.0	&	4460	&	3250	&	23.0	&	15000	&	12200	&	23.0	&	36900	&	30600	\\
Sgr B1 I	&	35.0	&	$<$5.12	&	con	&	35.0	&	7410	&	3920	&	35.0	&	21500	&	16800	&	35.0	&	49200	&	39000	\\
Sgr B1 J	&	28x80	&	$<$2.03	&	UD	&	28x80	&	$<$2.13	&	UD	    &	28x80	&	$<$6.67	&	UD	    &	28x80	&	13900	&	8710		
\enddata
\tablecomments{Entries with ``sat'' means the sources are themselves saturated in that band or are affected by array saturation effects from nearby bright sources. Entries with ``con'' means that the total measured emission is dominated by contamination from (most likely foreground) stars. For upper limits, UR means the source is unresolved from a much brighter nearby source or extended emission, and UD means the source is not undetected. For all undetected sources (as well as unresolved sources), there is significant background emission or contamination from nearby sources that the flux upper limit given is the measured flux in the indicated aperture. For saturated sources, we use the point source saturation fluxes of 190, 200, 1400, and 740\,mJy at 3.6, 4.5, 5.8, and 8.0\,$\mu$m, respectively (from the Spitzer Observer’s Manual, Version 7.1.), as lower limits in the SED modeling.}
\end{deluxetable*}

\vspace{-5mm}

\section{Data release}\label{sec:appendixA}

The FITS images used in this study are publicly available in the Dataverse \citep{DVN/POYMK5_2019}. This repository includes the versions of the SOFIA-FORCAST 25 and 37\,$\mu$m final image mosaics (of both Sgr~B and Sgr~C and their exposure maps) that were used in our analyses presented here. The data for Sgr~B were cropped from the larger Galactic Center mosaic of \citet{2020ApJ...894...55H} with additional background correction (see Section \ref{sec:obs}), however the Sgr~C mosaics were created specifically for this study, and cover a larger area around Sgr~C than was covered in the \citet{2020ApJ...894...55H} mosaic.  

\begin{deluxetable*}{lrrrrrrrrrrrr}
\tabletypesize{\scriptsize}
\tablecaption{Spitzer-IRAC Observational Parameters of Sources in Sgr~B2 \label{tb:B2IRAC}}
\tablehead{\colhead{  }&
           \multicolumn{3}{c}{${\rm 3.6\mu{m}}$}&
           \multicolumn{3}{c}{${\rm 4.5\mu{m}}$}&
           \multicolumn{3}{c}{${\rm 5.8\mu{m}}$}&
           \multicolumn{3}{c}{${\rm 8.0\mu{m}}$}\\[-4pt]
           \cmidrule(lr){2-4} \cmidrule(lr){5-7} \cmidrule(lr){8-10} \cmidrule(lr){11-13}\\[-12pt]
           \colhead{ Source }&
           \colhead{ $R_{\rm int}$ } &
           \colhead{ $F_{\rm int}$ } &
           \colhead{ $F_{\rm int-bg}$ } &
                      \colhead{ $R_{\rm int}$ } &
           \colhead{ $F_{\rm int}$ } &
           \colhead{ $F_{\rm int-bg}$ } &
                      \colhead{ $R_{\rm int}$ } &
           \colhead{ $F_{\rm int}$ } &
           \colhead{ $F_{\rm int-bg}$ } &
                      \colhead{ $R_{\rm int}$ } &
           \colhead{ $F_{\rm int}$ } &
           \colhead{ $F_{\rm int-bg}$ } \\[-6pt]
	   \colhead{  } &
	   \colhead{ ($\arcsec$) } &
	   \colhead{ (mJy) } &
	   \colhead{ (mJy) } &
	   \colhead{ ($\arcsec$) } &
	   \colhead{ (mJy) } &
	   \colhead{ (mJy) } &
	   \colhead{ ($\arcsec$) } &
	   \colhead{ (mJy) } &
	   \colhead{ (mJy) } &
	   \colhead{ ($\arcsec$) } &
	   \colhead{ (mJy) } &
	   \colhead{ (mJy) } \\[-12pt]
}
\startdata
\multicolumn{13}{c}{Compact Sources} \\
\hline
Sgr B2 1	&	4.8	&	$<$10.8	&	UD	    &	4.8	&	14.9	&	1.43	&	4.8	&	70.5	&	10.7	&	4.8	&	167	    &	34.4	\\
Sgr B2 2	&	6.0	&	$<$77.3	&	UD	    &	6.0	&	$<$105	&	UD	    &	6.0	&	$<$174	&	UD	    &	6.0	&	$<$237	&	UD	\\
Sgr B2 3	&	4.8	&	$<$16.1	&	UD	    &	4.8	&	$<$18.8	&	UD	    &	4.8	&	$<$78.8	&	UD	    &	4.8	&	$<$152	&	UD	\\
Sgr B2 AA	&	4.8	&	$<$18.0	&	UD	    &	4.8	&	30.2	&	17.2	&	6.0	&	155	    &	62.3	&	6.0	&	336	    &	153	\\
Sgr B2 4 	&	8.4	&	$<$104	&	UD	    &	8.4	&	$<$88.5	&	UD	    &	8.4	&	$<$190	&	UD	    &	8.4	&	$<$242	&	UD	\\
Sgr B2 H	&	6.0	&	69.3	&	40.7	&	6.0	&	216	    &	181	    &	6.0	&	496	    &	359	    &	6.0	&	772	    &	517	\\
Sgr B2 5	&	4.8	&	29.5	&	11.6	&	4.8	&	32.9	&	13.9	&	4.8	&	126	    &	32.1	&	4.8	&	273	    &	88.0	\\
Sgr B2 6	&	4.8	&	43.6	&	26.2	&	4.8	&	30.2	&	17.3	&	4.8	&	75.1	&	13.4	&	4.8	&	138	    &	17.0	\\
Sgr B2 BB	&	9.6	&	$<$117	&	UD	    &	9.6	&	148	    &	91.1	&	12.0&	610	    &	273	    &	12.0&	1440	&	840	\\
Sgr B2 L	&	6.0	&	$<$26.7	&	UD	    &	6.0	&	32.2	&	12.8	&	7.2	&	147	    &	41.5	&	7.2	&	237	    &	69.4	\\
Sgr B2 7	&	6.0	&	$<$33.2	&	UD	    &	6.0	&	$<$30.6	&	UD	    &	6.0	&	131	    &	11.8	&	6.0	&	281	    & 	33.8	\\
Sgr B2 O	&	7.2	&	$<$35.7	&	UD	    &	7.2	&	52.6	&	17.2	&	8.4	&	255	    &	58.9	&	8.4	&	581	    &	166	\\
Sgr B2 8	&	7.2	&	45.9	&	13.5	&	7.2	&	60.4	&	24.4	&	8.4	&	280	    &	94.6	&	8.4	&	657	    &	268	\\
Sgr B2 9	&	4.8	&	$<$28.8	&	UD	    &	4.8	&	38.7	&	11.1	&	4.8	&	109	    &	29.4	&	6.0	&	312	    &	90.9	\\
Sgr B2 P	&	4.8	&	29.5	&	8.75	&	4.8	&	63.5	&	45.3	&	7.2	&	278	    &	131	    &	7.2	&	715	    &	397	\\
Sgr B2 R	&	9.6	&	86.1	&	16.1	&	9.6	&	196	    &	115	    &	10.8&	729	    &	362	    &	10.8&	2070	&	1330	\\
Sgr B2 10	&	4.8	&	54.8	&	26.1	&	6.0	&	128	    &	68.5	&	6.0	&	328	    &	149	    &	6.0	&	729	    &	303	\\
Sgr B2 11	&	4.8	&	44.6	&	24.6	&	6.0	&	57.6	&	22.9	&	6.0	&	150	    &	28.4	&	6.0	&	333	    &	90.9	\\
Sgr B2 12	&	4.8	&	35.8	&	7.42	&	4.8	&	49.5	&	10.5	&	7.2	&	334	    &	285	    &	7.2	&	742	    &	93.2	\\
Sgr B2 13	&	4.8	&	27.1	&	4.69	&	4.8	&	36.1	&	10.7	&	4.8	&	159	    &	18.6	&	7.2	&	674	    &	165	\\
Sgr B2 14	&	9.6	&	217	    &	141	    &	9.6	&	296	    &	186	    &	9.6	&	721	    &	347	    &	9.6	&	1540	&	586	\\
Sgr B2 15	&	6.0	&	781	    &	729	    &	6.0	&	1020	&	958	    &	6.0	&	1960	&	1730	&	6.0	&	1430	&	967	\\
Sgr B2 16	&	6.0	&	78.4	&	35.8	&	6.0	&	89.2	&	46.0	&	8.4	&	614	    &	305	    &	8.4	&	1190	&	519	\\
Sgr B2 17	&	8.4	&	120	    &	17.6	&	9.6	&	257	    &	84.5	&	9.6	&	683	    &	145	    &	9.6	&	1640	&	497	\\
\hline
\multicolumn{13}{c}{Extended Sources} \\
\hline
Sgr B2 V	&	27.0	&	$<$1.00	    &	con	&	27.0	&	1760	    &	1310	&	27.0	&	5290	&	3220	&	27.0	&	12800	&	10400	\\
Sgr B2 Main	&	27.0	&	$<$0.918	&	UD	&	27.0	&	$<$0.983	&	con	    &	27.0	&	$<$3.18	&	con	    &	27.0	&	4860	&	2510	\\
Sgr B2 Ext1	&	28x50	&	$<$0.831	&	UD	&	28x50	&	$<$1.02	    &	con	    &	28x50	&	3250	&	1910	&	28x50	&	7720	&	5260	\\
Sgr B2 Ext2	&	197.0	&	$<$1.07	    &	con	&	197.0	&	1540	    &	1080	&	197.0	&	3720	&	2140	&	197.0	&	8010	&	6270			
\enddata
\tablecomments{See table notes from Table \ref{tb:BIRAC}.}
\end{deluxetable*}

\vspace{-4mm}

\section{Additional Photometry of Sources in Sgr~B1, Sgr~B2, and Sgr~C}\label{sec:appendixB}

As stated in Section \ref{sec:cps}, in addition to the fluxes derived from the SOFIA-FORCAST data, we derived additional aperture photometry values for all compact sources using archival Spitzer-IRAC data at 3.6, 4.5, 5.8, and 8.0\,$\mu$m, as well as Herschel-PACS data at 70 and 160\,$\mu$m. Like the FORCAST data, we applied the same optimal extraction technique to the Spitzer data. However, performing the optimal extraction technique on the Herschel-PACS data failed for all sources due to contamination from extended emission from other nearby sources and/or bright environmental emission. For the contaminated compact sources in the Herschel-PACS data, we used an aperture that best fit the largest size of the source at any wavelength to derive flux estimates within the aperture, but do not derive background-subtracted photometry values.

Table \ref{tb:BIRAC}, Table \ref{tb:B2IRAC}, and Table \ref{tb:CIRAC} list the photometry values we derived from the Spitzer-IRAC data for all sources within Sgr~B1, Sgr~B2, and Sgr~C, respectively. Table \ref{tb:BPACS}, Table \ref{tb:B2PACS}, and Table \ref{tb:CPACS} give the photometry values from the Herschel-PACS data for all sources within Sgr~B1, Sgr~B2, and Sgr~C, respectively.

\begin{deluxetable*}{lrrrrrrrrrrrr}
\tabletypesize{\scriptsize}
\tablecaption{Spitzer-IRAC Observational Parameters of Sources in Sgr~C \label{tb:CIRAC}}
\tablehead{\colhead{  }&
           \multicolumn{3}{c}{${\rm 3.6\mu{m}}$}&
           \multicolumn{3}{c}{${\rm 4.5\mu{m}}$}&
           \multicolumn{3}{c}{${\rm 5.8\mu{m}}$}&
           \multicolumn{3}{c}{${\rm 8.0\mu{m}}$}\\[-4pt]
           \cmidrule(lr){2-4} \cmidrule(lr){5-7} \cmidrule(lr){8-10} \cmidrule(lr){11-13}\\[-12pt]
           \colhead{ Source }&
           \colhead{ $R_{\rm int}$ } &
           \colhead{ $F_{\rm int}$ } &
           \colhead{ $F_{\rm int-bg}$ } &
                      \colhead{ $R_{\rm int}$ } &
           \colhead{ $F_{\rm int}$ } &
           \colhead{ $F_{\rm int-bg}$ } &
                      \colhead{ $R_{\rm int}$ } &
           \colhead{ $F_{\rm int}$ } &
           \colhead{ $F_{\rm int-bg}$ } &
                      \colhead{ $R_{\rm int}$ } &
           \colhead{ $F_{\rm int}$ } &
           \colhead{ $F_{\rm int-bg}$ } \\[-6pt]
	   \colhead{  } &
	   \colhead{ ($\arcsec$) } &
	   \colhead{ (mJy) } &
	   \colhead{ (mJy) } &
	   \colhead{ ($\arcsec$) } &
	   \colhead{ (mJy) } &
	   \colhead{ (mJy) } &
	   \colhead{ ($\arcsec$) } &
	   \colhead{ (mJy) } &
	   \colhead{ (mJy) } &
	   \colhead{ ($\arcsec$) } &
	   \colhead{ (mJy) } &
	   \colhead{ (mJy) } \\[-12pt]
}
\startdata
\multicolumn{13}{c}{Compact Sources} \\
\hline
Sgr C 1	    &	3.0	&	22.9	&	4.17	&	4.2	&	49.2	&	13.4	&	5.4	    &	413	    &	75.2	&	7.2	    &	1910	&	304	\\
Sgr C 2	    &	8.4	&	292	    &	128	    &	9.0	&	431	    &	211	    &	10.8	&	1680	&	441	    &	12.0	&	5130	&	1090	\\
Sgr C 3	    &	6.0	&	$<$101	&	UR	    &	6.0	&	118	    &	46.1	&	6.0	    &	445	    &	67.1	&	6.6	    &	1370	&	113	\\
Sgr C 4	    &	4.8	&	$<$58.6	&	UR	    &	4.8	&	$<$58.5	&	UR	    &	4.8	    &	322	    &	41.6	&	5.4	    &	1080	&	72.0	\\
Sgr C 5	    &	10.2&	$<$3000	&	UR	    &	10.2&	$<$355	&	UD	    &	10.2	&	1580	&	251	    &	10.8	&	4540	&	632	\\
Sgr C 6	    &	4.8	&	53.0    &	14.5	&	6.0	&	113	    &	46.9	&	7.8	    &	764	    &	128	    &	8.4	    &	2190	&	297	\\
Sgr C 7	    &	3.0	&	25.6	&	8.86	&	4.8	&	64.4	&	14.7	&	6.0	    &	471	    &	59.6	&	6.0	    &	1310	&	157	\\
Sgr C 8	    &	7.2	&	$<$159	&	UD	    &	7.2	&	$<$182	&	UD	    &	7.2	    &	668	    &	98.4	&	8.4	    &	2450	&	385	\\
Sgr C 9	    &	4.2	&	$<$31.7	&	UD	    &	4.2	&	$<$33.5	&	UD	    &	4.2	    &	$<$189	&	UD	    &	4.2	    &	521	    &	103	\\
Sgr C 10	&	4.2	&	35.4	&	14.3	&	4.2	&	64.2	&	32.5	&	4.8	    &	342	    &	87.4	&	4.8	    &	916	    &	200	\\
Sgr C 11	&	3.0	&	126	    &	109	    &	3.6	&	317	    &	289	    &	4.2	    &	797	    &	602	    &	5.4	    &	1750	&	903	\\
Sgr C 12	&	4.8	&	142	    &	81.4	&	4.8	&	151	    &	85.6	&	6.0	    &	631	    &	142	    &	6.0	    &	1620	&	423	\\
Sgr C 13	&	2.4	&	10.5	&	3.40    &	2.4	&	12.9	&	4.10	&	3.0	    &	102	    &	30.9	&	3.6	    &	367	    &	64.1	\\
Sgr.C 14	&	3.0	&	$<$9.91	&	UD	    &	3.0	&	$<$15.2	&	UR	    &	3.0	    &	71.7	&	15.7	&	3.0	    &	202	    &	31.4	\\
Sgr C H3	&	4.8	&	$<$42.4	&	UR	    &	4.8	&	192	    &	157	    &	4.8	    &	421	    &	274	    &	4.8	    &	610	    &	218	\\
Sgr C H4	&	3.6	&	$<$24.2	&	UR	    &	3.6	&	$<$81.3	&	UR	    &	3.6	    &	$<$129	&	UR	    &	3.6	    &	$<$215	&	UD	\\
Sgr C H1	&	3.0	&	$<$17.2	&	UR	    &	3.0	&	$<$20.4	&	UR	    &	3.0	    &	83.3	&	27.1	&	4.2	    &	384	    &	93.9	\\
Sgr C 15	&	7.2	&	$<$130	&	UR	    &	7.2	&	$<$124	&	UR	    &	7.2	    &	540	    &	25.1	&	8.4	    &	1970	&	188	\\
Sgr C 16	&	6.0	&	$<$80.9	&	UR	    &	6.0	&	$<$79.0	&	UR	    &	6.0	    &	446	    &	24.3	&	8.4	    &	2180	&	628	\\
\hline
\multicolumn{13}{c}{Extended Sources} \\
\hline
Sgr C HII 	&	83.6	&	20200	&	8480	&	83.6	&	21100	&	6560	&	83.6	&	82400	&	47300	&	83.6	&	212000	&	123000	\\
\enddata
\tablecomments{UR and UD have same meaning as discussion in caption of Table~\ref{tb:BIRAC}. For these sources, the $F_{\rm int}$ value is used as the upper limit in the SED modeling.}
\end{deluxetable*}

\begin{deluxetable*}{lrrrrrr}
\tabletypesize{\scriptsize}
\tablecolumns{7}
\tablewidth{0pt}
\tablecaption{Herschel-PACS Observational Parameters of Sources in Sgr~B1 \label{tb:BPACS}}
\tablehead{\colhead{  }&
           \multicolumn{3}{c}{${\rm 70\mu{m}}$}&
           \multicolumn{3}{c}{${\rm 160\mu{m}}$}\\[-4pt]
           \cmidrule(lr){2-4} \cmidrule(lr){5-7}\\[-12pt]
           \colhead{ Source }&
           \colhead{ $R_{\rm int}$ } &
           \colhead{ $F_{\rm int}$ } &
           \colhead{ $F_{\rm int-bg}$ } &
           \colhead{ $R_{\rm int}$ } &
           \colhead{ $F_{\rm int}$ } &
           \colhead{ $F_{\rm int-bg}$ } \\[-6pt]
	   \colhead{  } &
	   \colhead{ ($\arcsec$) } &
	   \colhead{ (Jy) } &
	  \colhead{ (Jy) } &
	   \colhead{ ($\arcsec$) } &
	   \colhead{ (Jy) } &
	   \colhead{ (Jy) } \\[-12pt]
}
\startdata
\multicolumn{7}{c}{Compact Sources} \\
\hline
Sgr B1 A	&	22.4	&	$<$963	&	UR	&	22.4	&	$<$1020	&	UR	\\
Sgr B1 C	&	16.0	&	$<$395	&	UR	&	22.4	&	$<$933	&	UD	\\
Sgr B1 D	&	16.0	&	$<$685	&	UR	&	22.4	&	$<$1100	&	UR	\\
Sgr B1 1	&	16.0	&	$<$699	&	UR	&	22.4	&	$<$1060	&	UD	\\
Sgr B1 2	&	16.0	&	$<$404	&	UR	&	22.4	&	$<$868	&	UD	\\
Sgr B1 F	&	16.0	&	$<$429	&	UR	&	22.4	&	$<$1070	&	UD	\\
Sgr B1 G	&	16.0	&	$<$579	&	UR	&	22.4	&	$<$1050	&	UD	\\
Sgr B1 H	&	22.4	&	$<$813	&	UR	&	22.4	&	$<$1300	&	UD	\\
Sgr B1 3	&	16.0	&	$<$419	&	UR	&	22.4	&	$<$1140	&	UD	\\
Sgr B1 4	&	16.0	&	$<$545	&	UR	&	22.4	&	$<$916	&	UR	\\
Sgr B1 5	&	16.0	&	$<$550	&	UR	&	22.4	&	$<$856	&	UR	\\
Sgr B1 6	&	16.0	&	$<$560	&	UR	&	22.4	&	$<$871	&	UR	\\
Sgr B1 7	&	16.0	&	$<$561	&	UR	&	22.4	&	$<$857	&	UD	\\
Sgr B1 8	&	16.0	&	$<$526	&	UD	&	22.4	&	$<$929	&	UD	\\
Sgr B1 9	&	16.0	&	$<$457	&	UR	&	22.4	&	$<$829	&	UD	\\
Sgr B1 10	&	16.0	&	$<$359	&	UR	&	22.4	&	$<$921	&	UD	\\
Sgr B1 11	&	16.0	&	$<$479	&	UD	&	22.4	&	$<$903	&	UD	\\
Sgr B1 12	&	16.0	&	$<$408	&	UR	&	22.4	&	$<$831	&	UR	\\
\hline
\multicolumn{7}{c}{Extended Sources} \\
\hline
Ionized Bar	&	143x40	&	4020	&	3180	&	143x40	&	$<$3580	&	UD	\\
Ionized Rim	&	64x114	&	5320	&	2120	&	64x114	&	$<$4650	&	UD	\\
Sgr B1 Ext1	&	26.1	&	1700	&	1280	&	26.1	&	$<$1450	&	UD	\\
Sgr B1 E 	&	23.0	    &	1340	&	744	    &	23.0	    &	$<$1070	&	UD	\\
Sgr B1 I	&	35.0	    &	2580	&	1800	&	35.0	    &	$<$2070	&	UD	\\
Sgr B1 J	&	28x80	&	1160	&	699	    &	28x80	&	$<$1390	&	UD			
\enddata
\tablecomments{UR and UD have same meaning as discussion in caption of Table~\ref{tb:BIRAC}. For these sources, the $F_{\rm int}$ value is used as the upper limit in the SED modeling.}
\end{deluxetable*}

\begin{deluxetable*}{lrrrrrr}
\tabletypesize{\scriptsize}
\tablecolumns{7}
\tablewidth{0pt}
\tablecaption{Herschel-PACS Observational Parameters of Sources in Sgr~B2 \label{tb:B2PACS}}
\tablehead{\colhead{  }&
           \multicolumn{3}{c}{${\rm 70\mu{m}}$}&
           \multicolumn{3}{c}{${\rm 160\mu{m}}$}\\[-4pt]
           \cmidrule(lr){2-4} \cmidrule(lr){5-7}\\[-12pt]
           \colhead{ Source }&
           \colhead{ $R_{\rm int}$ } &
           \colhead{ $F_{\rm int}$ } &
           \colhead{ $F_{\rm int-bg}$ } &
           \colhead{ $R_{\rm int}$ } &
           \colhead{ $F_{\rm int}$ } &
           \colhead{ $F_{\rm int-bg}$ } \\[-6pt]
	   \colhead{  } &
	   \colhead{ ($\arcsec$) } &
	   \colhead{ (Jy) } &
	  \colhead{ (Jy) } &
	   \colhead{ ($\arcsec$) } &
	   \colhead{ (Jy) } &
	   \colhead{ (Jy) } \\[-12pt]
}
\startdata
\multicolumn{7}{c}{Compact Sources} \\
\hline
Sgr B2 1	&	16.0	&	$<$325	&	UR	&	22.4	&	$<$2060	&	UD	\\
Sgr B2 2	&	16.0	&	$<$533	&	UR	&	22.4	&	$<$5320	&	UD	\\
Sgr B2 3	&	16.0	&	$<$528	&	UR	&	22.4	&	$<$4700	&	UR	\\
Sgr B2 AA	&	16.0	&	$<$411	&	UR	&	22.4	&	$<$4020	&	UR	\\
Sgr B2 4 	&	19.2	&	$<$2420	&	UR	&	sat	    &	sat	    &	sat	\\
Sgr B2 H	&	16.0	&	$<$3550	&	UR	&	22.4	&	$<$10900	&	UR	\\
Sgr B2 5	&	16.0	&	$<$2270	&	UD	&	22.4	&	$<$8180	&	UD	\\
Sgr B2 6	&	16.0	&	$<$310	&	UR	&	22.4	&	$<$3810	&	UR	\\
Sgr B2 BB	&	16.0	&	$<$1040	&	UR	&	22.4	&	$<$4230	&	UR	\\
Sgr B2 L	&	16.0	&	$<$695	&	UR	&	22.4	&	$<$4850	&	UR	\\
Sgr B2 7	&	16.0	&	$<$249	&	UR	&	22.4	&	$<$2300	&	UR	\\
Sgr B2 O	&	16.0	&	$<$1400	&	UR	&	22.4	&	$<$3410	&	UR	\\
Sgr B2 8	&	16.0	&	$<$661	&	UR	&	22.4	&	$<$3590	&	UD	\\
Sgr B2 9	&	16.0	&	$<$846	&	UR	&	22.4	&	$<$3150	&	UD	\\
Sgr B2 P	&	16.0	&	$<$715	&	UR	&	22.4	&	$<$3090	&	UD	\\
Sgr B2 R	&	19.2	&	$<$949	&	UR	&	22.4	&	$<$2600	&	UR	\\
Sgr B2 10	&	16.0	&	$<$375	&	UR	&	22.4	&	$<$1510	&	UR	\\
Sgr B2 11	&	16.0	&	$<$262	&	UR	&	22.4	&	$<$1470	&	UD	\\
Sgr B2 12	&	16.0	&	$<$391	&	UR	&	22.4	&	$<$1290	&	UD	\\
Sgr B2 13	&	16.0	&	$<$342	&	UR	&	22.4	&	$<$1460	&	UR	\\
Sgr B2 14	&	16.0	&	$<$325	&	UR	&	22.4	&	$<$1560	&	UD	\\
Sgr B2 15	&	16.0	&	$<$395	&	UD	&	22.4	&	$<$1300	&	UD	\\
Sgr B2 16	&	19.2	&	$<$470	&	UR	&	22.4	&	$<$1390	&	UR	\\
Sgr B2 17	&	16.0	&	$<$308	&	UR	&	22.4	&	$<$942	&	UR	\\
\hline
\multicolumn{7}{c}{Extended Sources} \\
\hline
Sgr B2 V	&	27.0	&	1250	&	905	&	27.0	&	2830	&	1100	\\
Sgr B2 Main	&	27.0	&	13400	&	13100	&	27.0	&	sat	&	sat	\\
Sgr B2 Ext1	&	28x50	&	783	&	545	&	28x50	&	$<$1150	&	UD	\\
Sgr B2 Ext2	&	19.2	&	454	&	257	&	19.2	&	$<$724	&	UD				
\enddata
\tablecomments{UR and UD have same meaning as discussion in caption of Table~\ref{tb:BIRAC}. For these sources, the $F_{\rm int}$ value is used as the upper limit in the SED modeling. Sgr~B2~K is saturated (``sat"), so for this target only we use the point source saturation flux of 1125~Jy at 160\,$\mu$m (from the PACS Observer’s Manual, Version 2.5.1), as a lower limit in the SED modeling.}
\end{deluxetable*}

\begin{deluxetable*}{lrrrrrr}
\tabletypesize{\scriptsize}
\tablecolumns{7}
\tablewidth{0pt}
\tablecaption{Herschel-PACS Observational Parameters of Sources in Sgr~C \label{tb:CPACS}}
\tablehead{\colhead{  }&
           \multicolumn{3}{c}{${\rm 70\mu{m}}$}&
           \multicolumn{3}{c}{${\rm 160\mu{m}}$}\\[-4pt]
           \cmidrule(lr){2-4} \cmidrule(lr){5-7}\\[-12pt]
           \colhead{ Source }&
           \colhead{ $R_{\rm int}$ } &
           \colhead{ $F_{\rm int}$ } &
           \colhead{ $F_{\rm int-bg}$ } &
           \colhead{ $R_{\rm int}$ } &
           \colhead{ $F_{\rm int}$ } &
           \colhead{ $F_{\rm int-bg}$ } \\[-6pt]
	   \colhead{  } &
	   \colhead{ ($\arcsec$) } &
	   \colhead{ (Jy) } &
	  \colhead{ (Jy) } &
	   \colhead{ ($\arcsec$) } &
	   \colhead{ (Jy) } &
	   \colhead{ (Jy) } \\[-12pt]
}
\startdata
\multicolumn{7}{c}{Compact Sources} \\
\hline
Sgr C 1	    &	16.0	&	$<$433	&	UR	&	22.4	&	$<$841	&	UD	\\
Sgr C 2	    &	16.0	&	$<$368	&	UD	&	22.4	&	$<$737	&	UD	\\
Sgr C 3	    &	16.0	&	$<$402	&	UR	&	22.4	&	$<$721	&	UR	\\
Sgr C 4	    &	16.0	&	$<$505	&	UD	&	22.4	&	$<$736	&	UD	\\
Sgr C 5	    &	16.0	&	$<$573	&	UR	&	22.4	&	$<$776	&	UR	\\
Sgr C 6	    &	16.0	&	$<$365	&	UR	&	22.4	&	$<$794	&	UD	\\
Sgr C 7	    &	16.0	&	$<$605	&	UR	&	22.4	&	$<$717	&	UD	\\
Sgr C 8	    &	16.0	&	$<$618	&	UR	&	22.4	&	$<$681	&	UD	\\
Sgr C 9	    &	16.0	&	$<$352	&	UR	&	22.4	&	$<$798	&	UR	\\
Sgr C 10	&	16.0	&	$<$382	&	UR	&	22.4	&	$<$736	&	UD	\\
Sgr C 11	&	16.0	&	$<$417	&	UR	&	22.4	&	$<$794	&	UR	\\
Sgr C 12	&	16.0	&	$<$470	&	UR	&	22.4	&	$<$674	&	UD	\\
Sgr C 13	&	16.0	&	$<$805	&	UR	&	22.4	&	$<$1850	&	UD	\\
Sgr C 14	&	16.0	&	$<$951	&	UR	&	22.4	&	$<$2000	&	UD	\\
Sgr C H3	&	16.0	&	$<$1020	&	UR	&	28.8	&	$<$2870	&	UR	\\
Sgr C H4	&	16.0	&	$<$696	&	UR	&	22.4	&	$<$2270	&	UR	\\
Sgr C H1	&	16.0	&	$<$369	&	UR	&	22.4	&	$<$1860	&	UR	\\
Sgr C 15	&	16.0	&	$<$350	&	UR	&	22.4	&	$<$703	&	UD	\\
Sgr C 16	&	16.0	&	$<$363	&	UR	&	22.4	&	$<$676	&	UD	\\
\hline
\multicolumn{7}{c}{Extended Sources} \\
\hline
HII Region	&	83.2	&	13000	&	12200	&	83.2	&	12000	&	11700	\\	
\enddata
\tablecomments{UR and UD have same meaning as discussion in caption of Table~\ref{tb:BIRAC}. For these sources, the $F_{\rm int}$ value is used as the upper limit in the SED modeling. }
\end{deluxetable*}

\vspace{-32mm}

\section{Observational and Derived Parameters For Other Regions Covered in the SOFIA Data}\label{sec:appendixD}

Within the confines of the FORCAST data covering the Sgr~B1 and Sgr~B2 G\ion{H}{2} regions, there is an additional region named G0.6-0.0 where we find several MYSO candidates. Since knowledge of the existence of any YSOs in the Galactic Central Molecular Zone is of interest, we have derived photometry values and run SED models for all MYSO candidates in G0.6-0.0. Likewise, we defined three regions likely not directly related to the Sgr~C G\ion{H}{2} region but close enough to be within the confines of our FORCAST maps: G359.38-0.08, G359.42-0.02, and G359.50-0.09. All four of these satellite regions also contain MYSO candidates, and so they too have had their photometry values measured and modeled. 

\begin{figure*}[t]
\epsscale{0.98}
\plotone{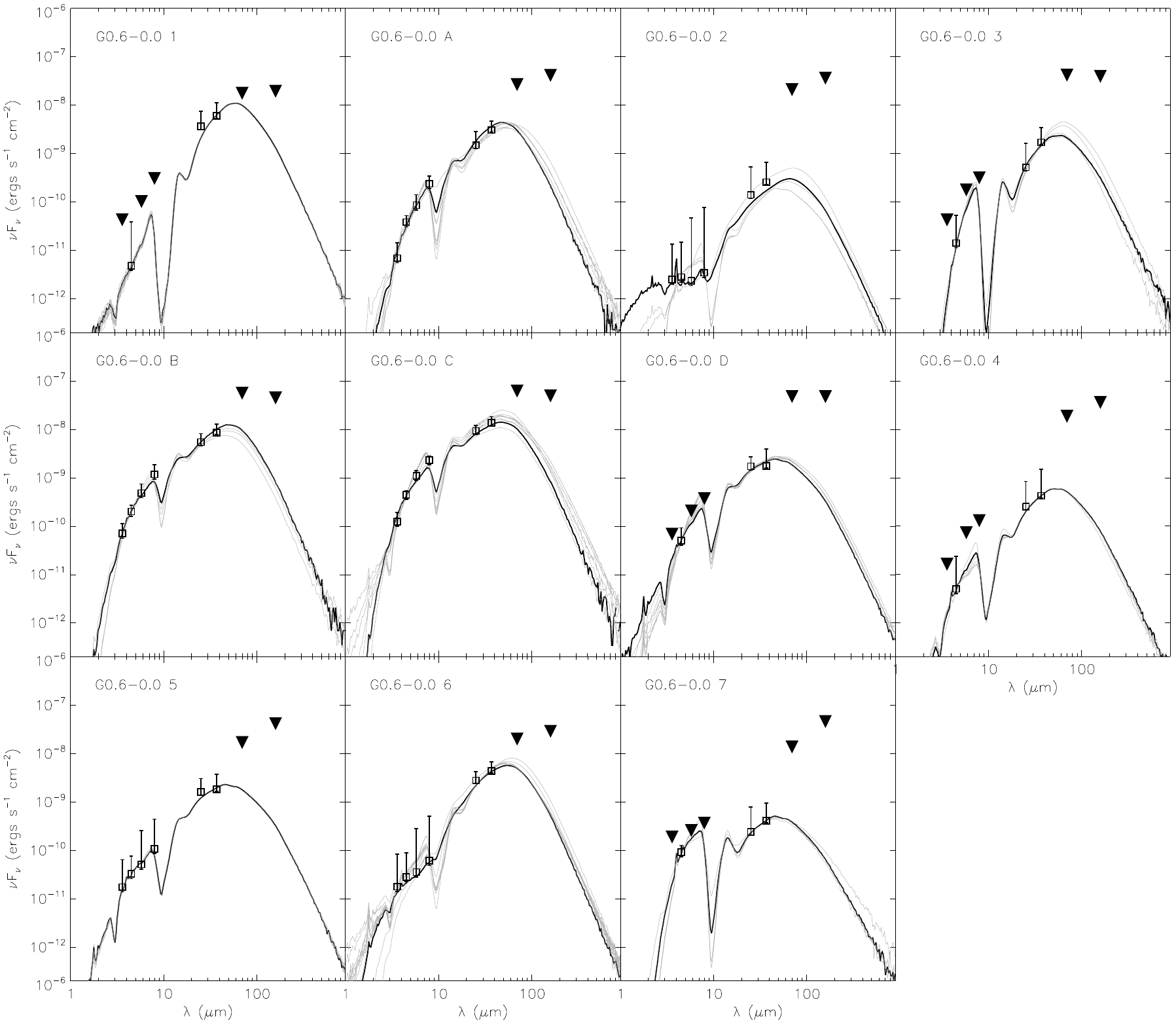}
\caption{SED fitting with ZT model for compact sources in G0.6-0.0. Black lines are the best fit model to the SEDs, and the system of gray lines are the remaining fits in the group of best fits (from Table \ref{tb:Cseds}). Upside-down triangles are data that are used as upper limits in the SED fits, and triangles are lower limits.\label{fig:GSED2}}
\end{figure*}

For all four regions, Table \ref{tb:others_S} gives the photometry data for all sources identified in the SOFIA-FORCAST data. Table \ref{tb:others_I} and Table \ref{tb:others_H} give the photometry information for those same sources from the Spitzer-IRAC data and Herschel-PACS data, respectively. While the resultant physical values of the SED model fitting are given in Table~\ref{tb:OTHERseds} for all satellite regions, Figure \ref{fig:GSED2} displays the SED plots for all compact mid-infrared sources in G0.6-0.0, whereas Figure \ref{fig:CSED2} displays them for all of the compact sources in G359.38-0.08, G359.42-0.02, and G359.50-0.09.

The results of these SED model fits imply that there is a significant level of massive star formation that may be occurring outside of but near the major sites of star formation (i.e., Sgr~B1, Sgr~B2, and Sgr~C). In total, 39 compact mid-infrared sources are found throughout the combined area of these four (non-G\ion{H}{2}) regions, with 1 confirmed MYSO, 8 Likely MYSOs, and 18 Possible MYSOs (see Table~\ref{tb:OTHERseds}).

\subsection{G0.6-0.0 MYSO Candidates}

The close (but likely separate) region just to the south of Sgr~B2, G0.6-0.0,  has 11 compact mid-infrared sources, of which 4 were previously identified in radio cm continuum \citep[e.g.,][]{1992ApJ...401..168M}. We identify 7 new compact infrared sources, one (G0.6-0.0~2) which is likely a low or intermediate mass YSO (based on the SED fitting) and one with radio continuum that was not previously identified in radio studies (G0.6-0.0~4). In this areas alone we find one confirmed MYSO, 4 Likely MYSOs, and 5 Possible MYSOs (Table~\ref{tb:OTHERseds}). In all, 91\% of the mid-infrared sources in G0.6-0.0 are MYSOs or MYSO Candidates. The most massive MYSO candidates are G0.6-0.0~B and G0.6-0.0~C, both with a best fit mass of 64\,$M_{\sun}$, with G0.6-0.0~C having the highest fit of all sources we studied in the CMZ with a value of 96\,$M_{\sun}$ in its range of best fit masses. 

\vspace{-3mm}
\vspace{1mm}

\subsection{G359.38-0.08 MYSO Candidates}

Southwest of the Sgr~C G\ion{H}{2} region, G359.38-0.08 has 10 compact mid-infrared sources among some faint and extended dust emission regions (interspersed with diffuse cm radio continuum emission as well), which may signal that this is a legitimate star-forming region. There are no confirmed MYSOs among the mid-infrared sources, but we do find 4 sources that we consider to be Likely MYSOs based on the presence of cm radio continuum emission (Table~\ref{tb:OTHERseds}). An additional 3 sources are listed as Possible MYSOs. We include G359.38-0.08~6 among the Possible MYSO candidates, even though its best fit model is only 2\,$M_{\sun}$, because the other 14 fits in the group of best fits are 8-16\,$M_{\sun}$. In all, 70\% of the mid-infrared sources in G359.38-0.08 are MYSOs or MYSO Candidates. The most massive MYSO candidate is G359.38-0.08~10 with a best fit mass of 48\,$M_{\sun}$.

\subsection{G359.42-0.02 MYSO Candidates}

G359.42-0.02 is located west of the Sgr~C G\ion{H}{2} region, and while we find 8 compact mid-infrared sources here, it does not contain any other star-formation region signatures (e.g., an IRDC, molecular cloud, extended cm radio continuum or mid-infrared continuum regions, or masers). This region has no confirmed MYSOs or Likely MYSOs, and only contains 4 Possible MYSOs. The other half (4) of the mid-infrared sources are not thought to be MYSOs. G359.42-0.02~1, though it is well fit with MYSO models and is the only source in the region with cm radio continuum emission, it was not found to have a 15.4\,$\mu$m ice feature in its spectrum, which is a YSO indicator \citep{2011ApJ...736..133A}. It may be a rare MYSO without this spectral feature, and thus it is labeled as not being an MYSO with a question mark in Table~\ref{tb:OTHERseds}. G359.42-0.02~2 has a decreasing flux with infrared wavelength, atypical of MYSOs, and indeed is coincident with a (likely foreground) KM giant star \citep{2022ApJ...930...16J}. Sources G359.42-0.02~6 and G359.42-0.02~8 have flat fluxes with wavelength, again atypical of MYSOs, and SED fits that are less than 8\,$M_{\sun}$, and thus are not likely to be MYSOs. In all, 50\% of the mid-infrared sources in G359.42-0.02 are MYSO Candidates, with the most massive being G359.42-0.02~3 with a mass of 32\,$M_{\sun}$.

\subsection{G359.50-0.09 MYSO Candidates}

Located northeast of the Sgr~C G\ion{H}{2} region, G359.50-0.09 hosts 10 compact mid-infrared sources here, but like G359.42-0.02, it does not contain very many other star-formation region signatures, except perhaps some faint and diffuse cm radio continuum emission. This region has no confirmed MYSOs or Likely MYSOs, and only contains 6 Possible MYSOs. The other 4 sources are believed to not be MYSOs (Table~\ref{tb:OTHERseds}). Both G359.50-0.09~2 and G359.50-0.09~8 were found to not have a 15.4\,$\mu$m ice feature in their spectra \citep{2011ApJ...736..133A}, indicating they are likely not YSOs. Indeed, G359.50-0.09~8 was found by \citet{2022ApJ...930...16J} to be a KM giant star. Both G359.50-0.09~5 and G359.50-0.09~9 were found to have decreasing flux with increasing infrared wavelength, atypical of MYSOs, with the latter source also having SiO maser emission more typical of a AGB star. In all, 60\% of the mid-infrared sources in G359.50-0.09 are MYSO Candidates, with the most massive being G359.50-0.09~4 with a mass of 32\,$M_{\sun}$.

\begin{deluxetable*}{lccccccccc}
\tabletypesize{\scriptsize}
\tablecolumns{10}
\tablewidth{0pt}
\tablecaption{SOFIA Observational Parameters of Sources in Other Regions \label{tb:others_S}}
\tablehead{\colhead{  }&
           \colhead{  }&
           \colhead{  }&
           \multicolumn{3}{c}{${\rm 25\mu{m}}$}&
           \multicolumn{3}{c}{${\rm 37\mu{m}}$} &
           \colhead{  }\\[-4pt]
           \cmidrule(lr){4-6} \cmidrule(lr){7-9}\\[-12pt]
           \colhead{ Source }&
           \colhead{ R.A.}&
           \colhead{ Decl. }&           
           \colhead{ $R_{\rm int}$ } &
           \colhead{ $F_{\rm int}$ } &
           \colhead{ $F_{\rm int-bg}$ } &
           \colhead{ $R_{\rm int}$ } &
           \colhead{ $F_{\rm int}$ } &
           \colhead{ $F_{\rm int-bg}$ } &
           \colhead{ Aliases }\\[-6pt]
	   \colhead{  } &
          \colhead{(J2000) }&
          \colhead{(J2000) }& 	   
	   \colhead{ ($\arcsec$) } &
	   \colhead{ (Jy) } &
	   \colhead{ (Jy) } &
	   \colhead{ ($\arcsec$) } &
	   \colhead{ (Jy) } &
	   \colhead{ (Jy) } &
           \colhead{  }\\[-12pt]
}
\startdata
\multicolumn{9}{c}{G0.6-0.0 Compact Sources} \\
\hline
G0.6-0.0 1	&	17 47 09.90	&	-28 27 54.2	&	14.6	&	62.1	&	30.8	&	16.9	&	139	    &	73.8 &	\\
G0.6-0.0 A	&	17 47 11.28	&	-28 26 31.2	&	8.4	    &	24.0	&	12.6	&	10.0	&	57.9	&	37.8 &	\\
G0.6-0.0 2	&	17 47 13.08	&	-28 26 05.0	&	6.1	    &	4.49	&	1.19	&	6.1	    &	8.19	&	3.17 & SSTGC773332$^{\ddagger}$ \\
G0.6-0.0 3	&	17 47 13.15	&	-28 27 00.3	&	5.4	    &	13.7	&	4.36	&	5.4	    &	42.5	&	21.0 &	\\
G0.6-0.0 B	&	17 47 13.73	&	-28 26 53.4	&	6.9	    &	68.9	&	45.9	&	6.9	    &	158	    &	106 &	\\
G0.6-0.0 C	&	17 47 14.67	&	-28 26 48.7	&	7.7	    &	103	    &	78.0	&	7.7	    &	228	    &	168 &	\\
G0.6-0.0 D	&	17 47 15.42	&	-28 26 42.5	&	6.1	    &	22.5	&	14.5	&	6.1	    &	48.7	&	21.6 &	\\
G0.6-0.0 4	&	17 47 15.76	&	-28 26 10.3	&	6.1	    &	7.11	&	2.14	&	6.9	    &	18.6	&	5.24 &	SSTGC779211$^{\ddagger}$\\
G0.6-0.0 5	&	17 47 16.81	&	-28 26 11.0	&	10.7	&	25.5	&	13.5	&	10.7	&	46.1	&	22.5 &	\\
G0.6-0.0 6	&	17 47 18.75	&	-28 27 28.5	&	11.5	&	35.8	&	23.4	&	12.3	&	83.3	&	53.9 &	\\
G0.6-0.0 7	&	17 47 20.13	&	-28 26 06.3	&	6.9	    &	6.57	&	2.01	&	8.4	    &	11.6	&	5.06 &	\\
\hline
\multicolumn{9}{c}{G359.38-0.08 Compact Sources} \\
\hline
G359.38-0.08 1	&	17 44 22.68	&	-29 32 09.5	&	12.3	&	19.0	&	8.38	&	15.3	&	43.0	&	32.2 &	\\
G359.38-0.08 2	&	17 44 23.33	&	-29 31 01.1	&	10.0	&	15.6	&	6.67	&	9.2	    &	25.3	&	9.24 &	\\
G359.38-0.08 3	&	17 44 23.92	&	-29 30 44.3	&	3.1	    &	1.48	&	0.403	&	5.4	    &	7.83	&	1.37 &	\\
G359.38-0.08 4	&	17 44 24.86	&	-29 29 34.4	&	4.6	    &	2.13	&	0.80	&	3.8	    &	1.37	&	0.659 &	\\
G359.38-0.08 5	&	17 44 25.80	&	-29 30 46.6	&	15.3	&	61.5	&	26.6	&	15.3	&	144	    &	59.3 &	\\
G359.38-0.08 6	&	17 44 26.57	&	-29 30 02.8	&	9.2	    &	14.6	&	3.03	&	10.0	&	38.3	&	6.99 &	SSTGC330325$^{\ddagger}$\\
G359.38-0.08 7	&	17 44 28.74	&	-29 29 55.1	&	12.3	&	33.3	&	13.6	&	13.8	&	73.1	&	22.0 &	\\
G359.38-0.08 8	&	17 44 29.10	&	-29 30 29.7	&	6.9	    &	5.31	&	1.09	&	6.9	    &	15.2	&	4.81 &	\\
G359.38-0.08 9	&	17 44 31.27	&	-29 29 36.7	&	9.2	    &	11.3	&	1.79	&	10.7	&	33.5	&	4.68 &	\\
G359.38-0.08 10	&	17 44 33.45	&	-29 29 50.5	&	10.7	&	20.2	&	8.72	&	10.7	&	37.6	&	13.4 &	\\	
\hline
\multicolumn{9}{c}{G359.38-0.08 Extended Sources} \\
\hline
G359.38-0.08 HII	&	17 44 28.03	&	-29 30 19.2	&	82.1	&	731	&	550	&	82.1	&	1300	&	1270 & \\
\hline
\multicolumn{9}{c}{G359.42-0.02 Compact Sources} \\
\hline
G359.42-0.02 1	&	17 44 11.22	&	-29 26 38.4	&	8.4	    &	7.14	&	7.12	&	8.4	    &	16.3	&	10.4 &	SSTGC293528$^{\dagger}$\\
G359.42-0.02 2	&	17 44 12.69	&	-29 26 56.1	&	5.4	    &	5.82	&	5.50    &	6.9	    &	7.59	&	5.02 & 	\\
G359.42-0.02 3	&	17 44 17.69	&	-29 27 12.3	&	10.7	&	23.3	&	20.1	&	11.5	&	53.2	&	44.6 &	\\
G359.42-0.02 4	&	17 44 19.10	&	-29 27 26.1	&	7.7	    &	3.34	&	1.83	&	7.7	    &	8.96	&	3.05 &	\\
G359.42-0.02 5	&	17 44 21.45	&	-29 27 36.9	&	13.8	&	32.2	&	21.5	&	15.3	&	81.0	&	59.3 &	\\
G359.42-0.02 6	&	17 44 22.57	&	-29 25 17.9	&	10.0	&	5.92	&	5.80	&	10.0	&	9.11	&	8.07 &	SSTGC320517$^{\ddagger}$\\
G359.42-0.02 7	&	17 44 22.98	&	-29 27 04.6	&	5.4	    &	3.01	&	1.67	&	6.9	    &	4.78	&	3.09 & 	\\
G359.42-0.02 8	&	17 44 28.21	&	-29 26 33.1	&	3.1	    &	1.44	&	0.833	&	3.8	    &	1.41	&	1.18 &	SSTGC335380$^{\ddagger}$\\
\hline
\multicolumn{9}{c}{G359.50-0.09 Compact Sources} \\
\hline																
G359.50-0.09 1	&	17 44 41.38	&	-29 25 12.4	&	10.0	&	10.0	&	4.44	&	10.0	&	30.2	&	10.2 &	\\
G359.50-0.09 2	&	17 44 41.55	&	-29 24 30.2	&	14.6	&	38.9	&	25.5	&	15.3	&	97.5	&	44.1 & SSTGC368854$^{\dagger}$	\\
G359.50-0.09 3	&	17 44 41.79	&	-29 23 31.8	&	6.1	    &	3.57	&	2.12	&	9.2	    &	14.4	&	6.97 & SSTGC370438$^{\ddagger,\dagger}$	\\
G359.50-0.09 4	&	17 44 42.73	&	-29 23 15.7	&	6.1	    &	3.00	&	1.89	&	6.9	    &	7.62	&	3.27 & SSTGC372630$^{\ddagger,\dagger}$\\
G359.50-0.09 5	&	17 44 44.49	&	-29 23 21.0	&	3.8	    &	0.887	&	0.518	&	3.8	    &	1.81	&	0.576 &	\\
G359.50-0.09 6	&	17 44 46.78	&	-29 23 44.8	&	6.1	    &	9.15	&	5.82	&	6.1	    &	16.1	&	11.5 &	\\
G359.50-0.09 7	&	17 44 47.02	&	-29 23 35.6	&	6.1	    &	8.15	&	5.35	&	6.1	    &	14.4	&	8.39 &	\\
G359.50-0.09 8	&	17 44 48.84	&	-29 23 41.7	&	7.7	    &	12.7	&	7.04	&	7.7	    &	22.0	&	12.7 & SSTGC388790$^{\dagger}$\\
G359.50-0.09 9	&	17 44 51.19	&	-29 24 52.4	&	5.4	    &	17.2	&	15.8	&	6.1	    &	9.16	&	8.24 &	\\
G359.50-0.09 10	&	17 44 53.13	&	-29 24 31.6	&	3.8	    &	3.27	&	1.55	&	6.1	    &	4.67	&	3.25 & SSTGC400062$^{\ddagger}$	\\
\enddata
\tablenotetext{\dagger}{From \citet{2011ApJ...736..133A}.}
\tablenotetext{\ddagger}{From \citet{2009ApJ...702..178Y}.}
\end{deluxetable*}

\begin{deluxetable*}{lrrrrrrrrrrrr}
\tabletypesize{\scriptsize}
\tablecaption{Spitzer-IRAC Observational Parameters of Sources in Other Regions \label{tb:others_I}}
\tablehead{\colhead{  }&
           \multicolumn{3}{c}{${\rm 3.6\mu{m}}$}&
           \multicolumn{3}{c}{${\rm 4.5\mu{m}}$}&
           \multicolumn{3}{c}{${\rm 5.8\mu{m}}$}&
           \multicolumn{3}{c}{${\rm 8.0\mu{m}}$}\\[-4pt]
           \cmidrule(lr){2-4} \cmidrule(lr){5-7} \cmidrule(lr){8-10} \cmidrule(lr){11-13}\\[-12pt]
           \colhead{ Source }&
           \colhead{ $R_{\rm int}$ } &
           \colhead{ $F_{\rm int}$ } &
           \colhead{ $F_{\rm int-bg}$ } &
                      \colhead{ $R_{\rm int}$ } &
           \colhead{ $F_{\rm int}$ } &
           \colhead{ $F_{\rm int-bg}$ } &
                      \colhead{ $R_{\rm int}$ } &
           \colhead{ $F_{\rm int}$ } &
           \colhead{ $F_{\rm int-bg}$ } &
                      \colhead{ $R_{\rm int}$ } &
           \colhead{ $F_{\rm int}$ } &
           \colhead{ $F_{\rm int-bg}$ } \\[-6pt]
	   \colhead{  } &
	   \colhead{ ($\arcsec$) } &
	   \colhead{ (mJy) } &
	   \colhead{ (mJy) } &
	   \colhead{ ($\arcsec$) } &
	   \colhead{ (mJy) } &
	   \colhead{ (mJy) } &
	   \colhead{ ($\arcsec$) } &
	   \colhead{ (mJy) } &
	   \colhead{ (mJy) } &
	   \colhead{ ($\arcsec$) } &
	   \colhead{ (mJy) } &
	   \colhead{ (mJy) } \\[-12pt]
}
\startdata
\multicolumn{13}{c}{G0.6-0.0 Compact Sources} \\
\hline
G0.6-0.0 1	&	4.8	&	51.1	&	7.31	&	4.8	&	58.3	&	7.17	&	4.8	&	197	    &	22.0	&	4.8	&	811	    &	309	\\
G0.6-0.0 A	&	4.8	&	17.2	&	8.26	&	6.0	&	77.7	&	57.4	&	6.0	&	272	    &	165	    &	6.0	&	907	    &	635	\\
G0.6-0.0 2	&	4.8	&	16.2	&	2.99	&	4.8	&	22.1	&	4.18	&	4.8	&	89.3	&	4.49	&	4.8	&	203	    &	9.18	\\
G0.6-0.0 3	&	4.8	&	50.8	&	12.9	&	4.8	&	79.0	&	21.1	&	4.8	&	341	    &	85.2	&	4.8	&	837	    &	156	\\
G0.6-0.0 B	&	6.0	&	137	    &	84.0	&	6.0	&	405	    &	301	    &	6.0	&	1430	&	930	    &	7.2	&	5000	&	3150	\\
G0.6-0.0 C	&	7.2	&	229	    &	148	    &	7.2	&	806	    &	662	    &	7.2	&	2720	&	2120	&	7.2	&	7490	&	6210	\\
G0.6-0.0 D	&	4.8	&	81.9	&	55.3	&	4.8	&	138	    &	75.9	&	4.8	&	398	    &	159	    &	4.8	&	985	    &	420	\\
G0.6-0.0 4	&	4.8	&	$<$19.6	&	UD	    &	4.8	&	35.7	&	7.58	&	4.8	&	141	    &	12.3	&	4.8	&	345	    &	63.2	\\
G0.6-0.0 5	&	7.2	&	76.6	&	20.8	&	7.2	&	115	    &	48.6	&	8.4	&	494	    &	98.5	&	8.4	&	1170	&	285	\\
G0.6-0.0 6	&	7.2	&	99.6	&	21.2	&	7.2	&	133	    &	41.9	&	7.2	&	546	    &	68.3	&	7.2	&	1360	&	161	\\
G0.6-0.0 7	&	8.4	&	225	    &	167	    &	8.4	&	190	    &	136	    &	8.4	&	494	    &	190	    &	8.4	&	970	    &	304	\\
\hline
\multicolumn{13}{c}{G359.38-0.08 Compact Sources} \\
\hline																								
G359.38-0.08 1	&	12.0	&	$<$388	&	UR	    &	12.0	&	$<$445	&	UR	    &	12.0	&	1480	&	335	    &	12.0	&	3910	&	683	\\
G359.38-0.08 2	&	3.0	    &	$<$17.4	&	UD	    &	3.0	    &	$<$22.6	&	UD	    &	5.4	    &	252	    &	34.9	&	6.6	    &	999	    &	82.5	\\
G359.38-0.08 3	&	6.0	    &	$<$55.4	&	UD	    &	6.0	    &	76.3	&	15.2	&	6.6	    &	359	    &	48.1	&	7.2	    &	1200	&	121	\\
G359.38-0.08 4	&	3.6	    &	700	    &	669	    &	3.6	    &	708	    &	678	    &	4.2	    &	1140	&	980	    &	4.2	    &	913	    &	539	\\
G359.38-0.08 5	&	4.8	    &	$<$87.3	&	UD	    &	5.4	    &	150	    &	90.3	&	7.8	    &	756	    &	246	    &	10.8	&	3480	&	603	\\
G359.38-0.08 6	&	3.0	    &	14.3	&	3.03	&	4.2	    &	34.9	&	6.80	&	6.6	    &	379	    &	44.7	&	7.2	    &	1230	&	141	\\
G359.38-0.08 7	&	12.0	&	$<$4900	&	UR	    &	12.0	&	$<$556	&	UR	    &	12.0	&	$<$1310	&	UD	    &	12.0	&	$<$2830	&	UD	\\
G359.38-0.08 8	&	6.6	    &	$<$73.1	&	UR	    &	6.6	    &	$<$70.3	&	UD	    &	6.6	    &	417	    &	93.4	&	7.2	    &	1320	&	333	\\
G359.38-0.08 9	&	9.0	    &	$<$1600	&	UR	    &	9.0	    &	$<$169	&	UD	    &	9.0	    &	$<$579	&	UD	    &	9.0	    &	$<$1680	&	UD	\\
G359.38-0.08 10	&	5.4	    &	75.3	&	20	    &	10.8	&	348	    &	90.9	&	10.8	&	1310	&	321	    &	10.8	&	3520	&	835	\\
\hline
\multicolumn{13}{c}{G359.38-0.08 Extended Sources} \\
\hline
G359.38-0.08 HII	&	82.2	&	14800	&	3460	&	82.2	&	16100	&	2580	&	82.2	&	55600	&	20300	&	82.2	&	141000	&	51300	\\
\hline
\multicolumn{13}{c}{G359.42-0.02 Compact Sources} \\
\hline																								
G359.42-0.02 1	&	3.6	    &	32.8	&	10.4	&	7.8	    &	141	    &	31.5	&	7.8	    &	588	    &	95.5	&	8.4	    &	1860	&	360	\\
G359.42-0.02 2	&	sat	    &	sat	    &	sat	    &	sat	    &	sat	    &	sat	    &	6.6	    &	5330	&	4860	&	sat	    &	sat	    &	sat	\\
G359.42-0.02 3	&	sat	    &	sat	    &	sat	    &	sat	    &	sat	    &	sat	    &	6.0	    &	2270	&	1920	&	6.6	    &	3670	&	2610	\\
G359.42-0.02 4	&	3.6	    &	$<$29.0	&	UR	    &	3.6	    &	38.6	&	13.6	&	4.2	    &	197	    &	27.4	&	4.2	    &	546	    &	84.2	\\
G359.42-0.02 5	&	10.8	&	$<$385	&	UR	    &	10.8	&	396	    &	184	    &	12.0	&	1630	&	423	    &	13.8	&	5470	&	852	\\
G359.42-0.02 6	&	9.6	    &	883	    &	UR	    &	9.6	    &	$<$761	&	UR	    &	9.6	    &	$<$1380	&	UR	    &	9.6	    &	2660	&	709	\\
G359.42-0.02 7	&	3.6	    &	25.1	&	9.85	&	4.2	    &	38.1	&	15.9	&	4.2	    &	138	    &	27.7	&	5.4	    &	629	    &	63.5	\\
G359.42-0.02 8	&	4.8	    &	$<$38.8	&	UR	    &	4.8	    &	66.2	&	40.0	&	4.8	    &	209	    &	90.6	&	4.8	    &	626	    &	243	\\
\hline
\multicolumn{13}{c}{G359.50-0.09 Compact Sources} \\
\hline																									
G359.50-0.09 1	&	10.2	&	$<$330	&	UR	    &	10.2	&	$<$320	&	UR	    &	10.2	&	1790	&	384	    &	10.2	&	3860	&	327	\\
G359.50-0.09 2	&	12.0	&	$<$598	&	UR	    &	12.0	&	$<$595	&	UR	    &	12.0	&	2360	&	433	    &	14.4	&	8690	&	1520	\\
G359.50-0.09 3	&	5.4	    &	$<$59.4	&	UR	    &	5.4	    &	71.0	&	21.3	&	5.4	    &	344	    &	39.7	&	6.0	    &	1190	&	194	\\
G359.50-0.09 4	&	4.2	    &	62.7	&	23.8	&	6.0	    &	150	    &	64.8	&	7.2	    &	693	    &	164	    &	7.2	    &	1850	&	402	\\
G359.50-0.09 5	&	3.6	    &	318	    &	282	    &	4.2	    &	558	    &	509	    &	4.8	    &	1100	&	869	    &	5.4	    &	1490	&	656	\\
G359.50-0.09 6	&	5.4	    &	$<$500	&	UR	    &	5.4	    &	519	    &	406	    &	5.4	    &	1040	&	552	    &	5.4	    &	2030	&	1060	\\
G359.50-0.09 7	&	4.2	    &	107	    &	51.2	&	4.2	    &	129	    &	70.1	&	5.4	    &	691	    &	345	    &	5.4	    &	1670	&	660	\\
G359.50-0.09 8	&	3.6	    &	82.8	&	45.4	&	4.8	    &	170	    &	115	    &	4.8	    &	479	    &	223	    &	5.4	    &	1470	&	579	\\
G359.50-0.09 9	&	sat	    &	sat	    &	sat	    &	sat	    &	sat	    &	sat	    &	sat	    &	sat	    &	sat	    &	sat	    &	sat	    &	sat	\\
G359.50-0.09 10	&	4.8	    &	46.0	&	7.46	&	4.8	    &	59.5	&	17.2	&	6.6	    &	416	    &	73.0	&	6.0	    &	972	    &	156	\\
\enddata
\tablecomments{See table notes from Table \ref{tb:BIRAC}.}
\end{deluxetable*}

\begin{deluxetable*}{lrrrrrr}
\tabletypesize{\scriptsize}
\tablecolumns{7}
\tablewidth{0pt}
\tablecaption{Herschel-PACS Observational Parameters of Sources in Other Regions \label{tb:others_H}}
\tablehead{\colhead{  }&
           \multicolumn{3}{c}{${\rm 70\mu{m}}$}&
           \multicolumn{3}{c}{${\rm 160\mu{m}}$}\\[-4pt]
           \cmidrule(lr){2-4} \cmidrule(lr){5-7}\\[-12pt]
           \colhead{ Source }&
           \colhead{ $R_{\rm int}$ } &
           \colhead{ $F_{\rm int}$ } &
           \colhead{ $F_{\rm int-bg}$ } &
           \colhead{ $R_{\rm int}$ } &
           \colhead{ $F_{\rm int}$ } &
           \colhead{ $F_{\rm int-bg}$ } \\[-6pt]
	   \colhead{  } &
	   \colhead{ ($\arcsec$) } &
	   \colhead{ (Jy) } &
	  \colhead{ (Jy) } &
	   \colhead{ ($\arcsec$) } &
	   \colhead{ (Jy) } &
	   \colhead{ (Jy) } \\[-12pt]
}
\startdata
\multicolumn{7}{c}{G0.6-0.0 Compact Sources} \\
\hline
G0.6-0.0 1	&	16	&	$<$409	&	UR	&	22.4	&	$<$1030	&	UD	\\
G0.6-0.0 A	&	16	&	$<$618	&	UR	&	22.4	&	$<$2190	&	UR	\\
G0.6-0.0 2	&	16	&	$<$486	&	UD	&	22.4	&	$<$1920	&	UD	\\
G0.6-0.0 3	&	16	&	$<$972	&	UR	&	22.4	&	$<$2090	&	UD	\\
G0.6-0.0 B	&	16	&	$<$1300	&	UR	&	22.4	&	$<$2370	&	UD	\\
G0.6-0.0 C	&	16	&	$<$1440	&	UR	&	22.4	&	$<$2610	&	UR	\\
G0.6-0.0 D	&	16	&	$<$1110	&	UR	&	22.4	&	$<$2520	&	UR	\\
G0.6-0.0 4	&	16	&	$<$434	&	UR	&	22.4	&	$<$1910	&	UD	\\
G0.6-0.0 5	&	16	&	$<$392	&	UR	&	22.4	&	$<$2200	&	UD	\\
G0.6-0.0 6	&	16	&	$<$462	&	UR	&	22.4	&	$<$1530	&	UR	\\
G0.6-0.0 7	&	16	&	$<$317	&	UR	&	22.4	&	$<$2440	&	UR	\\
\hline
\multicolumn{7}{c}{G359.38-0.08 Compact Sources} \\
\hline											
G359.38-0.08 1	&	22.4	&	$<$491	&	UR	&	22.4	&	$<$585	&	UD	\\
G359.38-0.08 2	&	16.0	&	$<$306	&	UR	&	22.4	&	$<$766	&	UD	\\
G359.38-0.08 3	&	16.0	&	$<$342	&	UR	&	22.4	&	$<$786	&	UR	\\
G359.38-0.08 4	&	16.0	&	$<$281	&	UR	&	22.4	&	$<$688	&	UD	\\
G359.38-0.08 5	&	16.0	&	$<$378	&	UR	&	22.4	&	$<$786	&	UD	\\
G359.38-0.08 6	&	16.0	&	$<$345	&	UR	&	22.4	&	$<$688	&	UD	\\
G359.38-0.08 7	&	16.0	&	$<$317	&	UR	&	22.4	&	$<$650	&	UD	\\
G359.38-0.08 8	&	16.0	&	$<$331	&	UR	&	22.4	&	$<$710	&	UR	\\
G359.38-0.08 9	&	16.0	&	$<$316	&	UR	&	22.4	&	$<$641	&	UD	\\
G359.38-0.08 10	&	16.0	&	$<$356	&	UR	&	22.4	&	$<$670	&	UD	\\
\hline
\multicolumn{7}{c}{G359.38-0.08 Extended Sources} \\
\hline
HII Region	&	83.2	&	8250	&	7520	&	83.2	&	9710	&	8910	\\
\hline
\multicolumn{7}{c}{G359.42-0.02 Compact Sources} \\
\hline												
G359.42-0.02 1	&	16.0	&	$<$269	&	UR	&	22.4	&	$<$674	&	UD	\\
G359.42-0.02 2	&	16.0	&	$<$284	&	UR	&	22.4	&	$<$699	&	UD	\\
G359.42-0.02 3	&	19.2	&	$<$379	&	UR	&	22.4	&	$<$659	&	UR	\\
G359.42-0.02 4	&	16.0	&	$<$264	&	UD	&	22.4	&	$<$628	&	UD	\\
G359.42-0.02 5	&	16.0	&	$<$312	&	UR	&	22.4	&	$<$686	&	UD	\\
G359.42-0.02 6	&	16.0	&	$<$210	&	UR	&	22.4	&	$<$474	&	UD	\\
G359.42-0.02 7	&	16.0	&	$<$252	&	UR	&	22.4	&	$<$720	&	UD	\\
G359.42-0.02 8	&	16.0	&	$<$233	&	UR	&	22.4	&	$<$853	&	UD	\\
\hline
\multicolumn{7}{c}{G359.50-0.09 Compact Sources} \\
\hline												
G359.50-0.09 1	&	16.0	&	$<$362	&	UR	&	22.4	&	$<$723	&	UD	\\
G359.50-0.09 2	&	22.4	&	$<$737	&	UR	&	22.4	&	$<$758	&	UR	\\
G359.50-0.09 3	&	22.4	&	$<$648	&	UR	&	22.4	&	$<$877	&	UD	\\
G359.50-0.09 4	&	16.0	&	$<$345	&	UR	&	22.4	&	$<$871	&	UD	\\
G359.50-0.09 5	&	16.0	&	$<$320	&	UD	&	22.4	&	$<$745	&	UD	\\
G359.50-0.09 6	&	16.0	&	$<$339	&	UR	&	22.4	&	$<$617	&	UD	\\
G359.50-0.09 7	&	16.0	&	$<$332	&	UR	&	22.4	&	$<$618	&	UD	\\
G359.50-0.09 8	&	16.0	&	$<$307	&	UR	&	22.4	&	$<$642	&	UD	\\
G359.50-0.09 9	&	16.0	&	$<$254	&	UR	&	22.4	&	$<$898	&	UD	\\
G359.50-0.09 10	&	16.0	&	$<$268	&	UR	&	22.4	&	$<$883	&	UR
\enddata
\tablecomments{UR and UD have same meaning as discussion in caption of Table~\ref{tb:BIRAC}. For these sources, the $F_{\rm int}$ value is used as the upper limit in the SED modeling. }
\end{deluxetable*}

\begin{figure*}[t]
\epsscale{0.98}
\plotone{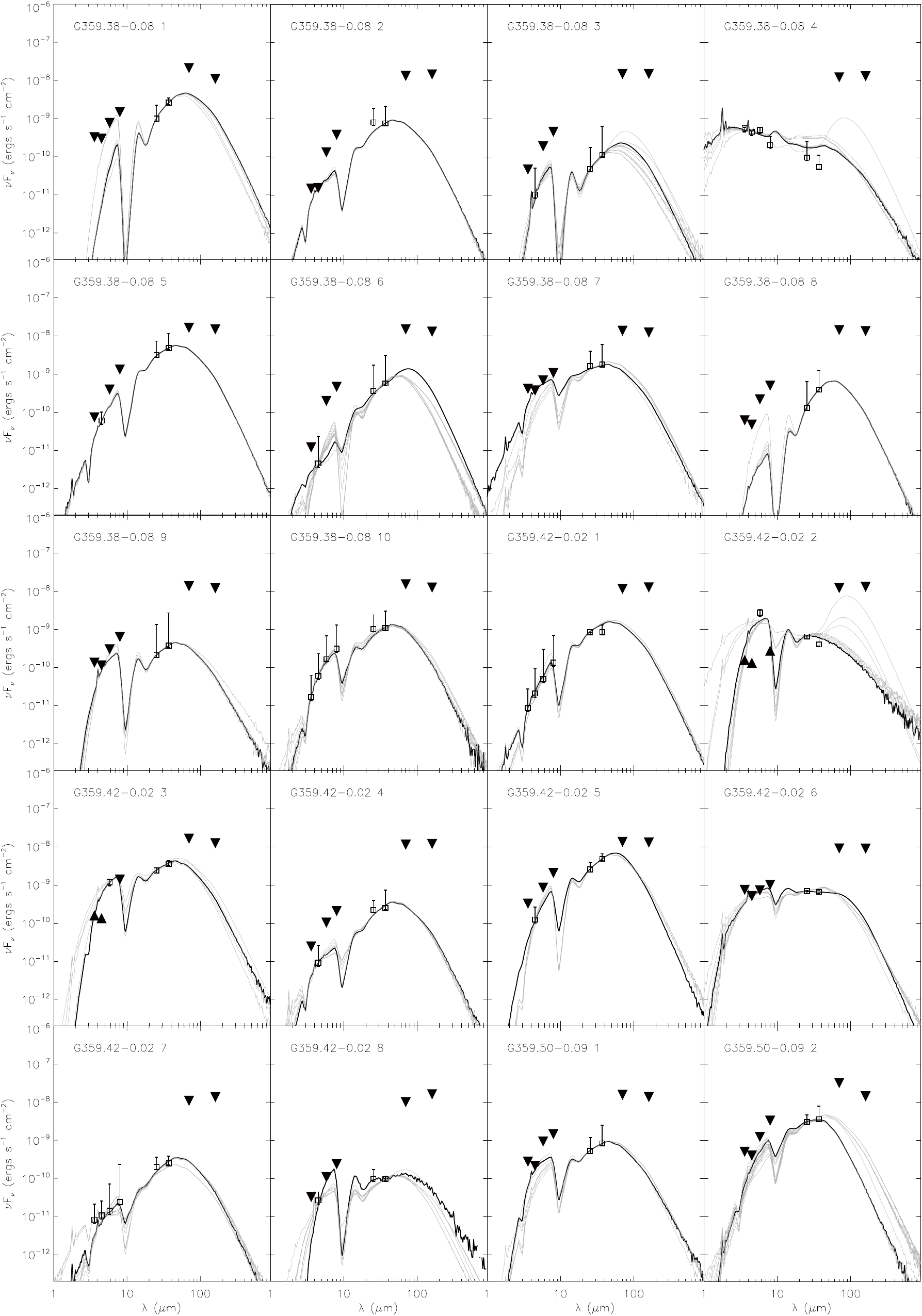}
\caption{SED fitting with ZT model for compact sources in G359.38-0.08, G359.42-0.02, and G359.50-0.09. Black lines are the best fit model to the SEDs, and the system of gray lines are the remaining fits in the group of best fits (from Table \ref{tb:Cseds}). Upside-down triangles are data that are used as upper limits in the SED fits, and triangles are lower limits.\label{fig:CSED2}}
\end{figure*}

\begin{figure*}[t]
\figurenum{18}
\epsscale{1.00}
\plotone{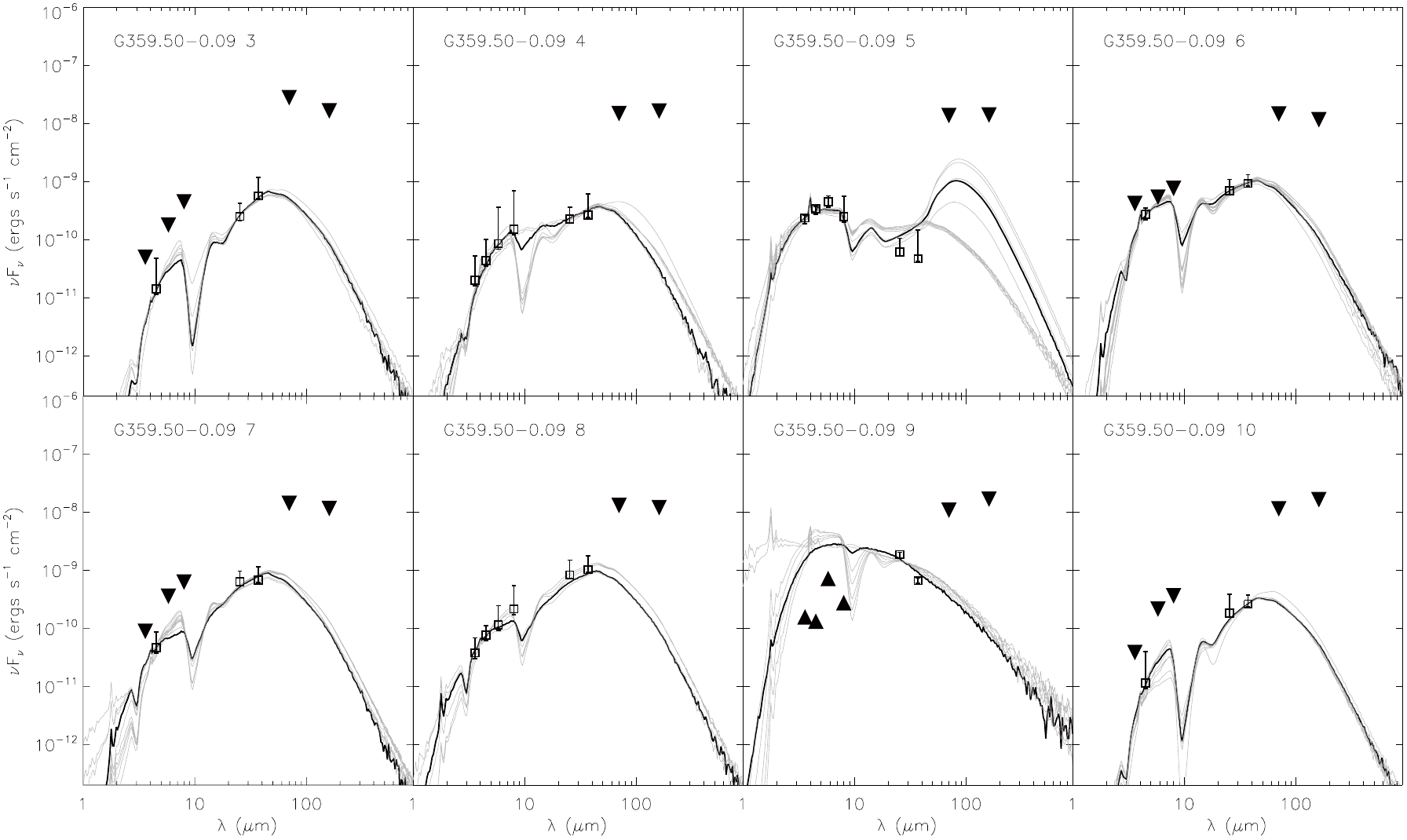}
\caption{\textit{Continued.}}
\end{figure*}

\begin{deluxetable*}{lccccccccccc}[ht]
\tabletypesize{\scriptsize}
\tablecolumns{1}
\tablewidth{0pt}
\tablecaption{SED Fitting Parameters of Compact Infrared Sources in Other Region\label{tb:OTHERseds}}
\tablehead{\colhead{Source   }  &
           \colhead{  $L_{\rm obs}$   } &
           \colhead{  $L_{\rm tot}$   } &
           \colhead{ $A_v$ } &
           \colhead{  $M_{\rm star}$  } &
           \colhead{$A_v$ Range}&
           \colhead{$M_{\rm star}$ Range}&
           \colhead{ Best }&
           \colhead{ Well }&
           \colhead{ Reso- }&
           \colhead{  }&
           \colhead{  }\\[-6pt]
	   \colhead{        } &
	   \colhead{ ($\times 10^3 L_{\sun}$) } &
	   \colhead{ ($\times 10^3 L_{\sun}$) } &
	   \colhead{ (mag.) } &
	   \colhead{ ($M_{\sun}$) } &
       \colhead{(mag.)}&
       \colhead{($M_{\sun}$)}&
       \colhead{  Models   }&
       \colhead{ Fit? }&
           \colhead{ lved? }&
           \colhead{ Features }&
           \colhead{ MYSO? }
}
\startdata
\multicolumn{12}{c}{G0.6-0.0} \\
\hline
G0.6-0.0 1 &    23.3 &     38.1 &      7.9 &     12 &   5.3 -  13 &  12 - 12 & 12 &Y      &Y &70            &Possible   \\ 
G0.6-0.0 A &    10.2 &      189 &       53 &     24 &    27 -  92 &   8 - 32 &  8 &Y      &Y &ice,cm,70     &Yes  \\ 
G0.6-0.0 2 &    0.78 &     1.69 &      8.4 &      4 &   8.4 -  63 &   4 -  8 &  6 &Y      &Y &              &No$^{e}$      \\ 
G0.6-0.0 3 &    5.49 &      158 &      130 &     32 &    98 - 212 &  12 - 32 &  5 &Y      &Y &70            &Possible      \\ 
G0.6-0.0 B &    31.4 &      858 &       53 &     64 &   50  -  80 &  16 - 64 &  5 &Y      &Y &cm,70         &Likely    \\ 
G0.6-0.0 C &    39.1 &      858 &       53 &     64 &   1.7 -  67 &  24 - 96 &  9 &Y      &Y &cm,70         &Likely      \\ 
G0.6-0.0 D &    6.87 &     17.0 &      3.4 &     12 &   3.4 -  16 &  12 - 16 &  6 &Y      &Y &cm,70?        &Likely     \\ 
G0.6-0.0 4 &    1.46 &     19.6 &       40 &     12 &    27 -  76 &   8 - 12 &  6 &Y      &Y &cm,70         &Likely   \\ 
G0.6-0.0 5 &    6.22 &     17.0 &      9.2 &     12 &   8.4 -  13 &  12 - 12 &  6 &Y      &Y &70            &Possible    \\ 
G0.6-0.0 6 &    13.8 &     49.1 &       27 &     12 &   2.6 -  80 &  12 - 24 & 13 &Y      &Y &ice?,70       &Possible    \\ 
G0.6-0.0 7 &    1.55 &     31.4 &      109 &     16 &    55 - 151 &  12 - 32 &  5 &Y      &Y &70            &Possible \\ 
\hline
\multicolumn{12}{c}{G359.38-0.08} \\
\hline
G359.38-0.08 1  &      13.3 &      22.9 &       74 &     12 &    62 - 183   &  12 - 24 &  8 &Y$^a$  &Y &cm,70    &Likely    \\
G359.38-0.08 2  &      2.80 &      9.48 &       24 &      8 &   20  -  27   &   8 -  8 &  7 &N$^a$  &Y &70       &No   \\
G359.38-0.08 3  &      0.69 &      1.69 &      159 &      4 &    85 - 204   &   2 - 32 & 12 &Y      &Y &70       &No$^{e}$   \\ 
G359.38-0.08 4  &      4.34 &      24.7 &      1.7 &      4 &   0.8 -   8.4 &   2 - 16 &  7 &N$^d$  &N &70       &No$^{e}$   \\ 
G359.38-0.08 5  &      17.6 &      34.2 &      4.2 &     16 &   3.4 -   7.5 &  16 - 16 &  7 &Y      &Y &70       &Possible    \\ 
G359.38-0.08 6  &      3.83 &      4.76 &       27 &      2 &    25 - 106   &   2 - 16 & 15 &Y      &Y &70       &Possible?$^{e,f}$  \\ 
G359.38-0.08 7  &      7.52 &      16.1 &      5.9 &     12 &   5.9 -  46   &   8 - 48 &  7 &Y$^a$  &Y &cm,70    &Likely    \\
G359.38-0.08 8  &      1.67 &      11.7 &       64 &      8 &    64 - 176   &   8 - 16 & 10 &Y$^a$  &Y &cm,70    &Likely    \\
G359.38-0.08 9  &      1.62 &      31.4 &      103 &     16 &    98 - 143   &  16 - 32 &  7 &Y$^a$  &Y &cm,70    &Likely    \\
G359.38-0.08 10 &      4.16 &       344 &       37 &     48 &   6.7 -  53   &   8 - 48 & 11 &Y      &Y &70       &Possible    \\
\hline
\multicolumn{12}{c}{G359.42-0.02} \\
\hline
G359.42-0.02 1 &      5.20 &      9.67 &      5.9 &      8 &   1.7 -  23   &   8 -  8 &  7 &Y      &Y &no\,ice,cm,70 &No?$^g$ \\
G359.42-0.02 2 &      5.28 &      1200 &       93 &     96 &   8.4 - 130   &   2 - 96 &  9 &Y$^{a,d}$  &Y &70        &No$^{e,i}$  \\ 
G359.42-0.02 3 &      14.9 &       147 &       63 &     32 &    29 -  82   &  12 - 32 &  5 &Y$^a$  &Y &70            &Possible   \\ 
G359.42-0.02 4 &      1.07 &      13.6 &       38 &     12 &   7.9 -  38   &  12 - 16 &  7 &Y      &Y &              &Possible  \\ 
G359.42-0.02 5 &      19.8 &      47.3 &       59 &     12 &    59 -  78   &  12 - 32 &  5 &Y      &Y &70            &Possible   \\ 
G359.42-0.02 6 &      4.94 &      3.33 &       27 &      4 &    13 -  42   &   4 - 16 &  9 &Y$^{a,d}$  &Y &70          &No$^{e}$   \\ 
G359.42-0.02 7 &      1.04 &      31.4 &       16 &     16 &   0.8 -  52   &   8 - 16 & 11 &Y      &Y &70            &Possible   \\ 
G359.42-0.02 8 &      0.69 &      1200 &       90 &     96 &    37 - 106   &   1 - 96 &  8 &Y$^d$      &N &70        &No$^{e}$  \\ 
\hline
\multicolumn{12}{c}{G359.50-0.09} \\
\hline
G359.50-0.09 1  &      3.36 &     71.4 &       46 &     24 &    24 -  83   &   8 - 24 &  7 &Y$^a$  &Y &70            &Possible   \\ 
G359.50-0.09 2  &      11.6 &      102 &       27 &     16 &   4.2 -  34   &  12 - 32 & 15 &Y$^a$  &Y &no\,ice,70    &No?$^g$ \\
G359.50-0.09 3  &      1.90 &     71.4 &       70 &     24 &    32 -  80   &   8 - 24 &  7 &Y$^a$  &Y &ice?,70       &Possible   \\ 
G359.50-0.09 4  &      1.41 &      158 &       25 &     32 &   8.4 -  53   &   12 - 32 &  8 &Y$^a$  &Y &ice?,70       &Possible  \\ 
G359.50-0.09 5  &      3.41 &     2.31 &       25 &      2 &    17 -  34   &   1 - 16 & 10 &N$^d$  &N &              &No$^{e}$  \\ 
G359.50-0.09 6  &      4.05 &     71.4 &       29 &     24 &    13 -  92   &   8 - 24 & 14 &Y      &Y &70            &Possible   \\ 
G359.50-0.09 7  &      2.76 &     71.4 &       21 &     24 &   1.7 -  38   &   8 - 24 & 10 &Y      &Y &70?           &Possible   \\ 
G359.50-0.09 8  &      3.19 &     71.4 &       11 &     24 &   2.5 -  25   &  16 - 24 &  6 &Y      &Y &no\,ice,70    &No?$^i$ \\
G359.50-0.09 9  &      12.7 &     1200 &       16 &     96 &   2.6 -  72   &  32 - 96 &  8 &Y$^{a,d}$  &N &S,70        &No \\ 
G359.50-0.09 10 &      1.03 &     13.6 &       55 &     12 &    34 - 140   &  12 - 16 &  9 &Y      &Y &70            &Possible   
\enddata
\tablecomments{Symbols same as for Table \ref{tb:Bseds}. However, in this table S=SiO masers from \citet{2006PASJ...58..529F}.}
\tablenotetext{i}{Spectroscopically determined to be a K or M red giant by \citet{2022ApJ...930...16J}.}
\end{deluxetable*}

\end{document}